\documentclass{ptephy_v1}

\preprintnumber{XXXX-XXXX} 
\usepackage{hyperref}

\begin{document}

\title{On the well-posed variational principle in degenerate point particle systems using embeddings of the symplectic manifold}


\author[1]{Kyosuke \sc{Tomonari}}
\affil[1]{Department of Physics, Tokyo Institute of Technology, 2-12-1 Ookayama, Meguro-ku, Tokyo 152-8551, Japan. \email{ktomonari.phys@gmail.com}}

\begin{abstract}
A methodology on making the variational principle well-posed in degenerate systems is constructed. In the systems including higher-order time derivative terms being compatible with Newtonian dynamics, we show that a set of position variables of a coordinate system of a given system has to be fixed on the boundaries and that such systems are always Ostrogradski stable
. For these systems, Frobenius integrability conditions are derived in explicit form. Relationships between integral constants indicated from the conditions and boundary conditions in a given coordinate system are also investigated by introducing three fundamental correspondences between Lagrange and Hamilton formulation. Based on these ingredients, we formulate problems that have to be resolved to realize the well-posedness in the degenerate systems. To resolve the problems, we compose a set of embbedings that extract a subspace holding the symplectic structure of the entire phase space in which the variational principle should be well-posed. Using these embeddings, we establish a methodology to set appropriate boundary conditions that the well-posed variational principle demands. Finally, we apply the methodology to examples and summarize this work as three-step procedure such that one can use just by following it.
\end{abstract}
\subjectindex{A00, A34, E00}
\maketitle

\section{Introduction}\label{Sec01}
The variational principle plays a crucial role in the modern physics to derive the equations of motion for a given system, provided that the boundary term vanishes under certain boundary conditions\cite{Teitelboim1992}. When the system is non-degenerate, the variational principle is applied under imposing the Dirichlet boundary conditions as usual, leading to no problematic situations. However, in degenerate systems, blindly fixing all position coordinates at the boundaries would lead to cumbersome situations since the boundary conditions determine the dynamics of the given system\cite{DyerHinterbichler2009}. 

For instance, let us consider two systems: $L_{1}=\dot{q}^{1}\dot{q}^{3}+q^{2}(q^{3})^{2}/2$, which is the so-called Cawley model\cite{Cawley1979}, and $L_{2}=q^{1}\dot{q}^{2}-q^{2}\dot{q}^{1}-(q^{1})^{2}-(q^{2})^{2}$, which is an imitation of the Dirac system\cite{Dirac1928}. Hamilton-Dirac analysis reveals that $L_{1}$ and $L_{2}$ have three first-class constraints in the six-dimensional phase space and two second-class constraints in the four-dimensional phase space, respectively\cite{Dirac1950,Dirac1958,Bergmann1949,BergmannBrunings1949,Bergmann1950,AndersonBergmann1951}. Therefore, when applying the variational principle to each system, on one hand, $L_{1}$ and $L_{2}$ need to impose up to three boundary conditions and two boundary conditions, respectively. On the other hand, direct computations indicates that, to make the variational principle well-posed, $L_{1}$ and $L_{2}$ need to fix $q^{1}$ and $q^{3}$, and $q^{1}$ and $q^{2}$ on the boundaries, respectively. These conditions, however, are over-imposing. This implies that the integral constants in the solutions that are indicated by Frobenius integrability of the system do not uniquely determine: the dynamics of the system cannot be consistent with the boundary conditions. Furthermore, let us consider another example: $L_{4}:=-q\ddot{q}/2-q^{2}/2$. This example is a modification of the system discussed in \cite{DyerHinterbichler2009} in the literature of Gibbons-Hawking-York term\cite{York1972,GibbonsHawking1977,York1986,HawkingHorowitz1996}. We can find this sort of systems in many gravity theories\cite{Einstein1916,Lovelock1971,HehlMcCreaMielkeNeeman1995,Rosen1940,Rosen1940-2,HassanRosen2012,Buchdahl1970,SotiriouFaraoni2010,Horndeski1974,DeffayetGaoSteerZahariade2011,KobayashiYamaguchiYokoyama2011,GleyzesLangloisPiazzaVernizzi2015,GleyzesLangloisPiazzaVernizzi2015-2,LangloisNoui2016,AchourCrisostomiKoyamaLangloisNouiTasuato2016,DeFeliceLangloisMukohyamaNouiWang2018,Moffat2006}, which contain higher-order derivative terms. Hamilton-Dirac analysis upgraded by \cite{Ostrogradsky1850,Woodard2015,SatoSuganoOhtaKimura1989,SatoSuganoOhtaKimura1989-2,Pons1989} reveals that this system has two second-class constraints in the four-dimensional phase space. However, since the boundary term of the first-order variation of the action integral of $L_{4}$ includes both $\delta q$ and $\delta \dot{q}$, if we blindly fix all configurations then the boundary condition becomes over-imposing; we are stuck with the same situation again. The purpose of this paper is to provide a methodology to resolve these cumbersome situations.

In a previous work\cite{KKKM}, we established a five-step procedure to compose well-posed boundary terms, meaning that the variational principle leads to the equations of motion in the well-posed manner, provided that there are boundary conditions that fix the configurations for the physical degrees of freedom on the boundaries after imposing all of the constraints of the system. Applying the five-step procedure, for instance, $L_{1}$, $L_{2}$ and $L_{4}$ need to fix no configuration, $Q:=(q^{1}+p_{2})/\sqrt{2}$, and $q$, respectively, on the boundaries, where $p_{2}$ is the canonical momentum to $q^{2}$ of $L_{2}$. In this paper, we reconsider the same problem but based on a different philosophy; as many integral constants in the solutions of a given system as possible should be uniquely determined through the boundary conditions.

The construction of this paper is as follows. In Sect. 2, we show that any system, which includes higher-order time derivative systems, being compatible with Newtonian dynamics needs to fix a set of position variables, not velocity or momentum variables, on the boundaries when the variational principle is applied. We also show that such systems are always stable: the Hamiltonian is bounded from below in the sense of Ostrogradski's framework\cite{Ostrogradsky1850,Woodard2015}. In Sect. 3, we review a method to derive Frobenius integrability conditions in a constraint system based on a novel work \cite{SuganoKamo1982} and relate the integral constants in the solutions to the boundary conditions. We then introduce three fundamental correspondences as maps to describe the well-posedness of the variational principle. In Sect. 4, first, we formulate problems to consider the well-posedness based on these maps. Second, we show two lemmas and a theorem in an explicit way by using the concept of a function group that states the existence of a canonical coordinate system being decomposed into constraint and physical coordinates. The former and the latter coordinates are composed only of the constraint conditions of the system, which are derived by using Hamilton-Dirac analysis, and only of the physical degrees of freedom, respectively. These explicit proofs would give a clue to construct such canonical coordinate systems in an explicit manner. Third, we introduce embeddings that restrict the entire phase space to a subspace holding the symplectic structure on which the variational principle should be well-posed. Finally, we construct a methodology to make the variational principle well-posed. In Sect. 5, we apply the methodology to examples including $L_{1}$, $L_{2}$, and $L_{4}$ given above. Finally, we summarize this work and give future perspectives. 

\section{Boundary conditions in the variational principle on the ground of Newtonian dynamics}\label{Sec02}
\subsection{The characteristics of Newton's laws of motion}\label{Sec02:01}
Newtonian dynamics is established upon three fundamental laws. The first law is the {\it{law of inertia}} stated as {\textit{every body continues in its state of rest, or uniform motion in a straight line unless compelled to change that state by forces impressed upon it}}\cite{Fitzpatrick2021}. This law implies that position coordinates which describe the trajectory of a point particle depend only on the first-power of time variable and constant parameters. Quantitatively, $\vec{x}(t)=\vec{A}t+\vec{B}$ where $t$ is a time variable, $\vec{x}(t)$ is three-dimensional position vector of the point particle at time $t$, and $\vec{A}$ and $\vec{B}$ are three-dimensional constant vectors composed of constant parameters. Note that the mass of the point particle is absent here. That is, the first law indicates the existence of an inertial frame. 

If a force exists the situation gets changed; we need the second law of motion, the {\it{equations of motion}}, stated as {\textit{the change of motion (i.e., momentum) of an object is proportional to the force impressed upon it, and is made in the direction of the straight-line in which the force is impressed}}\cite{Fitzpatrick2021}. Quantitatively, the second law, of course, is written as $d\vec{p}/dt=\vec{F}$ where $\vec{p}$ is the three-dimensional momentum vector of the particle with mass $m$, $\vec{p}=md\vec{x}/dt$, and $\vec{F}$ is a force. When $\vec{F}=\vec{0}$, the law of inertia is expressed in the equation of motion. That is, the equation of motion is established upon the existence of inertial frame. Therefore, the constant parameters in $\vec{A}$ and $\vec{B}$ are none other than integral constants that are demanded from the equations of motion. This law implies the crucial fact that the equation of motion is described as the second-order derivative differential equation with respect to the time variable. 

To make the second law viable, it needs to clarify what is the force. First, as a general statement, there is the third law of motion, the {\it{law of action and reaction}}, stated as {\it{to every action there is always opposed an equal reaction; or, the mutual actions of two bodies upon each other are always equal and directed to contrary parts}}\cite{Fitzpatrick2021}. Second, forces are classified into two types: conservative forces and non-conservative forces. In this paper, we treat only conservative forces; for a force $\vec{F}$ there exists the potential $U(\vec{x})$ such that $\vec{F}=-{\text{grad}}\ U(\vec{x})$ where ${\text{'grad'}}$ is the gradient operator with respect to $\vec{x}$. That is, the force depends only on the position coordinates. 

\subsection{Compatibility of the variational principle with Newtonian dynamics and boundary conditions}\label{Sec02:02}
From the previous section, to ensure the compatibility of a given theory with Newtonian dynamics, we have to verify whether or not the following three conditions are satisfied; (i). The existence of Lagrangian which expresses the law of inertial, (ii). Euler-Lagrange equations include up to second-order time derivative terms, and (iii). Conservative forces are taken into account correctly: the equations of motion under the conservative force is recovered. 

\subsubsection{First-order time derivative systems}\label{Sec02:02:01}
The action integral of the system is given as follows:
\begin{equation}
S^{(1)}=\int^{t_{2}}_{t_{1}}L^{(1)}(\dot{q}^{i},q^{i},t)dt
\label{AI in 1st syetem}
\end{equation}
where $q^{i}$s are position coordinates, $\dot{q}^{i}$s are first-order time derivative of $q^{i}$s, $L^{(1)}$ is the Lagrangian for the system, $t_{2}>t_{1}$, and $i=1,2,\cdots,n$. The first-order variation with respect to the position coordinates is 
\begin{equation}
\delta S^{(1)}=\int^{t_{2}}_{t_{1}}\sum_{i=1}^{n}\left[\frac{\partial L^{(1)}}{\partial q^{i}}-\frac{d}{dt}\left(\frac{\partial L^{(1)}}{\partial \dot{q}^{i}}\right)\right]\delta q^{i}dt+\left[\sum_{i=1}^{n}\left(\frac{\partial L^{(1)}}{\partial \dot{q}^{j}}\right)\delta q^{j}\right]^{t_{2}}_{t_{1}}.
\label{}
\end{equation}
If the system is non-degenerate, i.e. the kinetic matrix of the system $K^{(1)}:=\partial^{2} L^{(1)}/\partial\dot{q}^{i}\partial\dot{q}^{j}$ is full rank, the Euler-Lagrange equations, of course, are 
\begin{equation}
\frac{\partial L^{(1)}}{\partial q^{i}}-\frac{d}{dt}\left(\frac{\partial L^{(1)}}{\partial \dot{q}^{i}}\right)=0
\label{E-L eqs in 1st system}
\end{equation}
from the variational principle under position-fixing boundary conditions, in particular, in this case, the Dirichlet boundary conditions:
\begin{equation}
\delta q^{i}(t_{1})=\delta q^{i}(t_{2})=0.
\label{BC in 1st-order theory}
\end{equation}
This system satisfies all conditions: (i), (ii), and (iii). For (i), the Lagrangian is $L^{(1)}=\sum_{i}(\dot{q}^{i})^{2}/2$. For (ii), the second term in the left hand-side of Eq.~$(\ref{E-L eqs in 1st system})$ leads to $\ddot{q}^{i}$-terms. For (iii), the first-term in the left hand-side of Eq.~$(\ref{E-L eqs in 1st system})$ realizes the correct force terms for the Lagrangian: $L^{(1)}=\sum_{i}(\dot{q}^{i})^{2}/2-U(q^{i})$. 

There is another possibility to take the first-order variation of the original action integral~$(\ref{AI in 1st syetem})$. That is, the variation with respect to the velocities $\dot{q}^{i}$:
\begin{equation}
\delta S_{1}=\int^{t_{2}}_{t_{1}}\sum_{i=1}\frac{\partial L^{(1)}}{\partial \dot{q}^{i}}\delta \dot{q}^{i}.
\label{}
\end{equation}
Remark that this manipulation implies that the variation $\delta$ does not commute with the time derivative $d/dt$: $\delta(d/dt)\cdot\neq(d/dt)\delta\cdot$. It can be interpreted as taking a variation while {\it{in advance}} fixing all configurations: $\delta q^{i}:=0$ throughout all time. In this case, without any boundary conditions, the variational principle is applicable, and the Euler-Lagrange equations are 
\begin{equation}
\frac{\partial L^{(1)}}{\partial \dot{q}^{i}}=0.
\label{}
\end{equation}
However, these equations do not satisfy any of the conditions: (i), (ii), and (iii). Hence, this theory is ruled out; it makes sense since we cannot determine {\it{in advance}} the trajectory of the system without equations of motion. 

Throughout these considerations, the first-order time derivative systems being compatible with Newtonian dynamics are viable only for the variational principle varying with respect to position coordinates. In this case, the variational principle needs the position-fixing (Dirichlet) boundary conditions.

\subsubsection{Second-order time derivative systems}\label{Sec02:02:02}
The action integral for the systems is given as follows:
\begin{equation}
S^{(2)}=\int^{t_{2}}_{t_{1}}L^{(2)}(\ddot{q}^{i},\dot{q}^{i},q^{i},t)dt,
\label{}
\end{equation}
where $\ddot{q}^{i}$s are second-order time derivatives of $q^{i}$s. There are three possibilities to take the first-order variation of the action integral with respect to (a) the position coordinates $q^{i}$, (b) the velocity coordinates $\dot{q}^{i}$, and (c) the acceleration coordinates $\ddot{q}^{i}$. 

Let us consider the first case (a). The first-order variation is computed as follows:  
\begin{equation}
\begin{split}
\delta S^{(2)}=&\int^{t_{2}}_{t_{1}}\sum_{i=1}^{n}\left[\frac{\partial L^{(2)}}{\partial q^{i}}-\frac{d}{dt}\left(\frac{\partial L^{(2)}}{\partial \dot{q}^{i}}\right)+\frac{d^{2}}{dt^{2}}\left(\frac{\partial L^{(2)}}{\partial\ddot{q}^{i}}\right)\right]\delta q^{i}dt \\
&+\left[\sum_{i=1}^{n}\left\{\frac{\partial L^{(2)}}{\partial \dot{q}^{i}}-\frac{d}{dt}\left(\frac{\partial L^{(2)}}{\partial \ddot{q}^{i}}\right)\right\}\delta q^{i}+\sum^{n}_{i=1}\left(\frac{\partial L^{(2)}}{\partial \ddot{q}^{i}}\right)\delta\dot{q}^{i}\right]^{t_{2}}_{t_{1}}
\end{split}
\label{AI in 2nd system}
\end{equation}
The variational principle leads to the following Euler-Lagrange equations:
\begin{equation}
\frac{\partial L^{(2)}}{\partial q^{i}}-\frac{d}{dt}\left(\frac{\partial L^{(2)}}{\partial \dot{q}^{i}}\right)+\frac{d^{2}}{dt^{2}}\left(\frac{\partial L^{(2)}}{\partial\ddot{q}^{i}}\right)=0
\label{E-L eqs in 2nd system}
\end{equation}
under fixing both the position and velocity coordinates:
\begin{equation}
\begin{split}
\delta q^{i}(t_{1})=&\delta q^{i}(t_{2})=0,\\
\delta \dot{q}^{i}(t_{1})=&\delta \dot{q}^{i}(t_{2})=0
\end{split}
\label{BC in 2nd system}
\end{equation}
if the system is non-degenerate, i.e. $K^{(2)}=\partial^{2}L^{(2)}/\partial\ddot{q}^{i}\ddot{q}^{j}$ is full rank, but this condition is not satisfied. This is just because the following conditions are imposed due to the compatibility with Newtonian dynamics: the condition (ii). That is, the Euler-Lagrange equations $(\ref{E-L eqs in 2nd system})$ are rewritten as follows:
\begin{equation}
K^{(2)}_{ij}\ddddot{q}^{j}+E^{(2)}_{ij}\dddot{q}^{j}+{\textrm{(up\ to\ 2nd-order\ terms)}}=0
\label{}
\end{equation}
where we defined 
\begin{equation}
K^{(2)}_{ij}:=\frac{\partial^{2} L^{(2)}}{\partial\ddot{q}^{i}\partial\ddot{q}^{j}},E^{(2)}_{ij}:=\frac{\partial^{2}L^{(2)}}{\partial\ddot{q}^{i}\partial\dot{q}^{j}}-\frac{\partial^{2}L^{(2)}}{\partial\ddot{q}^{j}\partial\dot{q}^{i}}.
\label{}
\end{equation}
To satisfy condition (ii), the matrices $K^{(2)}$ and $E^{(2)}$ have to be zero:
\begin{equation}
K^{(2)}=0, E^{(2)}=0.
\label{condition for K and E}
\end{equation}
Then the Euler-Lagrange equations are up to second-order time derivative and now satisfy condition (ii). The first condition in Eq.~$(\ref{condition for K and E})$ indicates that this system has to be degenerate as per the second-order derivative theory. 

In addition, on one hand, the first condition of Eq.~$(\ref{condition for K and E})$ leads to a specific form of Lagrangian\cite{MotohashiSuyama2015,MotohashiNouriSuyamaYamaguchiLanglois2016}:
\begin{equation}
L^{(2)}=\sum_{i=1}^{n}f_{i}(\dot{q}^{j},q^{j})\ddot{q}^{i}+g(\dot{q}^{j},q^{j}).
\label{L in 2nd system}
\end{equation}
Conditions (i) and (iii) are now satisfied for the Lagrangian $L^{(2)}=-\sum_{i}q^{i}\ddot{q}^{i}/2$ and $L^{(2)}=-\sum_{i}q^{i}\ddot{q}^{i}/2-U(q^{i})$, respectively. On the other hand, since we consider these under the compatibility with Newtonian dynamics, the second-order time derivative systems should be equivalent to the first-order time derivative systems discussed in Sect.~$\ref{Sec02:02:01}$. This indicates that the Lagrangian in this theory should be equivalent up to surface terms. That is,
\begin{equation}
L^{(2)}\rightarrow L'^{(1)}=L^{(2)}+\frac{dW}{dt},
\label{transformed L in 2nd system}
\end{equation}
where $W=W(\dot{q}^{i},q^{i})$. If $W$ satisfies the following conditions:
\begin{equation}
f_{i}+\frac{\partial W}{\partial \dot{q}^{i}}=0
\label{}
\end{equation}
or
\begin{equation}
W=-\sum_{i=1}^{n}\int f_{i}d\dot{q}^{i}+C(q^{j}),
\label{counter-term in 2nd system}
\end{equation}
where $C$ is an arbitrary function of position coordinates, the first terms in the Lagrangian $(\ref{L in 2nd system})$ vanish, and the system turns into a first-order time derivative system. Note, here, that Eq.~$(\ref{counter-term in 2nd system})$ is none other than a counter-term in the second-order time derivative system. Furthermore, the variation of the action integral of $L'^{(1)}$ becomes as follows:
\begin{equation}
\delta S'^{(1)}={\textrm{ the same terms to }} \delta S^{(2)}
+\left[\sum^{n}_{i=1}\left\{\frac{\partial L^{(2)}}{\partial \dot{q}^{i}}-\frac{d}{dt}\left(\frac{\partial L^{(2)}}{\partial \ddot{q}^{i}}\right)+\frac{\partial W}{\partial q^{i}}\right\}\delta q^{i}\right]^{t_{2}}_{t_{1}}
\label{transformed AI in 2nd system}
\end{equation}
Therefore, the boundary condition $(\ref{BC in 2nd system})$ for the variational principle now turns into position-fixing boundary conditions:
\begin{equation}
\delta q^{i}(t_{1})=\delta q^{i}(t_{2})=0
\label{p-fixing BC in 2nd-order systems}
\end{equation}
if the kinetic matrix $K^{(1)}_{ij}:=\partial^{2}L'^{(1)}/\partial{\dot{q}^{i}}\partial{\dot{q}^{j}}$ is non-degenerate. This coincides with the boundary conditions for the first-order time derivative systems: Eq.~$(\ref{BC in 1st-order theory})$. It makes sense since boundary conditions uniquely determine the dynamics; it should have the same Lagrangian up to surface terms under Newtonian dynamics. In fact, for the Lagrangian $L^{(2)}=-\sum_{i}q^{i}\ddot{q}^{i}/2-U(q^{i})$, the counter-term is $W=\sum_{i}q^{i}\dot{q}^{i}/2$. Then the Lagrangian turns into $L'^{(1)}=\sum_{i}(\dot{q}^{i})^{2}/2-U(q^{i})$. From these considerations, case (a) is compatible with Newtonian dynamics, and the boundary conditions for the variational principle are position-fixing boundary conditions. 

In case (b), the first-order variation is computed as follows: 
\begin{equation}
\delta S^{(2)}=\int^{t_{2}}_{t_{1}}\sum_{i=1}^{n}\left[\frac{\partial L^{(2)}}{\partial \dot{q}^{i}}-\frac{d}{dt}\left(\frac{\partial L^{(2)}}{\partial \ddot{q}^{i}}\right)\right]\delta \dot{q}^{i}dt+\left[\sum_{i=1}^{n}\left(\frac{\partial L^{(2)}}{\partial \dot{q}^{i}}\right)\delta\dot{q}^{i}\right]^{t_{2}}_{t_{1}}.
\label{}
\end{equation}
The variational principle under velocity-fixing boundary conditions:
\begin{equation}
\delta\dot{q}^{i}(t_{1})=\delta\dot{q}^{i}(t_{2})=0
\label{}
\end{equation}
leads to the following Euler-Lagrange equations:
\begin{equation}
\frac{\partial L^{(2)}}{\partial \dot{q}^{i}}-\frac{d}{dt}\left(\frac{\partial L^{(2)}}{\partial \ddot{q}^{i}}\right)=0
\label{}
\end{equation}
or
\begin{equation}
K^{(2)}_{ij}\dddot{q}^{j}+{\textrm{(up to 2nd-order terms)}}=0.
\label{}
\end{equation}
$K^{(2)}_{ij}=\partial^{2}L/\partial\ddot{q}^{i}\partial\ddot{q}^{j}=0$ leads to the same Lagrangian to $(\ref{L in 2nd system})$. This indicates that these equations satisfy condition (ii) but does not satisfy conditions (i) and (iii). Therefore, the case (b) is ruled out. 

Finally, in case (c), the first-order variation is computed as follows:
\begin{equation}
\delta S^{(2)}=\int^{t_{2}}_{t_{1}}\sum_{i=1}^{n}\left(\frac{\partial L^{(2)}}{\partial \ddot{q}^{i}}\right)\delta \ddot{q}^{i}dt.
\label{}
\end{equation}
The variational principle leads to the following Euler-Lagrange equations without any boundary condition:
\begin{equation}
\frac{\partial L^{(2)}}{\partial \ddot{q}^{i}}=0.
\label{EL in acceleration in 2nd-order}
\end{equation}
This system satisfies all conditions: (i), (ii), and (iii) for $L^{(2)}=\sum_{i=1}(\ddot{q}^{i})^{2}/2$, an arbitrary Lagrangian, and $L^{(2)}=\sum_{i}(\ddot{q}^{i})^{2}/2-\sum_{i}\ddot{q}^{i}(\partial U(q)/\partial q^{i})$, respectively. However, this manipulation does not make sense since the form of $L_{2}$ implies that the Newtonian equations of motion of the system, $\ddot{q}^{i}=\partial U(q)/\partial q^{i}$, are already known; the Euler-Lagrange equations~$(\ref{EL in acceleration in 2nd-order})$ are identically satisfied. In other words, we already know the trajectory of the system with a set of initial conditions. 

Throughout these considerations, the second-order time derivative systems in cases (a) and (c) are possible systems to be compatible with Newtonian mechanics. Boundary conditions are necessary only for the case (a), and these are position-fixing boundary conditions. For case (c), the variational principle itself is always well-posed but does not have any ability to predict dynamics. 

\subsubsection{Higher-order time derivative systems}\label{Sec02:02:03}
The action integral for the systems is given as follows:
\begin{equation}
S^{(d)}=\int^{t_{f}}_{t_{i}}L^{(d)}\left(D^{d}{q^{i}},\cdots,Dq^{i},q^{i},t\right)dt,
\label{L in dth system}
\end{equation}
where $D^{\alpha}q^{i}$ denote $\alpha(\geq2)$th-order time derivative of the position coordinates $q^{i}$: $D^{\alpha}q^{i}=(d/dt)^{\alpha}q^{i}$. A similar consideration leads to that the cases of first-order variation with respect to $Dq^{i},D^{2}q^{i},\cdots,D^{d-1}q^{i}$ are ruled out. For the case of $q^{i}$s, the first-order variation of the action integral is computed as follows:
\begin{equation}
\delta S^{(d)}=\int^{t_{2}}_{t_{1}}dt\sum_{i=1}^{n}\left[\sum^{d}_{\alpha=0}(-D)^{\alpha}\frac{\partial L^{(d)}}{\partial (D^{\alpha}q^{i})}\right]\delta q^{i}
+\left[\sum_{i=1}^{n}\sum^{d}_{\beta=\alpha\geq1}(-D)^{\beta-\alpha}\frac{\partial L^{(d)}}{\partial (D^{\beta}q^{i})}\delta(D^{\alpha-1}q^{i})\right]^{t_{2}}_{t_{1}}.
\label{}
\end{equation}
If the following differential equations for $W=W(q^{i},Dq^{i},\cdots,D^{d-1}q^{i})$:
\begin{equation}
\sum^{d}_{\beta=\alpha}\left[(-D)^{\beta-\alpha}\frac{\partial L^{(d)}}{\partial (D^{\beta}q^{i})}+\frac{\partial W}{\partial(D^{\alpha-1}q^{i})}\right]=0\ \ \ \ \ (\alpha\geq 2)
\label{}
\end{equation}
are solvable, the Euler-Lagrange equations are derived as follows:
\begin{equation}
\sum^{d}_{\alpha=0}(-D)^{\alpha}\frac{\partial L^{(1)}}{\partial (D^{\alpha}q^{i})}=\frac{\partial L^{(1)}}{\partial q^{i}}-\frac{d}{dt}\frac{\partial L^{(1)}}{\partial \dot{q}^{i}}=0,
\label{}
\end{equation}
where $L^{(1)}=L^{(d)}+dW/dt$, under position-fixing boundary conditions:
\begin{equation}
\delta q^{i}(t_{1})=\delta q^{i}(t_{2})=0.
\label{p-fixing BC in general systems}
\end{equation}
Then all conditions (i), (ii), and (iii) are satisfied.

Finally, for the case of $D^{d}q^{i}$s, the variational principle itself is always well-posed without any boundary conditions and all conditions (i), (ii), and (iii) are satisfied for the Lagrangian $L^{(d)}=\sum_{i}(D^{d}q^{i})(D^{2}q^{i})-\sum_{i}(D^{d}q^{i})(\partial U(q^{i})/\partial q^{i})$ but it does not has any ability to predict dynamics. 

\subsection{Stability of the systems}\label{Sec02:03}
Newtonian dynamics demands that Hamiltonian has to be bounded from below. For instance, let us consider two systems described by Lagrangian $L_{1}=\dot{q}^{2}/2-U(q)$ and $L_{2}=-\dot{q}^{2}/2-U(q)$ for some potential $U(q)\propto-1/q$ $(q>0)$. The corresponding Hamiltonians for $L_{1}$ and $L_{2}$ are $H=p^{2}/2+U(q)$ and $H=-p^{2}/2+U(q)$, respectively. The former system is bounded from below and stable, but the latter system is unstable due to the negative infiniteness of the Hamiltonian. For the higher-order time derivative systems, another type of instability occurs: Ostrogradski's instability\cite{Ostrogradsky1850,Woodard2015}.

In order to introduce Hamilton formulation for the Lagrangian $L=L\left(D^{d}{q^{i}},\cdots,Dq^{i},q^{i},t\right)$ in Eq.~$(\ref{L in dth system})$, the following variables:
\begin{equation}
Q_{(\alpha)}^{i}:=D^{\alpha-1}q^{i},\ \ \ P^{(\alpha)}_{i}:=\sum^{d}_{\beta=\alpha\geq1}(-D)^{\beta-\alpha}\frac{\partial L}{\partial (D^{\beta}q^{i})},
\label{}
\end{equation}
are defined as canonical variables. In addition, we assume that the non-degeneracy of the kinetic matrix $K^{(d)}_{ij}:=\partial^{2}L/\partial \dot{Q}_{(d)}^{i}\partial \dot{Q}_{(d)}^{j}=\partial P^{(d)}_{i}/\partial \dot{Q}_{(d)}^{i}$. Then the Legendre transformation of $L$ called as Ostrogradski transformation, i.e. the Ostrogradski Hamiltonian, is introduced as follows:
\begin{equation}
H=\sum_{i=1}^{n}\sum^{d-1}_{\alpha=1}P^{(\alpha)}_{i}Q^{i}_{(\alpha+1)}+\sum_{i=1}^{n}P^{(d)}_{i}F^{i}-L,
\label{}
\end{equation}
where $F^{i}=F^{i}(Q^{i}_{(1)},Q^{i}_{(2)},\cdots,Q^{i}_{(d)},P^{(d)}_{i})$, which corresponds to $\dot{Q}^{i}_{(d)}$, comes from the implicit function theorem for the kinetic matrix $K^{(d)}$. The linear dependency on $P^{(\alpha)}_{i}$ in the first term implies that the Hamiltonian is not bounded from below while $P^{(d)}_{i}$ in the second term is bounded from below through the function $F^{i}$. This unboundedness makes the system unstable and this is none other than the Ostrogradski's instability. 

In particular, the case of $d=2$ with the Lagrangian~$(\ref{L in 2nd system})$ is a constraint system since the canonical momenta $P^{(2)}_{i}$ are given by
\begin{equation}
P^{(2)}_{i}=f_{i}(Q_{(1)}^{j},Q_{(2)}^{j}).
\label{}
\end{equation}
That is, we have primary constraints:
\begin{equation}
\Phi^{(1)}_{i}:=P^{(2)}_{i}-f_{i}(Q_{(1)}^{j},Q_{(2)}^{j}):\approx0,
\label{}
\end{equation}
where we denote $:\approx$ as imposing the condition in the weak equallity. Total Hamiltonian is 
\begin{equation}
H_{T}=P^{(1)}_{i}Q_{(2)}^{i}-g+v^{s}\Phi^{(1)}_{s},
\label{}
\end{equation}
where $s=1,2,\cdots,n$ and $v^{s}$s are Lagrange multipliers. The Dirac procedure leads to secondary constraints:
\begin{equation}
\Phi^{(2)}_{i}:=\{\Phi^{(1)}_{i},H_{T}\}\approx-P^{(1)}_{i}-\frac{\partial f_{i}}{\partial Q_{(1)}^{j}}Q_{(2)}^{j}+\frac{\partial g}{\partial Q_{(2)}^{i}}\approx0,
\label{}
\end{equation}
where we denote $\approx$ as the weak equality. Depending on whether or not the Poisson brackets among $\Phi^{(1)}_{i}$ and $\Phi^{(1)}_{i}$ are weak equal to zero, some Lagrange multipliers are determined and others remain arbitrary. The latter case indicates that the consistency conditions for secondary constraints lead to tertiary constraints. Further analysis needs a specific Lagrangian. The above analysis implies that the physical degrees of freedom are equal to or less than $(2n+2n-2n)/2=n$. The Hamilton analysis described above is based on \cite{Dirac1950,Dirac1958,Ostrogradsky1850,Woodard2015,SatoSuganoOhtaKimura1989,SatoSuganoOhtaKimura1989-2}. The authors in  \cite{MotohashiSuyama2015,MotohashiNouriSuyamaYamaguchiLanglois2016,MotohashiSuyamaYamaguchi2018,MotohashiSuyamaYamaguchi2018-2} apply the method in \cite{Dirac1950,Dirac1958,Ostrogradsky1850,Woodard2015,Pons1989}. An example is given in the section \ref{Sec05:04}.

The point is that, after imposing these constraints, the system turns into a first-order time derivative system, and the unbounded momentum variables $P^{(2)}_{i}$s drop out as constraints. In addition, since $P^{(1)}_{i}$s either are bounded from below if the matrix $\Delta_{ij}:=\partial P^{(1)}_{i}/\partial \dot{Q}_{(2)}^{j}$ is non-degenerate or become primary constraints if $\Delta_{ij}$ is degenerate, the system is stable. This fact would also be deduced from the transformed Lagrangian~$(\ref{transformed L in 2nd system})$ under Eq.~$(\ref{counter-term in 2nd system})$. Therefore, the systems being compatible with Newtonian dynamics are stable.  

\subsection{Causality of the systems}\label{Sec02:04}
As mentioned in Sect.~\ref{Sec02:02}, the compatibility of the systems with Newtonian dynamics demands condition (ii). Euler-Lagrange equations include up to second-order time derivative terms, and this property implies that the corresponding Lagrangian is composed up to first-order time derivative terms and vice versa. In these systems, causality is, of course, satisfied. However, higher-order time derivative systems generically suffer from {\it{'acausality'}}\cite{Dirac1938,Kuti1993}. 

In order to see this problem more precisely, let us consider the following equation of motion\cite{Dirac1938}:
\begin{equation}
\ddot{q}=\frac{1}{\beta}\dddot{q}+\frac{\alpha}{\beta}\delta(t-t_{0})
\label{}
\end{equation}
where $\delta(\cdot)$ is delta function, $\alpha$ and $\beta$ are positive constants. The first term in the right-hand side is the so-called Abraham-Lorentz force, which was rediscovered by Dirac in the literature of self-radiation of an electron as a non-relativistic approximation. The second term in the right-hand side is an impact force given at the time $t=t_{0}$. This equation has a solution:
\begin{equation}
q(t)=\left\{\begin{array}{ll}
\frac{\alpha}{\beta^{2}}\exp{[\beta(t-t_{0})]}+q_{0} & (t<t_{0})\\
\frac{\alpha}{\beta}(t-t_{0})+\frac{\alpha}{\beta^{2}}+q_{0} & (t_{0}<t)
\end{array}
\right.
\label{}
\end{equation}
under the conditions: $q(t\rightarrow-\infty)=q_{0}$, $\dot{q}(t\rightarrow-\infty)=0$, the continuity of $\dot{q}$ at $t=t_{0}$ and $\ddot{q}(t=t_{0})=\alpha$. The last condition is caused by the impact force at $t=t_{0}$. This solution has a strange feature; the particle accelerates before the impact force affects. This is the acausal behavior firstly indicated by Dirac\cite{Dirac1938}. The Lagrangian of the system is given as follows:
\begin{equation}
L=\frac{1}{2\beta^{2}}\ddot{q}^{2}-\frac{1}{\beta}\ddot{q}\dot{q}+\frac{1}{2}\dot{q}^{2}+\frac{\alpha}{\beta^{2}}\dot{q}\delta(t-t_{0})-\frac{\alpha}{\beta}q\delta(t-t_{0}).
\label{}
\end{equation}
This Lagrangian actually contains the second-order time derivative, and it is a non-degenerate system. However, we show below that this acausality is none other than a paradox. 

We can easily rewrite the Lagrangian as follows:
\begin{equation}
L=\frac{1}{2}\dot{Q}^{2}-\frac{\alpha}{\beta}Q\delta(t-t_{0}),
\label{}
\end{equation}
where we set $Q:=q-\dot{q}/\beta$. Here, the answer for the paradox is almost trivial; $q$ is a {\it{gauge}} variable, but $Q$ is the {\it{physical}} variable. To see this, we only check the invariance of $Q$ for the following transformation: $q\rightarrow q'=q+\exp{(\beta t)}$. Therefore, when we use the physical variable $Q$, the paradox is removed out. In fact, the equation of motion turns into
\begin{equation}
\ddot{Q}=\frac{\alpha}{\beta}\delta(t-t_{0})
\label{}
\end{equation}
and the solution is given as follows:
\begin{equation}
Q(t)=\left\{\begin{array}{ll}
q_{0} & (t<t_{0})\\
\frac{\alpha}{\beta}(t-t_{0})+q_{0} & (t>t_{0})
\end{array}
\right.
\label{}
\end{equation}
under the same conditions. This is a causal behavior. The message of the example is that, as long as we treat a system in terms of physical variables, the equations of motion contain up to second-order time derivatives and the causality holds. The generalization of this statement would be a difficult subject and is out of the scope of this paper. However, as long as we restrict our investigation to the systems introduced in Sect.~\ref{Sec02:02}, the acausal problem never occurs since the equations of motion of the systems are restricted up to second-order time derivative. This implies also that the variational principle under the position-fixing boundary conditions $(\ref{p-fixing BC in 2nd-order systems})$ or, more generically, Eq~$(\ref{p-fixing BC in general systems})$, guarantees that the systems are causal. 

\section{Existence of solutions and difficulties in degenerate systems}\label{Sec03}
\subsection{Non-degenerate systems and three fundamental correspondences}\label{Sec03:01}
In this section, based on \cite{SuganoKamo1982}, we reveal the existence of solutions and the relations of objects in Lagrange and Hamilton formulation. Based on these investigations, we define three fundamental maps that describe correspondences to consider the well-posedness of the variational principle.

\subsubsection{Lagrange formulation}\label{Sec03:01:01}
The Euler-Lagrange equations $(\ref{E-L eqs in 1st system})$ can be split into a set of 1-forms, called a Pfaffian system, as follows:
\begin{equation}
\begin{split}
\rho^{i}:=&dq^{i}-v^{i}dt=0,\\
\theta_{i}:=&K^{(1)}_{ij}dv^{j}-S_{i}dt
\end{split}
\label{split E-L equation in 1st system}
\end{equation}
where $K^{(1)}_{ij}$ and $S_{i}$ are defined as follows:
\begin{equation}
\begin{split}
&K^{(1)}_{ij}:=\partial^{2}L/\partial\dot{q}^{i}\partial\dot{q}^{j},\\
&S_{i}:=-v^{j}\frac{\partial^{2}L}{\partial q^{i}\partial v^{j}}-\frac{\partial^{2}L}{\partial t\partial v^{i}}+\frac{\partial L}{\partial q^{i}}.
\end{split}
\label{Si}
\end{equation}
The existence of solutions for the Euler-Lagrange equations depends on whether or not a system of generators, i.e. a closed algebra composed of vector fields that are orthogonal to the Pfaffian system, exists. Since the Pfaffian system $(\ref{split E-L equation in 1st system})$ spans $2n$-dimensional 1-form space, which is a subspace of $T^{*}(TM\times \mathbb{R})$, the system of generator is 1-dimensional vector space; denote the generator as $X_{t}$, which is a subspace of $T(TM\times \mathbb{R})$. Where $M$ is a $n$-dimensional configuration space, $TM$ is the velocity-phase space of $M$, and $TM\times \mathbb{R}$ is the extended velocity-phase space. This vector field $X_{t}$ is a tangent vector to the trajectory of the system in the space $TM\times \mathbb{R}$. Therefore, the generator can be expressed as follows:
\begin{equation}
X_{t}:=a^{i}\frac{\partial}{\partial v^{i}}+b^{i}\frac{\partial}{\partial q^{i}}+\frac{\partial}{\partial t},
\label{generator in 1st system}
\end{equation}
where $\partial/q^{i}$, $\partial/v^{i}$ and $\partial/\partial t$ are the coordinates basis of $T(TM\times \mathbb{R})$. The dual basis is, of course, $dq^{i}$, $dv^{i}$, and $dt$. $a^{i}$s and $b^{i}$s are undetermined coefficients. If these coefficients exist, a system of generators is determined. The duality between $T(TM\times \mathbb{R})$ and $T^{*}(TM\times \mathbb{R})$ indicates the existence of the following inner products: $<dq^{i},\partial/\partial q^{j}>=\delta^{ij}$, $<dv^{i},\partial/\partial v^{j}>=\delta^{ij}$, $<dt,\partial/\partial t>=1$, and otherwise vanish. Therefore, the orthogonality among the Pfaffian equations $(\ref{split E-L equation in 1st system})$ and the generator $(\ref{generator in 1st system})$ determine all the coefficients $a^{i}$s and $b^{i}$s as follows:
\begin{equation}
\begin{split}
&a^{i}=({K^{(1)}}^{-1})^{ij}S_{j},\\
&b^{i}=v^{i}.
\end{split}
\label{}
\end{equation}
Notice, here, that the coefficients $a^{i}$s are determined uniquely by virtue of the non-degeneracy of the kinetic matrix $K^{(1)}$. Therefore, since $[X_{t},X_{t}]=0$, Frobenius theorem indicates the existence of $2n$ functions $f^{I}$ and integral constants $c^{I}$ such that
\begin{equation}
f^{I}(q^{i},v^{i},t)=c^{I},
\label{solution in non-degenerate 1st-derivative}
\end{equation}
where $I=1,2,\cdots,2n$. Remark that Frobenius theorem indicates only the existence of $f^{I}$; a set of functions $f^{I}$s are generically not uniquely determined. However, as long as we restrict our consideration to physical interests, it would be allowed to regard the case where the theorem gives unique functions $f^{I}$s. That is, Eq.~$(\ref{solution in non-degenerate 1st-derivative})$ implies that there exists a unique trajectory of which the tangent vector is $X_{t}$. For such $f^{I}$s, the implicit function theorem leads to the unique solutions:
\begin{equation}
q^{i}=q^{i}(t,c^{I}),v^{i}=v^{i}(t,c^{I}).
\label{configuration and velocity}
\end{equation}
Under the solutions of the first equation in Eq.~$(\ref{split E-L equation in 1st system})$, the second solutions above can be rewritten as $\dot{q}^{i}=\dot{q}^{i}(t,c^{I})$. The integral constants $c^{I}$s play a crucial role for considering the well-posed variational principle.

\subsubsection{Hamilton formulation}\label{Sec03:01:02}
The Lagrangian in the phase space relating to the Hamiltonian is
\begin{equation}
{_{*}L}:=q^{i}dp_{i}-H(q^{i},p_{i})dt,
\label{inverse-Legendre of Hamiltonian}
\end{equation}
where the left-lowered star $*$ of ${_{*}L}$ denotes an object expressed in the phase space. This is none other than a Legendre transformation. Under the assumption of non-degeneracy, the Lagrangian can be equivalently expressed in both the velocity-phase space and the phase space. The total differentiation of ${_{*}L}$ is
\begin{equation}
d{_{*}L}={_{*}\theta}_{i}\wedge{_{*}\rho}^{i},
\label{}
\end{equation}
where
\begin{equation}
\begin{split}
&{_{*}\rho}^{i}:=dq^{i}-\frac{\partial H}{\partial p_{i}}dt,\\
&{_{*}\theta}_{i}:=dp_{i}+\frac{\partial H}{\partial q^{i}}dt.
\end{split}
\label{Pfaffian system for Hamilton formulation}
\end{equation}
For $d{_{*}L}$ to vanish, both ${_{*}\rho}^{i}$ and ${_{*}\theta}_{i}$ have to be zero: ${_{*}\rho}^{i}=0$ and ${_{*}\theta}_{i}=0$, resulting in the following Hamilton's equations:
\begin{equation}
\dot{q}^{i}=\frac{\partial H}{\partial p_{i}},\dot{p}^{i}=-\frac{\partial H}{\partial q_{i}}. 
\label{}
\end{equation}

The same considerations as in Sect.~\ref{Sec03:01:01} but in the phase space $T^{*}M\times \mathbb{R}$ leads to the generator:
\begin{equation}
{_{*}X_{t}}={_{*}a}^{i}\frac{\partial}{\partial q^{i}}+{_{*}b}_{i}\frac{\partial}{\partial p_{i}}+\frac{\partial}{\partial t}.
\label{}
\end{equation}
where
\begin{equation}
{_{*}a}^{i}=\frac{\partial H}{\partial p_{i}}, {_{*}b}_{i}=-\frac{\partial H}{\partial q^{i}}.
\label{}
\end{equation}
Since the generator satisfies $[{_{*}X_{t}},{_{*}X_{t}}]=0$, Frobenius theorem leads to
\begin{equation}
{_{*}f^{I}}(q^{i},p_{i},t)={_{*}c}^{I},
\label{}
\end{equation}
where $a$ runs from $1$ to $2n$, ${_{*}f^{I}}$s and ${_{*}c}^{I}$s are functions and integral constants, respectively, which are implied from this theorem. Assuming that the functions ${_{*}f^{I}}$ are uniquely determined by physical interests, the implicit function theorem gives the solutions:
\begin{equation}
q^{i}={_{*}q}^{i}(t,{_{*}c}^{I}), p_{i}=p_{i}(t,{_{*}c}^{I}).
\label{}
\end{equation}
Comparing with the result from Sect.~\ref{Sec03:01:01}, we already know that the configurations are given by the first formulas in Eq.~$(\ref{configuration and velocity})$. In addition, both the formulations have a common configuration space. That is, ${_{*}q}^{i}(t,{_{*}c}^{I})$s should exactly be the same as $q^{i}(t,c^{I})$. This fact leads to ${_{*}c}^{I}=c^{I}$. Therefore, the solutions become:
\begin{equation}
q^{i}={q}^{i}(t,c^{I}), p_{i}=p_{i}(t,c^{I}).
\label{}
\end{equation}

The non-degeneracy of the kinetic matrix $K^{(1)}$ is crucial for the existence and uniqueness of the inverse Legendre transformation in Eq.~$(\ref{inverse-Legendre of Hamiltonian})$. This means that a one-to-one correspondence between $X_{t}$ and ${_{*}X}_{t}$ exists. The correspondence is established using the relations:
\begin{equation}
\begin{split}
&\frac{\partial}{\partial q^{i}}\leftrightarrow\frac{\partial}{\partial q^{i}}+\frac{\partial p_{j}}{\partial q^{i}}\frac{\partial}{\partial p_{j}},\\
&\frac{\partial}{\partial v^{i}}\leftrightarrow\frac{\partial p_{j}}{\partial v^{i}}\frac{\partial}{\partial p_{j}}=K^{(1)}_{ij}\frac{\partial}{\partial p_{j}},\\
&\frac{\partial}{\partial t}\leftrightarrow\frac{\partial p_{i}}{\partial t}\frac{\partial}{\partial p_{i}}+\frac{\partial}{\partial t},
\end{split}
\label{vector fields correspondence between vps and ps}
\end{equation}
where the left-hand side and the right-hand side are composed of the coordinate basis of $T(TM\times\mathbb{R})$ and $T(T^{*}M\times\mathbb{R})$, respectively. This one-to-one correspondence holds only when the kinetic matrix is non-degenerate. With this correspondence, it can be shown that $X_{t}$ and ${_{*}X}_{t}$ are also in one-to-one correspondence:
\begin{equation}
X_{t}\leftrightarrow{_{*}X}_{t},
\label{}
\end{equation}
where we used $p_{i}=\partial L/\partial \xi^{i}$. The non-degeneracy of $K^{(1)}$ guarantees that the Lagrange and Hamilton formulations are related through the Legendre transformation in a unique manner, meaning that the velocity-phase space $TM\times \mathbb{R}$ and the phase space $T^{*}M\times \mathbb{R}$ have a one-to-one correspondence.

\subsubsection{Three fundamental correspondences}\label{Sec03:01:03}
In Sect.~\ref{Sec02}, we showed that the compatibility with Newtonian dynamics requires position-fixing boundary conditions to be imposed when applying the variational principle. Taking the configurations $q^{i}={q}^{i}(t,c^{I})$ into account, the boundary conditions uniquely determine all the integral constants at least in non-degenerate systems. This leads to a map $\iota$ that maps the configurations at times $t_{1}$ and $t_{2}$ to a parameter space $C$ which is spanned by all the integral constants:
\begin{equation}
\iota:M[t_{1}]\times M[t_{2}]\rightarrow C; (q^{i}(t_{1}),q^{i}(t_{2}))\mapsto c^{I},
\label{iota}
\end{equation}
where $M[t]$ is the space spanned by the configurations at a time $t$. The space $M[t]$ is usually a finite region since the variational principle leads to a local minimum of the action integral. In other words, giving boundary conditions indicates the possible regions for the configurations at each boundary are determined. $q^{i}(t_{1})$s are the values of configuration coordinates at the time $t_{1}$. $q^{i}(t_{2})$s are defined in the same manner. Conversely, if all the integral constants $c^{I}$s are given, through the solutions $q^{i}={q}^{i}(t,c^{I})$, the values of configuration coordinates are uniquely determined. Therefore, map $\iota$ is invertible. In other words, the position-fixing boundary conditions can be replaced by the integral constants in the solutions and determine the dynamics uniquely when the physical degrees of freedom matches the number of the independent integral constants. This is the first correspondence we have to state.

The second correspondence is introduced between the velocity and canonical momentum variables, expressed in a map as
\begin{equation}
\kappa:TM\times\mathbb{R}\rightarrow T^{*}M\times\mathbb{R};\dot{q}^{i}(t)\mapsto p_{i}(t).
\label{kappa}
\end{equation}
The non-degeneracy guarantees the invertibility of $\kappa$, meaning that the velocity variables can be restored from the canonical momentum variables as
\begin{equation}
\kappa^{-1}:T^{*}M\times\mathbb{R}\rightarrow TM\times\mathbb{R};p_{i}(t)\mapsto\dot{q}^{i}(t)=v^{i}(q^{j},p_{j},t).
\label{}
\end{equation}

The final correspondence is already mentioned in Sect.~\ref{Sec03:01:02}, expressed in a map as 
\begin{equation}
\mathfrak{O}:\mathfrak{G}[T(TM\times \mathbb{R})]\rightarrow\mathfrak{G}[T(T^{*}M\times \mathbb{R})];X_{t}\mapsto {_{*}X}_{t},
\label{frak O}
\end{equation}
where $\mathfrak{G}[T(TM\times \mathbb{R})]$ and $\mathfrak{G}[T^{*}(TQ\times \mathbb{R})]$ denote spaces of the system of generator for the Pfaffian system $(\ref{split E-L equation in 1st system})$ and Eq.~$(\ref{Pfaffian system for Hamilton formulation})$, respectively. The non-degeneracy guarantees that this map is invertible. 

\subsection{Degenerate systems and difficulties}\label{Sec03:02}
In Sect.~\ref{Sec03:01:03}, we introduced the three fundamental maps: $\iota$, $\kappa$, and $\mathfrak{O}$. In this section, based also on Ref.~\cite{SuganoKamo1982}, we explore solutions in a degenerate system and the alteration of the maps: $\iota$, $\kappa$ and $\mathfrak{O}$. 

\subsubsection{Lagrange formulation}\label{Sec03:02:01}
In order to find a system of generators for the Pfaffian system~$(\ref{split E-L equation in 1st system})$ in a degenerate system, we have to consider the degeneracy  ${\text{det}K^{(1)}}=0$, or equivalently, 
\begin{equation}
K^{(1)}_{ij}\tau^{j}_{\alpha}=0,
\label{zero-engenvalue vector}
\end{equation}
where $\alpha=1,2,\cdots,n-r$, $r={\textrm{rank }K^{(1)}}$, and $\tau^{i}_{\alpha}$s are zero-eigenvalue vectors. This property and the second equations of Eq.~$(\ref{split E-L equation in 1st system})$ lead to secondary constraints\footnote{Primary constraints are identically zero in the velocity-phase space: $\Phi^{(1)}_{\alpha}\equiv0$}:
\begin{equation}
\Phi^{(2)}_{\alpha}:=\tau^{i}_{\alpha}S_{i}:\approx0.
\label{}
\end{equation}
Then there exists a set of vectors, $\eta^{i}$s, such that
\begin{equation}
S_{i}\approx K^{(1)}_{ij}\eta^{j}
\label{eta}    
\end{equation}
by virtue of the completeness of the zero-eigenvalue vectors $\tau^{i}_{\alpha}$s, where we denotes $\approx$ as the weak equality. Based on the same consideration as for the non-degenerate case in Sect.~\ref{Sec03:01}, we can derive the following operators:
\begin{equation}
\begin{split}
&X_{t}:=\eta^{i}\frac{\partial}{\partial v^{i}}+v^{i}\frac{\partial}{\partial q^{i}}+\frac{\partial}{\partial t},\\
&Z_{\alpha}:=-\tau^{i}_{\alpha}\frac{\partial}{\partial v^{i}}.
\end{split}
\label{}
\end{equation}
These operators are not closed in the commutation relation since we have
\begin{equation}
Y^{(1)}_{\alpha}:=[X_{t},Z_{\alpha}]=\tau^{i}_{\alpha}\frac{\partial}{\partial q^{i}}+\left(-Z_{\alpha}\eta^{i}-X_{t}\tau^{i}_{\alpha}\right)\frac{\partial}{\partial v^{i}}.
\label{}
\end{equation}
$X_{t}$ and $Y^{(1)}_{\alpha}$s, $Y^{(1)}_{\alpha}$s themselves, and $Y^{(1)}_{\alpha}$s and $Z_{\alpha}$s are in the same situation. However, if the conditions:
\begin{equation}
Z_{\alpha}\tau_{\beta}^{i}-Z_{\beta}\tau_{\alpha}^{i}=T_{\alpha\beta}^{\gamma}\tau^{i}_{\gamma},
\label{}
\end{equation}
are imposed, $Z_{\alpha}$s forms a closed algebra, where $T_{\alpha\beta}^{\gamma}$s are some coefficients. The conditions are, for instance, satisfied if $\tau^{i}_{\alpha}$ is independent of the velocity variables $v^{i}$s. The authors in Ref.~\cite{SuganoKamo1982} do not impose the conditions on the ground that $[Z_{\alpha},Z_{\beta}]$ is automatically closed within $Z_{\alpha}$s. Then $Y^{(1)}_{\alpha}$ themselves, and $Y^{(1)}_{\alpha}$ and $Z_{\alpha}$ form a closed algebra by virtue of Jacobi identity. However, $X_{t}$ and $Y^{(1)}_{\alpha}$s still do not close. To form a system of generators that is compatible with total Hamiltonian formulation, the following operator is introduced~\cite{SuganoKamo1982}:
\begin{equation}
X_{T}:=X_{t}+\zeta^{\alpha}Y^{(1)}_{\alpha}+\xi^{\alpha}Z_{\alpha},
\label{time evolution operator in Lagrange formulation}
\end{equation}
where both $\zeta^{\alpha}$s and $\xi^{\alpha}$s are undetermined functions. Remark that the Dirac conjecture is not considered here~\cite{SuganoKamo1982}: $X_{E}=X_{t}+\zeta^{\alpha}_{s}Y^{(s)}_{\alpha}+\xi^{\alpha}Z_{\alpha}$. Instead, for $X_{T}$ and $Y^{(1)}_{\alpha}$, we introduce the following procedure to comprise a closed algebra:
\begin{equation}
\begin{split}
&Y^{(s+1)}_{\alpha}:=[X_{T},Y^{(s)}_{\alpha}]\ \ \ (s=1,2,\cdots,y_{\alpha}-1),\\
&Y^{(y_{\alpha}+1)}_{\alpha}:=[X_{T},Y^{(y_{\alpha})}_{\alpha}]=D_{\alpha s}^{\beta}Y^{(s)}_{\beta},
\end{split}
\label{higher-order Ys}
\end{equation}
where $D_{\alpha s}^{\beta}$s are constant coefficients. We will see Sect.~\ref{Sec03:02:02} that $Y^{(s)}_{\alpha}$s correspond to the constraints in the phase space. Then the operators $X_{T}$, $Y^{(s)}_{\alpha}$, and $Z_{\alpha}$, or equivalently $X_{t}$, $Y^{(s)}_{\alpha}$, and $Z_{\alpha}$, form a closed algebra. Therefore, if $X_{T}$ is orthogonal to $\rho_{i}$s and $\theta_{i}$s, the Pfaffian system is complete integrable by virtue of $[X_{T},X_{T}]=0$ and Frobenius theorem, and there exists a unique trajectory of which the tangent vector is $X_{T}$ in the space coordinated by $q^{i}$s, $v^{i}$s, and $t$ under these constraints. To achieve this, we add a set of constraints based on the secondary constraints: $\Phi^{(2)}_{\alpha}\approx0$ and impose the orthogonality of $X_{T}$ to $\rho_{i}$s and $\theta_{i}$s under these constraints. 

Since the secondary constraints have to be static, the time evolution of the secondary constraints has to be weak equal to zero:
\begin{equation}
\dot{\Phi}^{(2)}_{\alpha}=d\Phi^{(2)}_{\alpha}(X_{T}):\approx0.
\label{}
\end{equation}
$d\Phi^{(2)}_{\alpha}(Z_{\beta})=0$ are, on one hand, always held, which means that $\xi^{\alpha}$s remain arbitrary. On the other hand, there is a case $d\Phi^{(2)}_{\alpha}(Y^{(1)}_{\beta})\neq0$; the corresponding $\zeta^{\alpha}$ is determined. Otherwise, we get new constraints. Repeating this procedure, we obtain
\begin{equation}
\begin{split}
&\Phi^{(s+1)}_{\alpha}:=d\Phi^{(s)}_{\alpha}(X_{T}):\approx0\ \ \ (s=2,3,\cdots,m_{\alpha}-1),\\
&\Phi^{(m_{\alpha}+1)}_{\alpha}:=d\Phi^{(m_{\alpha})}_{\alpha}(X_{T})=C_{\alpha s}^{\beta}\Phi^{(s)}_{\beta},
\end{split}
\label{stationary equations in Lagrange formulation}
\end{equation}
where $C_{\alpha s}^{\beta}$s are some coefficients. Under these constraints, we question whether or not the following conditions are satisfied:
\begin{equation}
\theta_{i}(X_{T})\approx-\zeta^{\alpha}K^{(1)}_{ij}(Z_{\alpha}\eta^{j}+X_{t}\tau^{j}_{\alpha}):\approx0.
\label{integrability in 1st system in Lagrange formulation}
\end{equation}
If these conditions are satisfied, the Pfaffian system is Frobenius integrable. $\rho_{i}(X_{T})\approx0$s are automatically satisfied by virtue of $\rho_{i}:=K^{(1)}_{ij}\rho^{j}$.
In addition, if the total number of the constraints is equal to or less than the number of all variables:
\begin{equation}
\sum_{\alpha=1}^{r}m_{\alpha}+{\textrm{the number of the primary constraints}}\leq 2n,
\label{the number of constraints}
\end{equation}
the procedure derives the unique solutions satisfying all the constraints based on the same consideration as for the non-degenerate case in Sect.~\ref{Sec03:01}. Here, the authors in Ref.~\cite{SuganoKamo1982} do not impose this condition.

Although the unique solutions exist, map $\iota$ introduced by Eq.~$(\ref{iota})$ in the section \ref{Sec03:01:03} may not be well-defined due to inconsistent integral constants that are not determined by position-fixing boundary conditions. Furthermore, even if map $\iota$ is well-defined, it may not be invertible, presenting a challenge in the Lagrange formulation of degenerate systems and a crucial aspect in formulating a well-defined variational principle.

\subsubsection{Hamilton formulation}\label{Sec03:02:02}
In degenerate systems, the invertibility of the second transformation in Eq.~$(\ref{vector fields correspondence between vps and ps})$ disappears:
\begin{equation}
\frac{\partial}{\partial v^{i}}\rightarrow\frac{\partial p_{j}}{\partial v^{i}}\frac{\partial}{\partial p_{j}}=K^{(1)}_{ij}\frac{\partial}{\partial p_{j}}.
\label{vector fields correspondence between vps and ps in degenerate}
\end{equation}
As a result, while map $\mathfrak{O}$ defined in Eq.~$(\ref{frak O})$ remains well-defined, its inverse would generically not exist.

In order to translate the Lagrange formulation in Sect.~\ref{Sec03:02:01} into Hamilton formulation, the time evolution operator $X_{T}$ in Eq.~$(\ref{time evolution operator in Lagrange formulation})$ and consistency conditions in Eq.~$(\ref{stationary equations in Lagrange formulation})$ need to be translated into $T(T^{*}M\times \mathbb{R})$ from $T(TM\times \mathbb{R})$. First of all, using the transformation in Eq.~$(\ref{vector fields correspondence between vps and ps})$ replacing the second law by Eq.~$(\ref{vector fields correspondence between vps and ps in degenerate})$, $X_{t}$, $Z_{\alpha}$s, and $Y^{(1)}_{\alpha}$s are transformed as follows:
\begin{equation}
\begin{split}
&X_{t}\rightarrow{_{*}X}_{t}=\frac{\partial H}{\partial p_{i}}\frac{\partial}{\partial q^{i}}-\frac{\partial H}{\partial q^{i}}\frac{\partial }{\partial p_{i}}+\frac{\partial}{\partial t}:=X_{H}+\frac{\partial}{\partial t},\\
&Z_{\alpha}\rightarrow{_{*}Z}_{\alpha}=0,\\
&Y^{(1)}_{\alpha}\rightarrow{_{*}Y}^{(1)}_{\alpha}=\frac{\partial{_{*}\phi}^{(1)}_{\alpha}}{\partial p_{i}}\frac{\partial}{\partial q^{i}}-\frac{\partial{_{*}\phi}^{(1)}_{\alpha}}{\partial q^{i}}\frac{\partial}{\partial p_{i}}={_{*}X}_{{_{*}\phi}^{(1)}_{\alpha}}.
\end{split}
\label{}
\end{equation}
Therefore, $X_{T}\rightarrow{_{*}X}_{T}$ becomes
\begin{equation}
X_{T}\rightarrow{_{*}X}_{T}={_{*}X}_{H}+\zeta^{\alpha}{_{*}X}_{{_{*}\phi}^{(1)}_{\alpha}}+\frac{\partial}{\partial t}={_{*}X}_{H_{T}}+\frac{\partial}{\partial t},
\label{}
\end{equation}
where $H_{T}=H+\zeta^{\alpha}{_{*}\phi}^{(1)}_{\alpha}$ is none other than the total Hamiltonian of the system. Therefore, map $\mathfrak{O}$ is well-defined and pushes the consistency conditions $(\ref{stationary equations in Lagrange formulation})$ forward to the following formulas:
\begin{equation}
\begin{split}
&
{_{*}\Phi}^{(s+1)}_{\alpha}=\{{_{*}\Phi}^{(s)}_{\alpha},H_{T}\}+\frac{\partial {_{*}\Phi}^{(s)}_{\alpha}}{\partial t}:\approx0,\\
&{_{*}\Phi}^{(m_{\alpha}+1)}_{\alpha}:={_{*}X}_{T}{_{*}\Phi}^{(m_{\alpha})}_{\alpha}={_{*}C}_{\alpha s}^{\beta}{_{*}\Phi}^{(s)}_{\beta},
\end{split}
\label{}
\end{equation}
where $s=1,2,\cdots,m_{\alpha}-1$. These formulas reveal that
\begin{equation}
\begin{split}
&{_{*}X}_{{_{*}\phi}^{(s+1)}_{\alpha}}={_{*}X}_{\{{_{*}\phi}^{(s)}_{\alpha},H_{T}\}}=[{_{*}X}_{T},{_{*}Y}^{(s)}_{\alpha}]={_{*}Y}^{(s+1)}_{\alpha},\\
&{_{*}X}_{{_{*}\phi}^{(m_{\alpha}+1)}_{\alpha}}={_{*}C}_{\alpha s}^{\beta}{_{*}Y}^{(s)}_{\alpha},
\end{split}
\label{}
\end{equation}
by virtue of $[X_{f},X_{g}]=-X_{\{f,g\}}$, where $s=1,2,\cdots,m_{\alpha}-1$. These relations correspond to Eq.~$(\ref{higher-order Ys})$, and it indicates that $m_{\alpha}=y_{\alpha}$, and $C^{\beta}_{\alpha s}$, ${_{*}C}^{\beta}_{\alpha s}$, and $D^{\beta}_{\alpha s}$ correspond to each other, respectively. Remark that if the Dirac conjecture is allowed\cite{Dirac1950}: $X_{E}=X_{t}+\zeta^{\alpha}_{s}Y^{(s)}_{\alpha}+\xi^{\alpha}Z_{\alpha}$, we get $H_{E}=H+\zeta^{\alpha}_{s}{_{*}\phi}^{(s)}_{\alpha}$. This quantity is called the extended Hamiltonian.

Under these constraints, we question whether or not the following conditions hold:
\begin{equation}
{_{*}\theta}_{i}({_{*}X}_{T})\approx-\zeta^{\alpha}\frac{\partial {_{*}\Phi}^{(1)}_{\alpha}}{\partial q^{i}}:\approx0,
\label{integrability in 1st system in Hamilton formulation}
\end{equation}
${_{*}\rho}_{i}({_{*}X}_{T})\approx0\ $s are automatically satisfied, where 
\begin{equation}
{_{*}\rho}_{i}:=K^{(1)}_{ij}\left(dq^{j}-\frac{\partial H}{\partial p_{j}}dt\right), {_{*}\theta}_{i}:=dp_{i}+\frac{\partial H}{\partial q^{i}}dt.
\label{}
\end{equation}
Therefore, if Eqs.~$(\ref{integrability in 1st system in Hamilton formulation})$ and $(\ref{the number of constraints})$ are satisfied, the system has the unique solutions under these constraints. 

\subsubsection{Another difficulty in Lagrange and Hamilton formulation}\label{Sec03:02:03}
The degeneracy of the kinetic matrix results in the absence of inverse for the map $\kappa$. Although the corresponding canonical momentum variable, $p_{i}=\partial L/\partial \dot{q}^{i}$, always exists for a velocity variable $\dot{q}^{i}$, making map $\kappa$ well-defined, its inverse is generically not exist. To demonstrate this, let us consider the transformation of velocity variables $v^{i}$\cite{SuganoKamo1982}: $\tilde{v}^{i}=v^{i}+u^{\alpha}\tau^{i}_{\alpha}$, where $u^{\alpha}$s are arbitrary functions. Expanding around $(q^{i},v^{i})$ up to first-order terms, we can show that $p(q^{i},\tilde{v}^{i})=p(q^{i},v^{i})$, where we used Eq.~$(\ref{zero-engenvalue vector})$. This result indicates that the inverse of map $\kappa$ does generically not exist. That is, there is no one-to-one correspondence between the velocity-phase space and the phase space. This feature describes another aspect for the absence of an inverse of map $\mathfrak{O}$; the total Hamiltonian does not determine a unique corresponding Lagrangian, despite the dynamics being unique.

\subsection{Higher-order systems}\label{Sec03:03}
A higher-order time derivative system can be decomposed into a first-order time derivative system with additional second-class constraints using the method of Lagrange multiplier\cite{Pons1989}. That is, a Lagrangian $L=L(D^{d}q^{i},\cdots,Dq^{i},q^{i})$ is decomposed as follows:
\begin{equation}
L\rightarrow\tilde{L}=L(\dot{Q}_{(d-1)}^{i},Q_{(d-1)}^{i},\cdots,Q_{(2)}^{i},Q_{(1)}^{i},Q_{(0)}^{i})
+\sum_{\alpha=1}^{d-1}\lambda^{(\alpha)}_{i}\left(Q_{(\alpha)}^{i}-\dot{Q}_{(\alpha-1)}^{i}\right),
\label{}
\end{equation}
where $Q_{(0)}^{i}:=q^{i}$. Regarding the Lagrange multipliers $\lambda^{(\alpha)}_{i}$ also as position coordinates, the argument of the rewritten Lagrangian $\tilde{L}$ becomes
\begin{equation}
\tilde{L}=\tilde{L}(\dot{Q}_{(d-1)}^{i},\cdots,\dot{Q}_{(1)}^{i},\dot{Q}_{(0)}^{i};Q_{(d-1)}^{i},\cdots,Q_{(1)}^{i},Q_{(0)}^{i};
\dot{\lambda}^{(\alpha)}_{i};\lambda^{(\alpha)}_{i}).
\end{equation}
For this Lagrangian, all the discussions for first-order derivative systems in Sects.~\ref{Sec03:01} and \ref{Sec03:02} are applicable. The present work focuses on the systems which are compatible with Newtonian dynamics: the systems whose Lagrangian is given by Eq.~$(\ref{L in 2nd system})$. An example of such system is provided in Sect.~\ref{Sec05:04}.

\section{Conditions for well-posed variational principle}\label{Sec04}
\subsection{Problems and a strategy for resolution}\label{Sec04:01}
In Sect.~\ref{Sec02}, we verified that position-fixing boundary conditions are necessary for the variational principle to be compatible with Newtonian mechanics. We showed, in Sect.~\ref{Sec03:01:01} and~\ref{Sec03:01:02}, that non-degenerate systems have unique solutions up to integral constants that are indicated by the Frobenius integrability. Lastly, in Sect.~\ref{Sec03:01:03}, we introduced the three fundamental maps $\iota$, $\kappa$, and $\mathfrak{O}$ in a well-defined manner, and these maps had their inverses in non-degenerate systems, respectively. In particular, the invertible map $\iota$ is crucial since $\iota$ connects the integral constants with boundary conditions. Therefore, we would define the well-posedness of variational principle as follows:\\ \\
\hypertarget{definition01}{\textbf{Definition 1.}}\\
{\it{The variational principle is well-posed if and only if the three fundamental maps $\iota$, $\kappa$, and $\mathfrak{O}$ are well-defined and invertible on a phase (sub)space in which the physical phase space exists.}}
\begin{flushright}$\blacksquare$\end{flushright}
In Sect. \ref{Sec03:02}, we revealed that degenerate systems give rise to the following two difficulties:\\ \\
\textbf{Difficulty 1.}
\begin{enumerate}
\item {\it{Map $\iota$ is generically not introduced in a well-defined manner}}
\item {\it{Maps $\kappa$ and $\mathfrak{O}$ are always introduced in a well-defined manner, but their inverses generically do not exist, respectively}}
\end{enumerate}
\begin{flushright}$\blacksquare$\end{flushright}
Therefore, to establish a well-posed variational principle in degenerate systems, we have to remove these incompatibilities out from the theory. That is, we have to resolve the following two problems:\\ \\
\textbf{Problem 1.}
\begin{enumerate}
\item {\it{How to introduce a well-defined and invertible map $\iota$}}
\item {\it{How to restore the invertibility of the maps $\kappa^{-1}$ and $\mathfrak{O}^{-1}$}}
\end{enumerate}
{\it{when the Frobenius integrability conditions Eqs~$(\ref{integrability in 1st system in Lagrange formulation})$ and $(\ref{integrability in 1st system in Hamilton formulation})$ are satisfied under the conditions restricting the number of constraints~$(\ref{the number of constraints})$.}}
\begin{flushright}$\blacksquare$\end{flushright}
The strategy to tackle this problem is that we restrict the entire phase space to a subspace such that each invertible map for $\kappa$ and $\mathfrak{O}$ exists, and then we construct a well-defined map $\iota$ in the subspace of which the inverse map exists. The construction of subspaces of phase space, however, differs from the cases of ordinary differential manifolds. That is, arbitrary restrictions, or strictly speaking, embeddings, are not allowed unlike ordinary differential manifolds. 

Let us consider an embedding $\psi:S\rightarrow \mathbb{R}^{3}$ for some 2-dimensional differential manifold $S$. When $\mathbb{R}^{3}$ is equipped with polar-coordinates $(r,\theta, \varphi)$, we can identify an embedding by fixing the pullback of some coordinate functions by $\psi$. For instance, if we fix $\psi^{*}r$ as 1 then we get an embedding $\psi:S\rightarrow \mathbb{R}^{3};(\psi^{*}r=1,\psi^{*}\theta, \psi^{*}\varphi)\mapsto (r,\theta,\varphi)$; this is none other than the embedding of the unit sphere, $S^{2}$, where $\psi^{*}$ is the pullback operator of $\psi$. We can also identify an embedding for the case $S$ is a 1-dimensional differential manifold by fixing that both $\psi^{*}\theta$ and $\psi^{*}\varphi$ are constant, respectively, $\psi:S\rightarrow \mathbb{R}^{3};(\psi^{*}r,\psi^{*}\theta=constant, \psi^{*}\varphi=constant)\mapsto (r,\theta,\varphi)$; this is none other than a line. Similarly, fixing either $\psi^{*}\theta$ or $\psi^{*}\varphi$ leads to an embedding of a 2-dimensional plane. In the case of phase space, however, embeddings by fixing coordinate functions are generically restricted. For instance, any odd-number dimensional subspaces cannot be embedded into the entire phase space. Therefore, {\it{we have to realize this restriction as an embedding into the entire phase space such that the canonical structure holds at least for the subspace in which the dynamics lives.}} Let us call such embedding {\it{"canonical"}} if it exists. To consider canonical embeddings, we need to introduce the concept of symplectic manifolds.

Finally, note that once we compose such embeddings, we can freely map objects, such as equations and those solutions, from the original space to an embedded subspace. We will use this technique for specific computations.

\subsection{Symplectic manifold and peculiarity of its embedding}\label{Sec04:02}
At each time $t\in I$, where $I\subset\mathbb{R}$ is an open interval of the time variable, a space $T^{*}M\times\mathbb{R}$ is equivalent to $T^{*}M$. Then a symplectic manifold of $T^{*}M$ is defined as the structure $(T^{*}M, \omega)$ equipped with a 2-form $\omega$ on $T^{*}M$ represented by 
\begin{equation}
\omega:=\frac{1}{2}\omega_{mn}dz^{m}\wedge dz^{n}
\label{general symplectic 2-form}
\end{equation}
satisfying the following three conditions:
\begin{equation}
\begin{split}
&{\textrm{(i).   }} \omega_{mn}=-\omega_{nm}\\
&{\textrm{(ii).  }} {\textrm{det }}\omega_{mn}\neq0\\
&{\textrm{(iii). }} d\omega=0.
\end{split}
\label{symplectic 2-form}
\end{equation}
The $\omega$ plays a role like a metric tensor on the phase space by virtue that $\omega(v_{1},v_{2})=v_{1}^{m}v_{2}^{n}\omega_{mn}$ for vector fields $v_{1}=v_{1}^{m}(\partial/\partial z^{m})$, $v_{2}=v_{2}^{n}(\partial/\partial z^{n})$ on $T(T^{*}M)$. We call this $\omega$ symplectic 2-form.

Darboux theorem under the conditions in Eq.~$(\ref{symplectic 2-form})$ leads to the canonical structure:
\begin{equation}
\omega:=\frac{1}{2}J_{mn}dz^{m}\wedge dz^{n}=dq^{i}\wedge dp_{i}
\label{special symplectic 2-form}
\end{equation}
at least in a local region on the symplectic manifold $(T^{*}M, \omega)$, where $(z^{i},z^{n+i})=(q^{i},p_{i})$ and 
\begin{equation}
J=\begin{bmatrix}
0 & I_{n\times n}  \\
-I_{n\times n} & 0 
\end{bmatrix}.
\end{equation}
$I_{n\times n}$ is an $n\times n$ unit matrix. Then the Poisson bracket (P.b.) is defined as follows:
\begin{equation}
\{f,g\}:=\omega(X_{f},X_{g}),
\label{definition of P.b.}
\end{equation}
where $X_{f}$ and $X_{g}$ are the Hamiltonian vector fields of $f$ and $g$, respectively. Using these concepts, we can define a canonical transformation as a coordinate transformation $\phi:T^{*}M\rightarrow T^{*}M;(q^{i},p_{i})\mapsto(Q^{i},P_{i})$ such that the symplectic 2-form is invariant under the pullback operation $\phi^{*}$:
\begin{equation}
\phi^{*}\omega_{Q,P}=\omega_{q,p}
\label{definition of canonical transformation}
\end{equation}
where
\begin{equation}
\begin{split}
&\omega_{q,p}:=dq^{i}\wedge dp_{i}=\frac{1}{2}J_{mn}dz^{m}\wedge dz^{n},\\
&\omega_{Q,P}:=dQ^{i}\wedge dP_{i}=\frac{1}{2}J_{mn}dZ^{m}\wedge dZ^{n},
\end{split}
\label{}
\end{equation}
and $(Z^{i},Z^{n+i})=(Q^{i},P_{i})$. The condition $(\ref{definition of canonical transformation})$ is equivalent to 
\begin{equation}
S^{T}JS=J,
\label{necessary and sufficient condition for canonical transformation}
\end{equation}
where $S^{T}$ is the transpose of $S$, and S is given as follows:
\begin{equation}
\phi^{*}dZ^{m}=S^{m}_{n}dz^{n},
\label{}
\end{equation}
or more concretely,
\begin{equation}
S^{m}_{n}:=\frac{\partial Z^{m}}{\partial z^{n}}
\label{}
\end{equation}
for $Z^{m}=(\phi^{-1})^{*}z^{m}$. The definition of P.b. given in Eq.~$(\ref{definition of P.b.})$ implies that a canonical transformation does not change the P.b. . That is, the equations of motion are invariant; the dynamics remains unchanged before and after the transformation. Conversely, since the time evolution is decomposed into a series of infinitesimal canonical transformations, the symplectic 2-form Eq.~$(\ref{symplectic 2-form})$, or more general form Eq.~$(\ref{general symplectic 2-form})$, is invariant under the time evolution. 

Note that when we consider a subspace as an embedding into the symplectic manifold $(T^{*}M,\omega)$ by fixing some coordinates, a set of conjugate variables, i.e. pairs: $q^{i}$ and $p_{i}$, have to be fixed simultaneously. Fixing either $q^{i}$ or $p_{i}$ would destroy the symplectic structure of the phase space. That is, the first and/or second conditions in the definition~$(\ref{symplectic 2-form})$ would be violated. This is the peculiar property of symplectic manifolds differing from ordinary differential manifolds as mentioned at the end of Sect.~\ref{Sec04:01}. Remark that taking into account the dynamics, we have to consider $T^{*}M\times\mathbb{R}$ rather than $T^{*}M$ with the boundaries $T^{*}M\times\{t_{1}\}$ and $T^{*}M\times\{t_{2}\}$. 

\subsection{Phase space decomposition and canonical embedding}\label{Sec04:03}
The authors in Refs.~\cite{Shanmugadhasan1973,DominiciGomis1980,Dominici1982,MaskawaNakajima1976} proposed a novel theorem that states that a proper combination of constraints can form a part of a canonical coordinate system. On one hand, the authors in Refs.~\cite{Shanmugadhasan1973,DominiciGomis1980,Dominici1982} deduced the theorem without explicit proofs based on well-known facts in function group theory. The authors in Ref.~\cite{MaskawaNakajima1976}, on the other hand, provided rigorous proofs using their original methodology at least in the {\it{'weak'}} equality: '$\approx$'~\cite{Dirac1950,Dirac1958}. A mutual feature of these works is that the existence is smartly deduced/proved, but the explicit method to construct the coordinates is unclear. In this section, looking ahead to the application for specific models in Sect.~\ref{Sec05}, we provide the theorem together with rigorous and explicit proofs based on function group theory. The advance from the first set of previous works~\cite{Shanmugadhasan1973,DominiciGomis1980,Dominici1982} is that we provide a rigorous proof in terms of function groups and can guess an explicit method to compose a canonical coordinate system which is implied by the theorem. In addition, We will verify that the theorem holds in the sense of the {\it{'strong'}} equality: '$=$' \cite{Dirac1950,Dirac1958}, when a condition holds. (See \hyperlink{remark01}{Remark 1.}) The latter result is the advance from the previous work~\cite{MaskawaNakajima1976}. The theorem will be proved based on one proposition and two lemmas while introducing canonical embeddings.

\subsubsection{Function group and the existence of a reciprocal subgroup}\label{Sec04:03:01}
In order to decompose a phase space existing of constraints, we need the concept of the function group\cite{Eisenhart2003}.\\
\textbf{Definition 2.}\\ 
{\it{(i). Function group:\\
Let $\{f_{i}\}_{i=1,2,\cdots,r}$ be a set of functions on a symplectic manifold of which P.b.s are closed. Then a set of functions, $\{g_{a}\}_{a=1,2,\cdots,r}$, of $\{f_{i}\}_{i=1,2,\cdots,r}$, i.e. $g_{a}=g_{a}(f_{1},f_{2},\cdots,f_{r})$, is defined as a function group with the rank $r$ of the basis $\{f_{i}\}_{i=1,2,\cdots,r}$ if and only if all P.b.s among $\{g_{a}\}_{a=1,2,\cdots,r}$ are also closed. We denote $\{g_{a}\}_{a=1,2,\cdots,r}$ as $G_{r}(\{f_{i}\}_{i=1,2,\cdots,r})$. When the basis is apparent we abbreviate $G_{r}(\{f_{i}\}_{i=1,2,\cdots,r})$ by $G_{r}$.\\
(ii). (Non-)Commutative function group:\\
If all elements in a function group $G_{r}$ are commutative in P.b., then $G_{r}$ is called a commutative function group. If it is not the case, $G_{r}$ is called a non-commutative function group. \\
(iii). Subgroup:\\
For $s\leq r$, $G_{s}(\{f_{i}\}_{i=1,2,\cdots,s})\subset G_{r}(\{f_{i}\}_{i=1,2,\cdots,r})$ is called a subgroup of $G_{r}$.\\
(iv). Reciprocal subgroup:\\
Let $G_{n}$ be a non-commutative function group. For a subgroup $G_{r}$ $(r\leq n)$, if there exists a subgroup $G_{n-r}$ such that $G_{n}=G_{r}\sqcup G_{n-r}$ and all P.b.s between $G_{r}$ and $G_{n-r}$ vanish, then the subgroup $G_{n-r}$ is defined as a reciprocal subgroup of $G_{r}$, where $\sqcup$ is direct sum. We denote the reciprocal subgroup as $\bar{G}_{r}:=G_{n-r}$. Then, of course, $G_{r}$ is a reciprocal subgroup of $G_{n-r}$: $\bar{G}_{n-r}=G_{r}$. }}
\begin{flushright}$\blacksquare$\end{flushright}

We will use the existence of a reciprocal subgroup in Sect.~\ref{Sec04:03:03}\cite{Eisenhart2003}. \\
\hypertarget{proposition01}{\textbf{Proposition 1.}}\\
{\it{Let $G_{2n}$ be a non-commutative function group. Then, for any subgroup $G_{r}$, a reciprocal subgroup $\bar{G}_{r}=G_{2n-r}$ exists.\\
\textbf{Proof.}\\ 
Let $\{f_{a}\}_{a=1,2,\cdots,r}$ be a basis for $G_{r}$. Then we consider the following differential equations:
\begin{equation}
X_{f_{a}}g=0.
\label{equations for reciprocal subgroup}
\end{equation}
Using $X_{\{f_{a},f_{b}\}}=-[X_{f_{a}},X_{f_{b}}]$ and $\{f_{a},f_{b}\}=C_{ab}^{\ \ c}f_{c}$, where $C_{ab}^{\ \ c}$s are functions, we can verify that the operators, $X_{f_{a}}$s, form a complete set. Therefore, Frobenius theorem leads to $2n-r$ solutions for the equations: $\{g_{i}\}_{i=1,2,\cdots,2n-r}$. Using Jacobi identity, we can show that $\{g_{i}\}_{i=1,2,\cdots,2n-r}$ is also closed in P.b. . Taking into account the P.b.s between $\{f_{a}\}_{a=1,2,\cdots,r}$ and $\{g_{i}\}_{i=1,2,\cdots,2n-r}$ vanish by virtue of Eq.~$(\ref{equations for reciprocal subgroup})$, $\{g_{i}\}_{i=1,2,\cdots,2n-r}$ forms a basis for a reciprocal subgroup of $G_{r}$. (Q.E.D.)}}
\begin{flushright}$\blacksquare$\end{flushright}

\subsubsection{Systems with first-class constraint}\label{Sec04:03:02}
First, we consider a phase space only with first-class constraints. The phase space has a canonical coordinate system indicated as follows:\\
\hypertarget{lemma01}{\textbf{Lemma 1. }}\\
{\it{Let $\{\psi_{\alpha}\}_{\alpha=1,2,\cdots,r}$ be a set of first-class constraints in a $2n$-dimensional symplectic manifold. Then a canonical coordinate system: $\Xi^{\alpha}$s, $\Psi_{\alpha}$s, $Q^{a}$s, and $P_{a}$s such that $\{\Xi^{\alpha},\Psi_{\beta}\}=\delta^{\alpha}_{\ \beta}$, $\{Q^{a},P_{b}\}=\delta^{a}_{\ b}$, and otherwise vanish exists, and all of $\Xi^{\alpha}$s and $\Psi_{\alpha}$s satisfy those consistency conditions. Where $\alpha,\beta=1,2,\cdots,r$ and $a,b=1,2,\cdots,n-r$.\\
\textbf{Proof.}\\
The first-class constraints form a function subgroup with rank $r$ of $G_{2n}$ by its definition: $\{\psi_{\alpha},\psi_{\beta}\}=C(\psi_{\gamma})$ at least around the neighbor of the constraint space, where $C(\psi_{\gamma})$ is a function of which independent variables are the first-class constraints such that for all $\psi_{\gamma}\rightarrow0$ then $C(\psi_{\gamma})\rightarrow0$, or $\{\psi_{\alpha},\psi_{\beta}\}\approx0$. That is, $G_{r}=G_{r}(\{\psi_{i}\}_{i=1,2,\cdots,r})$. Then, for $\psi_{1}$, we consider the following differential equation:
\begin{equation}
X_{\psi_{1}}\xi=\{\xi,\psi_{1}\}\approx1
\label{step 0 in 1st-class into canonical}
\end{equation}
where $\xi$ is a function of which independent variables are belonging to $G_{2n-r}:=G_{2n}\setminus G_{r}$. This equation always has a solution at least in the weak equality, otherwise we find a new first-class constraint since $\xi$ is commutative with $\psi_{1}$ \footnote{If $\{\xi,\psi_{1}\}=f$, where $f$ is some function being non-zero in all region we consider, then just replacing $\psi_{1}$ by $\psi_{1}/f$, at least around the neighbor of the constraint space, we get $\{\xi,\psi_{1}/f\}\approx1$. Therefore, the absence of the solution of the equation implies $f=0$.}, but it contradicts that the Dirac procedure always takes the number of first-class constraints to the maximum. Let us denote the solution as $\Xi^{1}$ and set $\Psi_{1}:=\psi_{1}$. Then we get 
\begin{equation}
\{\Xi^{1},\Psi_{1}\}\approx1.
\label{step 1 in 1st-class into canonical}
\end{equation}
For these variables $\Xi^{1}$ and $\Psi_{1}$, we consider the following differential equations:
\begin{equation}
\begin{split}
&X_{\Xi_{1}}\Theta=0,\\
&X_{\Psi_{1}}\Theta=0,\\
\end{split}
\label{step 2 in 1st-class into canonical}
\end{equation}
where $\Theta$ is a function of which independent variables are belonging to $G_{2n-r-1}=G_{2n}\setminus(G_{r}\sqcup\{\Xi^{1}\})$. These equations implies that $X_{\Xi^{1}}$ and $X_{\Psi_{1}}$ form a complete set by virtue of $[X_{\Xi^{1}},X_{\Psi_{1}}]=-X_{\{\Xi^{1},\Psi_{1}\}}$ and Eq.~$(\ref{step 1 in 1st-class into canonical})$. Therefore, based on Frobenius theorem, there exists a set of solutions for Eq.~$(\ref{step 2 in 1st-class into canonical})$: $\Theta_{1},\Theta_{2},\cdots,\Theta_{2n-r-1}$, and these solutions form a function subgroup $G_{2n-r-1}=G_{2n-r-1}(\{\Theta_{a}\}_{a=1,2,\cdots,2n-r-1})$. Appending the two functions $\Xi^{1}$ and $\Psi_{1}$ to the basis of $G_{2n-r-1}$, we get $G_{2n-r+1}=G_{2n-r+1}(\{\Xi^{1},\Psi_{1},\Theta_{a}\}_{a=1,2,\cdots,2n-r-1})$ by virtue of $G_{2n-r-1}(\{\Theta_{a}\}_{a=1,2,\cdots,2n-r-1})\subset G_{2n-r+1}$. For other remaining elements $\psi_{\alpha}\in G_{r-1}$, the same processes lead to $G_{2n}=G_{2n}(\{\Xi^{\alpha}\}_{\alpha=1,2,\cdots,r},\{\Psi_{\alpha}\}_{\alpha=1,2,\cdots,r},\{\Theta_{a}\}_{a=1,2,\cdots,2n-2r})$. Since $G_{2n-2r}(\{\Theta_{a}\}_{a=1,2,\cdots,2n-2r})$ is a non-commutative subgroup, Eq.~$(\ref{step 0 in 1st-class into canonical})$ for $\Theta_{a}$s on $G_{2n-2r}=G_{2n-2r}(\{\Theta_{a}\}_{a=1,2,\cdots,2n-2r})$ gives a basis of $G_{2n-2r}=G_{2n-2r}(\{Q^{a}\}_{a=1,2,\cdots,n-r},\{P_{a}\}_{a=1,2,\cdots,n-r})$ such that $\{Q^{a},P_{b}\}=\delta^{a}_{\ b}$, and otherwise vanish. Therefore, we get $G_{2n}=G_{2n}(\{\Xi^{\alpha}\}_{\alpha=1,2,\cdots,r},\\ \{\Psi_{\alpha}\}_{\alpha=1,2,\cdots,r},\{Q^{a}\}_{a=1,2,\cdots,n-r},\{P_{a}\}_{a=1,2,\cdots,n-r})$. The statement is concluded. (Q.E.D.)}}
\begin{flushright}$\blacksquare$\end{flushright}
{\hypertarget{remark01}{\textbf{Remark 1.}}}\\
If the equation Eq.~$(\ref{step 0 in 1st-class into canonical})$ holds in the {\it{strong}} equality for all the processes, these statements hold also in the {\it{strong}} equality.
\begin{flushright}$\blacksquare$\end{flushright}

\hyperlink{lemma01}{Lemma 1} indicates that a canonical transformation $\phi:T^{*}M\rightarrow T^{*}M; (q^{i},p_{i})\mapsto (\Xi^{\alpha},\Psi_{\alpha},Q^{i},P_{i})$ under the necessary and sufficient condition Eq.~$(\ref{necessary and sufficient condition for canonical transformation})$ exists, and then we have
\begin{equation}
\omega=dq^{i}\wedge dp_{i}=d\Xi^{\alpha}\wedge d\Psi_{\alpha}+dQ^{i}\wedge dP_{i},
\label{decomposed symplectic 2-form in 1st-class}
\end{equation}
where the index $i$ in the domain of $\phi$ runs from $1$ to $n$. The indix $\alpha$ and $i$ in the range of $\phi$ run from $1$ to $r$ and from $1$ to $n-r$, respectively. In addition, from the proof of \hyperlink{lemma01}{Lemma 1}, since the subgroups $G_{2r}=G_{2r}(\Xi^{\alpha},\Psi_{\alpha})$ and $G_{2n-2r}=G_{2n-2r}(Q^{i},P_{i})$ relate with one another in the reciprocal manner, the phase space can be decomposed as follows:
\begin{equation}
T^{*}M=T^{*}M|_{\Xi,\Psi}\times T^{*}M|_{Q,P}
\label{}
\end{equation}
where we denote $T^{*}M|_{\Xi,\Psi}$ and $T^{*}M|_{Q,P}$ as the phase subspaces of $T^{*}M$, which are coordinated by $\Xi^{\alpha},\Psi_{\alpha}$, and, $Q^{i},P_{i}$, respectively. 

Next, let us consider an embedding of $T^{*}M|_{Q,P}$ into $T^{*}M$, i.e.  $\sigma_{1}:T^{*}M|_{Q,P}\rightarrow T^{*}M$, by fixing the pullback of the canonical coordinates $\Xi^{\alpha}$, $\Psi_{\alpha}$: $\sigma_{1}^{*}\Xi^{\alpha}$, $\sigma_{1}^{*}\Psi_{\alpha}$, where $\sigma_{1}^{*}$ denotes a pullback operator of $\sigma_{1}$. Then an embedding is given as follows:
\begin{equation}
\sigma_{1}:T^{*}M|_{Q,P}\rightarrow T^{*}M ;(\sigma_{1}^{*}\Xi^{\alpha}:=\epsilon^{\alpha},\sigma_{1}^{*}\Psi_{\alpha}:=\epsilon_{\alpha},\sigma_{1}^{*}Q^{i},\sigma_{1}^{*}P_{i})\mapsto (\Xi^{\alpha},\Psi_{\alpha},Q^{i},P_{i})
\label{}
\end{equation}
where $\epsilon^{\alpha}$s and $\epsilon_{\alpha}$s are constant parameters. The reason why we fixed $\Xi^{\alpha}$s and $\Psi_{\alpha}$s, not $Q^{i}$s and $P_{i}$s, will be revealed as we consider the time evolution of the system. Using this embedding, the symplectic manifold $(T^{*}M,\omega)$ can be decomposed into two submanifolds. That is, we consider the pullback of the symplectic 2-form Eq.~$(\ref{decomposed symplectic 2-form in 1st-class})$ by $\sigma_{1}$:
\begin{equation}
\sigma_{1}^{*}\omega=\omega_{Q,P},
\label{pullback of symplectic 2-form in 1st-class constraints}
\end{equation}
where
\begin{equation}
\begin{split}
&\omega_{\Xi,\Psi}:=d\sigma_{1}^{*}\Xi^{\alpha} \wedge d\sigma_{1}^{*}\Psi_{\alpha}=0,\\
&\omega_{Q,P}:=d\sigma_{1}^{*}Q^{a}\wedge d\sigma_{1}^{*}P_{a}
.
\end{split}
\label{pullback of symplectic 2-form in 1st-class constraints on each space}
\end{equation}
Then we obtain the two submanifolds $(T^{*}M|_{\Xi,\Psi},\omega_{\Xi,\Psi}=0)$ and $(T^{*}M|_{Q,P},\omega_{Q,P})$. The former submanifold does, on one hand, not have the symplectic structure; the second condition for the symplectic 2-form in Eq.~$(\ref{symplectic 2-form})$ is not satisfied. On the other hand, the latter submanifold holds the symplectic structure. That is, the embedding $\sigma_{1}$ is canonical.

Now, to take into account the time evolution of the system, we consider an embedding such that $\sigma_{1}(t):T^{*}M|_{Q,P}\times \mathbb{R}\rightarrow T^{*}M\times \mathbb{R}$; the embedding is now parameterized by a time variable $t$. In this case, on one hand, the first-class constraints, or equivalently $\Psi_{\alpha}$s, satisfy the consistency conditions:
\begin{equation}
\dot{\Psi}_{\alpha}=\{\Psi_{\alpha},H_{T}\}\approx0.
\end{equation}
On the other hand, $\Xi^{\alpha}$s evolve depending on time. That is, there are two possibilities to introduce the embedding $\sigma_{1}(t)$. First, fixing only $\sigma^{*}_{1}\Psi_{\alpha}$s, the embedding is introduced as follows:
\begin{equation}
\begin{split}
\tilde{\sigma}_{1}(t):&T^{*}M\times \mathbb{R}\rightarrow T^{*}M\times \mathbb{R}\\
&;({\sigma_{1}^{*}}(t)\Xi^{\alpha},{\sigma_{1}^{*}}(t)\Psi_{\alpha}:=\epsilon_{\alpha},{\sigma_{1}^{*}}(t)Q^{i},{\sigma_{1}^{*}}(t)P_{i},{\sigma_{1}^{*}}(t)u=t)\mapsto (\Xi^{\alpha},\Psi_{\alpha},Q^{i},P_{i},u).
\end{split}
\label{}
\end{equation}
${\sigma_{1}^{*}}(t)\Xi^{\alpha}$s are not fixed since the corresponding variables $\Xi^{\alpha}$s evolve in time. The pullback of the symplectic 2-form $\omega$ by $\tilde{\sigma}_{1}(t)$ turns Eqs.~$(\ref{pullback of symplectic 2-form in 1st-class constraints})$ and $(\ref{pullback of symplectic 2-form in 1st-class constraints on each space})$ into as follows:
\begin{equation}
{\sigma_{1}^{*}}(t)\omega=\omega_{Q,P}(t)+\omega_{\Xi,\Psi}(t),
\label{pullback of symplectic 2-form in 1st-class constraints without gauge-fixing}
\end{equation}
where
\begin{equation}
\begin{split}
&\omega_{\Xi,\Psi}(t):=d\Xi^{\alpha}(t)\wedge d\Psi_{\alpha}(t)=d\Xi^{\alpha}(t)\wedge0
,\\
&\omega_{Q,P}(t):=dQ^{a}(t)\wedge dP_{a}(t)
.
\end{split}
\label{pullback of symplectic 2-form in 1st-class constraints without gauge-fixing on each space}
\end{equation}
This result indicates that the case without the consistency conditions for $\Xi^{\alpha}$s does not reduce the phase space into the physical space. In addition, since the first and/or the second condition of the definition for symplectic 2-form $(\ref{symplectic 2-form})$ would be violated, this embedding is not canonical as it is. However, we will confirm that this embedding can be canonical as we impose gauge fixing conditions. This is a peculiar feature of systems existing of first-class constraints. Let us call this sort of embeddings being {\it{"quasi-canonical"}}. Notice that if we take the limit of $\epsilon_{\alpha}\rightarrow0$, then $\tilde{\sigma}_{1}(t)$ restores the constraint space wherein all the constraints are satisfied; this is another reason why ${\sigma_{1}^{*}}(t)\Xi^{\alpha}$s cannot be fixed in ${\sigma_{1}^{*}}(t)$. $\Xi^{\alpha}$s do not restrict the constraint space at all. Notice also that $\tilde{\sigma}_{1}(t)$ occupies $r$ integral constants that are indicated by the Frobenius integrability given in Sect.~\ref{Sec03}. 
 
Second, in contrast to the case of quasi-canonical, if the consistency conditions for $\Xi^{\alpha}$s are imposed:
\begin{equation}
\begin{split}
&\Xi^{\alpha}:\approx0,\\
&\dot{\Xi}^{\alpha}=\{\Xi^{\alpha},H_{T}\}:\approx0,
\end{split}
\label{gauge fixing}
\end{equation}
the phase subspace $T^{*}M|_{\Xi,\Psi}\times \mathbb{R}$ turns to be static. These conditions correspond to gauge fixing\cite{SuganoSaitoKimura1986,SuganoKagraokaKimura1992}. Therefore, a canonical embedding is given as follows:
\begin{equation}
\begin{split}
\sigma_{1}(t):&T^{*}M|_{Q,P}\times \mathbb{R}\rightarrow T^{*}M\times \mathbb{R}\\
&;({\sigma_{1}^{*}}(t)\Xi^{\alpha}:=\epsilon^{\alpha},{\sigma_{1}^{*}}(t)\Psi_{\alpha}:=\epsilon_{\alpha},{\sigma_{1}^{*}}(t)Q^{i},{\sigma_{1}^{*}}(t)P_{i},{\sigma_{1}^{*}}(t)u=t)\\
&\mapsto (\Xi^{\alpha},\Psi_{\alpha},Q^{i},P_{i},u).
\end{split}
\label{}
\end{equation}
For at a time $t$, $\sigma_{1}(t)$ restores, of course, $\sigma_{1}$. Hereinafter we abbreviate the pullback of a variable $X$, ${\sigma_{1}^{*}}(t)X$, as $X(t)$, when it is apparent in the context. Applying this embedding to the entire phase space $T^{*}M$ and the symplectic 2-form $\omega$ in the same manner to the case of $\sigma_{1}$, we get a decomposition of $(T^{*}M\times\mathbb{R},\omega(t)=dq^{i}(t)\wedge dp_{i}(t))$: $(T^{*}M|_{Q,P}\times\mathbb{R},\omega_{Q,P}(t)=dQ^{i}(t)\wedge dP_{i}(t))$ and $(T^{*}M|_{\Xi,\Psi}\times\mathbb{R},\omega_{\Xi,\Psi}(t)=d\Xi(t)\wedge d\Psi(t)=0)$. Note that $\omega_{\Xi,\Psi}(t)=0$ on $T^{*}M|_{\Xi,\Psi}\times\mathbb{R}$ throughout all time implies that the subspace $T^{*}M|_{\Xi,\Psi}\times\mathbb{R}$ can be removed from describing the dynamics. In contrast, $\omega_{Q,P}(t)\neq0$ on $T^{*}M|_{Q,P}\times\mathbb{R}$ holding with the symplectic structure describes the dynamics. Then $T^{*}M|_{Q,P}\times\mathbb{R}$ becomes the physical space. Notice that, taking the limit of $\epsilon^{\alpha}\rightarrow0$ and $\epsilon_{\alpha}\rightarrow0$, $\sigma_{1}(t)$ restores the constraint space. $\sigma_{1}(t)$ occupies $r+r=2r$ integral constants that are indicated by the Frobenius integrability in Sect.~\ref{Sec03}.

\subsubsection{Systems with second-class constraint}\label{Sec04:03:03}
Second, we consider a phase space only with second-class constraints. The phase space has a canonical coordinate system indicated as follows:\\
\hypertarget{lemma02}{\textbf{Lemma 2. }}\\
{\it{Let $\{\theta_{i}\}_{i=1,2,\cdots,2s}$ be a set of second-class constraints on a $2n$-dimensional symplectic manifold. Then a canonical coordinate system $\Theta_{\alpha}$s, $\Theta^{\alpha}$s, $Q^{a}$s, and $P_{a}$s such that $\{\Theta_{\alpha},\Theta^{\beta}\}=\delta^{\alpha}_{\ \beta}$, $\{Q^{a},P_{b}\}=\delta^{a}_{\ b}$, and otherwise vanish exists, and all of $\Theta_{\alpha}$s and $\Theta^{\alpha}$s satisfy those consistency conditions. Where $\alpha,\beta=1,2,\cdots,s$ and $a,b=1,2,\cdots,n-s$. The same remark to \hyperlink{remark01}{Lemma 1} holds.\\ 
\textbf{Proof.}\\
The second-class constraints do not directly form a function subgroup since the definition implies that $\{\theta_{i},\theta_{j}\}\neq C(\theta_{k})$: P.b.s among $\theta_{i}$s cannot be expressed only in $\theta_{i}$s themselves. However, we can reconstruct $\theta_{i}$s as follows:
\begin{equation}
\Theta_{i}=C_{i}^{\ j}\theta_{j},
\label{}
\end{equation}
where $C_{i}^{\ j}$s are arbitrary functions on the symplectic manifold. Then we impose the following conditions:
\begin{equation}
\{\Theta_{\alpha},\Theta_{s+\beta}\}=\delta_{\alpha\beta},
\label{step 1 in 2nd-class into canonical}
\end{equation}
and otherwise vanish, where $\alpha,\beta=1,2,\cdots,s$. The number of all the conditions above is $3s^{2}$, and this number is less than the number of independent components of $C_{i}^{\ j}$s: $4s^{2}$. Therefore, we can determine $C_{i}^{\ j}$s satisfying Eq.~$(\ref{step 1 in 2nd-class into canonical})$ although it is not unique. To form a function subgroup, let us consider the following equations:
\begin{equation}
\{\Theta_{i},\Theta_{j}\}=C_{ij}^{\ \ k}\Theta_{k},
\label{step 2 in 2nd-class into canonical}
\end{equation}
where $C_{ij}^{\ \ k}$s are functions on the symplectic manifold such that Eq.~$(\ref{step 1 in 2nd-class into canonical})$ is satisfied. It is possible to replace $C_{ij}^{\ \ k}\Theta_{k}$ by $f_{ij}(\Theta_{k})$ such that it is not weak equal to zero: for all $\Theta_{k}\rightarrow0$ then $f_{ij}(\Theta_{k})\nrightarrow0$, as follows:
\begin{equation}
\{\Theta_{i},\Theta_{j}\}=f_{ij}(\Theta_{k})
\label{step 3 in 2nd-class into canonical}
\end{equation}
This is none other than a generalization of Eq.~$(\ref{step 1 in 2nd-class into canonical})$. These equations lead to a function subgroup of $G_{2n}$: $G_{2s}=G_{2s}(\{\Theta_{i}\}_{i=1,2,\cdots,2s})$. Therefore, from \hyperlink{proposition01}{Proposition 1}, we get a reciprocal subgroup of $G_{2s}=G_{2s}(\{\Theta_{i}\}_{i=1,2,\cdots,2s})$: $\bar{G}_{2s}=G_{2n-2s}$. For each subgroup $G_{2s}$ and $G_{2n-2s}$, we can construct canonical variables in the same manner as the proof of \hyperlink{lemma01}{Lemma 1}. That is, $G_{2s}=G_{2s}(\{\Theta^{\alpha}\}_{\alpha=1,2,\cdots,s},\{\Theta_{\alpha}\}_{\alpha=1,2,\cdots,s})$ with $\{\Theta^{\alpha},\Theta_{\beta}\}=\delta^{\alpha}_{\ \beta}$, and otherwise vanish, and, $\bar{G}_{2s}=G_{2n-2s}(\{Q^{a}\}_{a=1,2,\cdots,n-s},\{P_{a}\}_{a=1,2,\cdots,n-s})$ with $\{Q^{a},P_{b}\}=\delta^{a}_{\ b}$, and otherwise vanish. The former recovers Eq.~$(\ref{step 1 in 2nd-class into canonical})$. Since these subgroups are reciprocal subgroups to one another, we get $G_{2s}\sqcup G_{2n-2s}=G_{2n}(\{\Theta^{\alpha}\}_{\alpha=1,2,\cdots,s},\{\Theta_{\alpha}\}_{\alpha=1,2,\cdots,s},\{Q^{a}\}_{a=1,2,\cdots,n-s},\{P_{a}\}_{a=1,2,\cdots,n-s})$. (Q.E.D.)}}
\begin{flushright}$\blacksquare$\end{flushright}

\hyperlink{lemma02}{Lemma 2} indicates the existence of a canonical transformation $\phi:T^{*}M\rightarrow T^{*}M; (q^{i},p_{i})\mapsto (\Theta^{\alpha},\Theta_{\alpha},Q^{i},P_{i})$ under the necessary and sufficient condition Eq.~$(\ref{necessary and sufficient condition for canonical transformation})$. The symplectic 2-form $\omega$ and the phase space $T^{*}M$ are decomposed as follows:
\begin{equation}
\omega=d\Theta^{\alpha}\wedge d\Theta_{\alpha}+dQ^{i}\wedge dP_{i}
\label{}
\end{equation}
and 
\begin{equation}
T^{*}M=T^{*}M|_{\Theta}\times T^{*}M|_{Q,P},
\label{}
\end{equation}
respectively, where $T^{*}M|_{\Theta}$, and, $T^{*}M|_{Q,P}$ are the phase subspaces of $T^{*}M$ of which canonical coordinates are given by $\Theta^{\alpha}$, $\Theta_{\alpha}$ and $Q^{i}$, $P_{i}$, respectively. The index $i$ in the domain of $\phi$ runs from $1$ to $n$. The index $\alpha$ runs from 1 to $s$ and the index $i$ in the range of $\phi$ runs from $1$ to $n-s$. Then the canonical embedding is given as follows:
\begin{equation}
\sigma_{2}:T^{*}M|_{Q,P}\rightarrow T^{*}M;(\sigma_{2}^{*}\Theta^{\alpha}:=\epsilon^{\alpha},\sigma_{2}^{*}\Theta_{\alpha}:=\epsilon_{\alpha},\sigma_{2}^{*}Q^{i},\sigma_{2}^{*}P_{i})\mapsto (\Theta^{\alpha},\Theta_{\alpha},Q^{i},P_{i}),
\label{}
\end{equation}
where $\sigma_{2}^{*}$ is the pullback operator of $\sigma_{2}$, $\epsilon^{\alpha}$s and $\epsilon_{\alpha}$s are constant parameters. Applying the embedding, $(T^{*}M,\omega)$ is decomposed into $(T^{*}M|_{\Theta},\omega_{\Theta})$ and $(T^{*}M|_{Q,P},\omega_{Q,P})$, where
\begin{equation}
\begin{split}
&\omega_{\Theta}:=d\sigma^{*}\Theta^{\alpha}\wedge d\sigma^{*}\Theta_{\alpha}=0,\\
&\omega_{Q,P}:=d\sigma^{*}Q^{i}\wedge d\sigma^{*}P_{i}
.
\end{split}
\label{}
\end{equation}
The pullback of $\omega$ by $\sigma_{2}$ is, of course, given as follows:
\begin{equation}
\sigma_{2}^{*}\omega=\omega_{Q,P}.
\label{}
\end{equation}
Therefore, only the submanifold $(T^{*}M|_{Q,P},\omega_{Q,P})$ holds the symplectic structure.

Taking into account the time evolution of the system, since the second-class constraints, or equivalently $\Theta^{\alpha}$s and $\Theta_{\alpha}$s, satisfy the consistency conditions, we lead straightforwardly to the canonical embedding with the time evolution as follows:
\begin{equation}
\begin{split}
\sigma_{2}(t):&T^{*}M|_{Q,P}\times \mathbb{R}\rightarrow T^{*}M\times \mathbb{R}\\
&;({\sigma_{2}^{*}}(t)\Theta^{\alpha}:=\epsilon^{\alpha},{\sigma_{2}^{*}}(t)\Theta_{\alpha}:=\epsilon_{\alpha},{\sigma_{2}^{*}}(t)Q^{i},{\sigma_{2}^{*}}(t)P_{i},{\sigma_{2}^{*}}(t)u=t)\\
&\mapsto (\Theta^{\alpha},\Theta_{\alpha},Q^{i},P_{i},u),
\end{split}
\label{}
\end{equation}
without any conditions in contrast to the case only existing of first-class constraints. Then we can decompose $(T^{*}M\times \mathbb{R},\omega(t)=dq^{i}(t)\wedge dp_{i}(t))$ into the same structure as in the case of $\sigma_{2}$: $(T^{*}M|_{Q,P}\times \mathbb{R},\omega_{Q,P}(t)=d{\sigma_{2}^{*}}(t)Q^{i}(t)\wedge d{\sigma_{2}^{*}}(t)P_{i}(t))$ and $(T^{*}M|_{\Theta}\times \mathbb{R},\omega_{\Theta}=d{\sigma_{2}^{*}}(t)\Theta^{\alpha}(t)\wedge d{\sigma_{2}^{*}}(t)\Theta_{\alpha}(t)=0)$. $\omega_{\Theta}(t)=0$ on $T^{*}M|_{\Theta}\times \mathbb{R}$ throughout all time implies that it does not relate to the dynamics, but $\omega_{Q,P}(t)\neq0$ on $T^{*}M|_{Q,P}\times \mathbb{R}$ with the symplectic structure describes the dynamics, and this subspace is none other than the physical one. Notice that taking limits of $\epsilon^{\alpha}\rightarrow0$ and $\epsilon_{\alpha}\rightarrow0$, $\sigma_{2}(t)$ restores the constraint space. $\sigma_{2}(t)$ occupies $2s$ integral constants that are indicated by the Frobenius integrability in Sect.~\ref{Sec03}.

\subsubsection{Systems with first- and second-class constraint}\label{Sec04:03:04}
Finally, we consider a phase space with both first- and second-class constraints. Combining \hyperlink{lemma01}{Lemma 1} and \hyperlink{lemma02}{2}, the following theorem holds\cite{Shanmugadhasan1973,DominiciGomis1980,Dominici1982,MaskawaNakajima1976}.\\
\hypertarget{theorem01}{\textbf{Theorem 1.}}\\
{\it{Let $\{\psi_{\alpha}\}_{\alpha=1,2,\cdots,r}$ and $\{\theta_{i}\}_{i=1,2,\cdots,2s}$ be a set of first-class and second-class constraints, respectively. Then a canonical coordinate system $\Xi^{a}$s, $\Psi_{a}$s, $\Theta^{\alpha}$s, $\Theta_{\alpha}$s, $Q^{i}$s, and $P_{i}$s such that $\{\Xi^{a},\Psi_{b}\}=\delta^{a}_{\ b}$, $\{\Theta^{\alpha},\Theta_{\beta}\}=\delta^{\alpha}_{\ \beta}$, $\{Q^{i},P_{j}\}=\delta^{i}_{\ j}$, and otherwise vanish exists, and all of the $\Xi^{a}$s, $\Psi_{a}$s, $\Theta^{\alpha}$s, and $\Theta_{\alpha}$s satisfy those consistency conditions. Where $a,b=1,2,\cdots,r$, $\alpha,\beta=1,2,\cdots,s$ and $i,j=1,2,\cdots,n-r-s$. The same remark to \hyperlink{remark01}{Lemma 1} holds. \\
\textbf{Proof.}\\
Combining \hyperlink{lemma01}{Lemma 1} and \hyperlink{lemma02}{2}, we get the statement by virtue of the relation: $G_{2n}=G_{2r}\sqcup G_{2s}\sqcup G_{2n-2r-2s}=G_{2n}(\Xi^{a},\Psi_{b},\Theta^{\alpha},\Theta_{\beta},Q^{i},P_{j})$, where $G_{2r}=G_{2r}(\Xi^{a},\Psi_{b})$, $G_{2s}=G_{2s}(\Theta^{\alpha},\Theta_{\beta})$, and $G_{2n-2r-2s}=G_{2n-2r-2s}(Q^{i},P_{j})$. Remark that $\Theta^{\alpha}$s and $\Theta_{\alpha}$s are generically written as $\Theta^{\alpha}=f^{\alpha}(\psi_{\alpha},\theta_{i})$ and $\Theta_{\alpha}=f_{\alpha}(\psi_{\alpha},\theta_{i})$, respectively, but these generalizations do not affect the proof of \hyperlink{lemma02}{Lemma 2} except being valid only in the weak equality. (Q.E.D.)}}
\begin{flushright}$\blacksquare$\end{flushright}

\hyperlink{theorem01}{Theorem 1} indicates that the existence of a canonical transformation $\phi:T^{*}M\rightarrow T^{*}M; (q^{i},p_{i})\mapsto (\Xi^{\alpha},\Psi_{\alpha},\Theta^{a},\Theta_{a},Q^{i},P_{i})$ under the necessary and sufficient condition $(\ref{necessary and sufficient condition for canonical transformation})$, and it leads to the following decomposition: 
\begin{equation}
\omega=d\Xi^{a}\wedge d\Psi_{a}+d\Theta^{\alpha}\wedge d\Theta_{\alpha}+dQ^{i}\wedge d P_{i}
\label{}
\end{equation}
and
\begin{equation}
T^{*}M=T^{*}M|_{\Xi,\Psi}\times T^{*}M|_{\Theta}\times T^{*}M|_{Q,P}.
\label{}
\end{equation}
Then the canonical embedding is
\begin{equation}
\begin{split}
\sigma_{3}:&T^{*}M|_{Q,P}\rightarrow T^{*}M\\
&;(\sigma_{3}^{*}\Xi^{a}:=\epsilon^{a},\sigma_{3}^{*}\Psi_{a}:=\epsilon_{a},\sigma_{3}^{*}\Theta^{\alpha}:=\epsilon^{\alpha},\sigma_{3}^{*}\Theta_{\alpha}:=\epsilon_{\alpha},\sigma_{3}^{*}Q^{i},\sigma_{3}^{*}P_{i})\\
&\mapsto (\Xi^{a},\Psi_{a},\Theta^{\alpha},\Theta_{\alpha},Q^{i},P_{i})
\end{split}
\label{}
\end{equation}
where $\sigma_{3}^{*}$ is the pullback operator of $\sigma_{3}$. The pullback of the symplectic 2-form $\omega$ by $\sigma_{3}$ is
\begin{equation}
\sigma_{3}^{*}\omega=\omega_{Q,P}
\label{}
\end{equation}
where
\begin{equation}
\begin{split}
&\omega_{\Xi,\Psi}:=d\sigma_{3}^{*}\Xi^{a}\wedge d\sigma_{3}^{*}\Psi_{a}=0,\\
&\omega_{\Theta}:=d\sigma_{3}^{*}\Theta^{\alpha}\wedge d\sigma_{3}^{*}\Theta_{\alpha}=0,\\
&\omega_{Q,P}:=d\sigma_{3}^{*}Q^{i}\wedge d\sigma_{3}^{*}P_{a}
.
\end{split}
\label{}
\end{equation}
Therefore, $(T^{*}M,\omega)$ is decomposed into three subspaces: $(T^{*}M|_{\Xi,\Psi},\omega_{\Xi,\Psi}=0)$, $(T^{*}M|_{\Theta},\omega_{\Theta}=0)$, and $(T^{*}M|_{Q,P},\omega_{Q,P})$. Only the last one holds the symplectic structure. 

Of course, the following two embeddings are also canonical:
\begin{equation}
\begin{split}
\sigma_{3}^{(1)}:&T^{*}M|_{\Theta}\times T^{*}M|_{Q,P}\rightarrow T^{*}M\\
;&(\sigma_{3}^{*}\Xi^{a}:=\epsilon^{a},\sigma_{3}^{*}\Psi_{a}:=\epsilon_{a},\sigma_{3}^{*}\Theta^{\alpha},\sigma_{3}^{*}\Theta_{\alpha},\sigma_{3}^{*}Q^{i},\sigma_{3}^{*}P_{i})\mapsto(\Xi^{a},\Psi_{a},\Theta^{\alpha},\Theta_{\alpha},Q^{i},P_{i}).
\end{split}
\label{}
\end{equation}
and
\begin{equation}
\begin{split}
\sigma_{3}^{(2)}:&T^{*}M|_{\Xi,\Psi}\times T^{*}M|_{Q,P}\rightarrow T^{*}M\\
;&(\sigma_{3}^{*}\Xi^{a},\sigma_{3}^{*}\Psi_{a},\sigma_{3}^{*}\Theta^{\alpha}:=\epsilon^{\alpha},\sigma_{3}^{*}\Theta_{\alpha}:=\epsilon_{\alpha},\sigma_{3}^{*}Q^{i},\sigma_{3}^{*}P_{i})\mapsto (\Xi^{a},\Psi_{a},\Theta^{\alpha},\Theta_{\alpha},Q^{i},P_{i}).
\end{split}
\label{}
\end{equation}
Armed with these canonical embeddings, $(T^{*}M,\omega)$ is decomposed into $(T^{*}M|_{\Theta}\times T^{*}M|_{Q,P},{\sigma^{*}_{3}}^{(1)}\omega)$ and $(T^{*}M|_{\Xi,\Psi},{\sigma^{*}_{3}}^{(1)}\omega=0)$, and $(T^{*}M|_{\Xi,\Psi}\times T^{*}M|_{Q,P},{\sigma^{*}_{3}}^{(2)}\omega)$ and $(T^{*}M|_{\Theta},{\sigma^{*}_{3}}^{(2)}\omega=0)$, respectively. The pullback of $\omega$ is computed in the same manner to the case of $\sigma_{3}$. 

Taking into account the time evolution of the system, under imposing the consistency condition $(\ref{gauge fixing})$, we obtain the canonical embedding with the time evolution as follows:
\begin{equation}
\begin{split}
\sigma_{3}(t):&T^{*}M|_{Q,P}\times \mathbb{R}\rightarrow T^{*}M\times \mathbb{R}\\
;&({\sigma_{3}^{*}}(t)\Xi^{a}:=\epsilon^{a},{\sigma_{3}^{*}}(t)\Psi_{a}:=\epsilon_{a},{\sigma_{3}^{*}}(t)\Theta^{\alpha}:=\epsilon^{\alpha},{\sigma_{3}^{*}}(t)\Theta_{\alpha}:=\epsilon_{\alpha},\\
&{\sigma_{3}^{*}}(t)Q^{i},{\sigma_{3}^{*}}(t)P_{i},{\sigma_{3}^{*}}(t)u=t)\\
&\mapsto (\Xi^{a},\Psi_{a},\Theta^{\alpha},\Theta_{\alpha},Q^{i},P_{i},u).
\end{split}
\label{}
\end{equation}
In this case, we can also consider an embedding such that either the first-class constraints, or equivalently $\Psi_{\alpha}$s, are fixed:
\begin{equation}
\begin{split}
\sigma_{3}^{(1)}(t):&T^{*}M|_{\Theta}\times T^{*}M|_{Q,P}\times \mathbb{R}\rightarrow T^{*}M\times \mathbb{R}\\
;&({\tilde{\sigma}_{3}^{*}}(t)\Xi^{a}:=\epsilon^{a},{\tilde{\sigma}_{3}^{*}}(t)\Psi_{a}:=\epsilon_{a},{\tilde{\sigma}_{3}^{*}}(t)\Theta^{\alpha},{\tilde{\sigma}_{3}^{*}}\Theta_{\alpha}(t),{\sigma_{3}^{*}}(t)Q^{i},{\sigma_{3}^{*}}(t)P_{i},{\sigma_{3}^{*}}(t)u=t)\\
&\mapsto (\Xi^{a},\Psi_{a},\Theta^{\alpha},\Theta_{\alpha},Q^{i},P_{i},u),
\end{split}
\label{}
\end{equation}
under the consistency condition $(\ref{gauge fixing})$ holding, or, the second-class constraints, or equivalently $\Theta^{\alpha}$s and $\Theta_{\alpha}$s, are fixed:
\begin{equation}
\begin{split}
\sigma_{3}^{(2)}(t):&T^{*}M|_{\Xi,\Psi}\times T^{*}M|_{Q,P}\times \mathbb{R}\rightarrow T^{*}M\times \mathbb{R}\\
;&({\tilde{\sigma}_{3}^{*}}(t)\Xi^{a},{\tilde{\sigma}_{3}^{*}}(t)\Psi_{a},{\tilde{\sigma}_{3}^{*}}(t)\Theta^{\alpha}:=\epsilon^{\alpha},{\tilde{\sigma}_{3}^{*}}(t)\Theta_{\alpha}:=\epsilon_{\alpha},{\sigma_{3}^{*}}(t)Q^{i},{\sigma_{3}^{*}}(t)P_{i},{\sigma_{3}^{*}}(t)u=t)\\
&\mapsto (\Xi^{a},\Psi_{a},\Theta^{\alpha},\Theta_{\alpha},Q^{i},P_{i},u).
\end{split}
\label{}
\end{equation}
For at a time $t$, $\sigma_{3}(t)$, $\sigma_{3}^{(1)}(t)$ and $\sigma_{3}^{(2)}$(t) restore, of course, $\sigma_{3}$, $\sigma_{3}^{(1)}$ and $\sigma_{3}^{(2)}$, respectively. The corresponding decompositions introduced by these embeddings are constructed in the same manner to $\sigma_{3}$, $\sigma_{3}^{(1)}$ and $\sigma_{3}^{(2)}$, respectively. Remark that $(T^{*}M|_{\Xi,\Psi}\times \mathbb{R},\omega_{\Xi,\Psi}(t)=0)$ and/or $(T^{*}M|_{\Theta}\times \mathbb{R},\omega_{\Theta}(t)=0)$ throughout all time implies that $T^{*}M|_{\Xi,\Psi}\times \mathbb{R}$ and/or $T^{*}M|_{\Theta}\times \mathbb{R}$ do not describe the dynamics. Only the symplectic submanifold $(T^{*}M|_{Q,P}\times \mathbb{R},\omega_{Q,P}(t))$ describe the dynamics. 

The cases that the additional consistency conditions $(\ref{gauge fixing})$ are not imposed lead to quasi-canonical embeddings. That is,
\begin{equation}
\begin{split}
\tilde{\sigma}_{3}(t):&T^{*}M|_{\Xi,\Psi}\times T^{*}M|_{Q,P}\times \mathbb{R}\rightarrow T^{*}M\times \mathbb{R}\\
;&({\tilde{\sigma}_{3}^{*}}(t)\Xi^{a},{\tilde{\sigma}_{3}^{*}}(t)\Psi_{a}:=\epsilon_{a},{\tilde{\sigma}_{3}^{*}}(t)\Theta^{\alpha}:=\epsilon^{\alpha},{\tilde{\sigma}_{3}^{*}}(t)\Theta_{\alpha}:=\epsilon_{\alpha},{\tilde{\sigma}_{3}^{*}}(t)Q^{i},{\tilde{\sigma}_{3}^{*}}(t)P_{i},{\tilde{\sigma}_{3}^{*}}(t)u=t)\\
&\mapsto (\Xi^{a},\Psi_{a},\Theta^{\alpha},\Theta_{\alpha},Q^{i},P_{i},u)
\end{split}
\label{}
\end{equation}
and
\begin{equation}
\begin{split}
\tilde{\sigma}_{3}^{(1)}(t):&T^{*}M\times \mathbb{R}\rightarrow T^{*}M\times \mathbb{R}\\
;&({\tilde{\sigma}_{3}^{*}}(t)\Xi^{a},{\tilde{\sigma}_{3}^{*}}(t)\Psi_{a}:=\epsilon_{a},{\tilde{\sigma}_{3}^{*}}(t)\Theta^{\alpha},{\tilde{\sigma}_{3}^{*}}(t)\Theta_{\alpha},{\tilde{\sigma}_{3}^{*}}(t)Q^{i},{\tilde{\sigma}_{3}^{*}}(t)P_{i},{\tilde{\sigma}_{3}^{*}}(t)u=t) \\
&\mapsto (\Xi^{a},\Psi_{a},\Theta^{\alpha},\Theta_{\alpha},Q^{i},P_{i},u).
\end{split}
\label{}
\end{equation}

Finally, remark that only for $\sigma_{3}(t)$ and $\tilde{\sigma}_{3}(t)$, taking limits all the constant parameters to zero, these embeddings restore the constraint space. $\sigma_{3}(t)$, $\sigma_{3}^{(1)}(t)$, $\sigma_{3}^{(2)}(t)$, $\tilde{\sigma}_{3}(t)$, and $\tilde{\sigma}_{3}^{(1)}(t)$ occupy $2r+2s$, $2r$, $2s$, $r+2s$, and $r$ integral constants, respectively, that are indicated by the Frobenius integrability in Sect.~\ref{Sec03}.

\subsection{An answer for the problems}\label{Sec04:04}
In this section, we reconstruct $\iota$, $\kappa$ and $\mathfrak{O}$ by using the concepts of canonical and quasi-canonical embedding assembled in the section \ref{Sec04:03}. 

\subsubsection{The canonical embeddings: $\sigma_{1}(t)$, $\sigma_{2}(t)$ and $\sigma_{3}(t)$}\label{Sec04:04:01}
The pullback of the entire symplectic manifold $(T^{*}M,\omega)$ by $\sigma_{\tau}(t)$, ${\sigma_{\tau}^{*}}(t)(T^{*}M\times \mathbb{R},\omega)=(TM|_{Q,P}\times \mathbb{R},\omega_{Q,P}(t)=dQ^{a}(t)\wedge dP_{a}(t))$, restores the existence of the inverse of $\kappa$ and $\mathfrak{O}$, respectively. Here, we denote the type of embeddings as $\tau=1,2,3$. First, we show this statement. 

The dual bundle of $T^{*}M|_{Q,P}\times \mathbb{R}={\sigma_{\tau}^{*}}(t)(T^{*}M\times\mathbb{R})$ is determined as $TM|_{Q,R}\times \mathbb{R}$ up to isomorphism, where $R^{a}(t)$s are a set of contra-variant vector components on $M|_{Q}$. Then assume that there is a function $L=L(Q^{a}(t),R^{a}(t),t)$ defined on $TM|_{Q,R}\times \mathbb{R}$ such that the following conditions are satisfied:
\begin{equation}
\begin{split}
&P_{a}(t)=\frac{\partial L}{\partial R^{a}(t)},\\
&{\textrm{det}}\left(\frac{\partial P_{a}(t)}{\partial R^{b}(t)}\right)\neq0.
\end{split}
\label{coordinates for v-p space in 2nd-class}
\end{equation}
Using the implicit function theorem, we get a set of functions: $R^{a}=R^{a}(Q^{a}(t),P^{a}(t),t)$. Since $\dot{Q}^{a}(t)$s are contra-variant vector components on $M|_{Q}$ as well, without any loss of generality, we can identify $R^{a}$s as $\dot{Q}^{a}(t)$s. Therefore, we acquire a coordinate system for the velocity phase space $TM$: $Q^{a}(t)$s and $\dot{Q}^{a}(t)$s. This construction leads to a one-to-one correspondence between $P_{a}(t)$s and $\dot{Q}^{a}(t)$s. That is, the following map is a well-defined and invertible map:
\begin{equation}
\kappa|_{\sigma_{\tau}(t)}:TM|_{Q,\dot{Q}}\times \mathbb{R}\rightarrow T^{*}M|_{Q,P}\times \mathbb{R};\dot{Q}^{a}(t)\mapsto P_{a}(t),
\label{}
\end{equation}
which restricts the domain $TM\times\mathbb{R}$ and the range $T^{*}M\times\mathbb{R}$ of $\kappa$ to ${\sigma_{\tau}^{*}(t)}(TM\times \mathbb{R})=TM|_{Q,\dot{Q}}\times \mathbb{R}$ and ${\sigma_{\tau}^{*}(t)}(T^{*}M\times \mathbb{R})=T^{*}M|_{Q,P}\times \mathbb{R}$, respectively. Of course, there is a relation: $TM|_{Q,\dot{Q}}\times \mathbb{R}\simeq T^{*}M|_{Q,P}\times \mathbb{R}$, and it implies that $TM|_{Q,\dot{Q}}\times T^{*}M|_{Q,P}\times\mathbb{R}$ is equivalent to $TM|_{Q,\dot{Q}}\times\mathbb{R}$ and $T^{*}M|_{Q,P}\times\mathbb{R}$.

Second, we define a Lagrangian in the space $TM|_{Q,\dot{Q}}\times T^{*}M|_{Q,P}\times\mathbb{R}$, denote $L_{T}$, which corresponds to the total Hamiltonian $H_{T}$, as follows:
\begin{equation}
\begin{split}
L_{T}&:={\sigma_{\tau}^{*}}(t)\left[P_{a}\dot{Q}^{a}+\Theta_{\alpha}\dot{\Theta}^{\alpha}-H_{T}(\Theta^{\alpha},\Theta_{\alpha},Q^{a},P_{a})\right]\\
&=P_{a}(t)\dot{Q}^{a}(t)-H_{T}(\Theta^{\alpha}(t)=\epsilon^{\alpha},\Theta_{\alpha}(t)=\epsilon_{\alpha},Q^{a}(t),P_{a}(t)).
\end{split}
\label{unique lagrangian to H_T in 2nd}
\end{equation}
That is, $L=L_{T}$ and $P_{a}(t)$s are introduced by Eq.~$(\ref{coordinates for v-p space in 2nd-class})$. Where, we used $\sigma^{*}_{\tau}(t)(dX/ds):=d(\sigma^{*}_{\tau}(t)X)/d(\sigma^{*}_{\tau}(t)s)=dX(t)/dt$, replacing $X$ by $Q^{a}$s and $\Theta^{\alpha}$s. This leads to ${\sigma_{\tau}^{*}}(t)\dot{\Theta}^{\alpha}={\sigma_{\tau}^{*}}(t)\{\Theta^{\alpha},H_{T}\}={\sigma_{\tau}^{*}}(t)F^{\alpha}(\Theta)=F^{\alpha}({\sigma_{\tau}^{*}}(t)\Theta)=constant$, where $F^{\alpha}(\Theta)$ denote a function depending only on the constraint coordinates: $\Theta^{\alpha}$s and $\Theta_{\alpha}$s. We used also that for $\tau=1,3$ all of $\Xi^{a}$s and $\Psi_{a}$s in \hyperlink{lemma01}{Lemma 1} and \hyperlink{theorem01}{Theorem 1} turn into second-class by virtue of $(\ref{gauge fixing})$. Based on this, we gathered these variables together into $\Theta^{\alpha}$s and $\Theta_{\alpha}$s and applied \hyperlink{lemma02}{Lemma 2}. The Lagrangian $L_{T}$ is uniquely determined by its construction up to surface terms. Let us define the pullback of {\it{"total Lagrangian"}} by ${\sigma_{\tau}^{*}}(t)$ as in the above. Taking into account the $\kappa|_{\sigma_{\tau}(t)}$, it suggests that the following map is a well-defined and invertible map:
\begin{equation}
\mathfrak{O}|_{\sigma_{\tau}(t)}:\mathfrak{G}[T(TM|_{Q,\dot{Q}}\times \mathbb{R})]\rightarrow\mathfrak{G}[T(T^{*}M|_{Q,P}\times \mathbb{R})];X_{t}|_{Q,\dot{Q}}\mapsto {_{*}X}_{t}|_{Q,P}
\label{}
\end{equation}
where $X_{t}|_{Q,\dot{Q}}$ and ${_{*}X}_{t}|_{Q,P}$ are Hamiltonian vector fields restricted to $T(TM|_{Q,\dot{Q}}\times \mathbb{R})$ and $T(T^{*}M|_{Q,P}\times \mathbb{R})$, respectively, and these corresponds in a one-to-one manner. 

Finally, let us consider the map $\iota$. Varying Eq.~$(\ref{unique lagrangian to H_T in 2nd})$, we get
\begin{equation}
\delta L_{T}=\left[\dot{Q}^{a}-\frac{\partial H_{T}}{\partial P_{a}}\right]\delta P_{a}+\left[-\dot{P}_{a}-\frac{\partial H_{T}}{\partial Q^{a}}\right]\delta Q^{a}+\frac{d}{dt}\left[P_{a}\delta Q^{a}\right]
\label{}
\end{equation}
where we abbreviated the argument "$t$" in $Q^{a}(t)$s and $P_{a}(t)$s. In the form of action integral, it can be written as follows:
\begin{equation}
\delta\left({\sigma_{\tau}^{*}}(t)I\right)=\int^{t_{2}}_{t_{1}}\left[\dot{Q}^{a}-\frac{\partial H_{T}}{\partial P_{a}}\right]\delta P_{a}dt+\left[-\dot{P}_{a}-\frac{\partial H_{T}}{\partial Q^{a}}\right]\delta Q^{a}dt+\left[P_{a}\delta Q^{a}\right]^{t_{2}}_{t_{1}}.
\label{}
\end{equation}
In the subspace $TM|_{Q,\dot{Q}}\times T^{*}M|_{Q,P}\times\mathbb{R}$, we can use the formulas in Eq.~$(\ref{coordinates for v-p space in 2nd-class})$ or the invertible map $\kappa|_{\sigma_{\tau}(t)}$. Therefore, the above formula becomes as follows:
\begin{equation}
\delta\left({\sigma_{\tau}^{*}}(t)I\right)=\int^{t_{2}}_{t_{1}}\left[\dot{Q}^{a}-\frac{\partial H_{T}}{\partial P_{a}}\right]\delta P_{a}dt+\left[-\frac{d}{dt}\left(\frac{\partial L_{T}}{\partial \dot{Q}^{a}}\right)-\frac{\partial H_{T}}{\partial Q^{a}}\right]\delta Q^{a}dt+\left[\frac{\partial L_{T}}{\partial \dot{Q}^{a}}\delta Q^{a}\right]^{t_{2}}_{t_{1}}.
\label{delta I in healthy}
\end{equation} 
The one-to-one correspondence between ${\sigma_{\tau}^{*}}(t)H_{T}$ and $L_{T}$ indicates that
\begin{equation}
{\sigma_{\tau}^{*}}(t)H_{T}:=P_{a}\dot{Q}^{a}-L_{T}(Q^{a},\dot{Q}^{a})
\label{unique total-Hamiltonian to L-T in 2nd}
\end{equation}
in the subspace $TM|_{Q,\dot{Q}}\times T^{*}M|_{Q,P}\times\mathbb{R}$ under Eq.~$(\ref{coordinates for v-p space in 2nd-class})$. Therefore, we get
\begin{equation}
\delta\left({\sigma_{\tau}^{*}}(t)I\right)=\int^{t_{2}}_{t_{1}}\left[-\frac{d}{dt}\left(\frac{\partial L_{T}}{\partial \dot{Q}^{a}}\right)+\frac{\partial L_{T}}{\partial Q^{a}}\right]\delta Q^{a}dt+\left[\frac{\partial L_{T}}{\partial \dot{Q}^{a}}\delta Q^{a}\right]^{t_{2}}_{t_{1}},
\label{}
\end{equation}
which is now defined in the subspace $TM|_{Q,\dot{Q}}\times\mathbb{R}$. The second condition of Eq.~$(\ref{coordinates for v-p space in 2nd-class})$ implies that $L_{T}$ is non-degenerate. Therefore, we can fix all positions $Q^{a}$s as boundary conditions:
\begin{equation}
\delta Q^{a}(t_{1})=\delta Q^{a}(t_{2})=0.
\label{well-posed boundary condition under healthy embeddings}
\end{equation}
Here, notice that in the subspace $TM|_{Q,\dot{Q}}\times\mathbb{R}$ this system always holds the Frobenius integrability based on the consideration given in Sect.~\ref{Sec03:01}. In this formulation, map $\iota$ can be introduced in a well-defined manner as an invertible map automatically:
\begin{equation}
\iota|_{\sigma_{\tau}}:M|_{Q}[t_{1}]\times M|_{Q}[t_{2}]\rightarrow C|_{\sigma_{\tau}}; (Q^{a}(t_{1}),Q^{a}(t_{2}))\mapsto c^{A},
\label{iota in healthy}
\end{equation}
where $M|_{Q}[t]$ is the configuration subspace restricted by the canonical embedding $\sigma_{\tau}$ and $C|_{\sigma_{\tau}}$ is a parameter space spanned by the independent integral constants restricted by all the constraints. $A$ runs from $1$ to the twice number of $Q^{a}$s. Therefore, based on \hyperlink{definition01}{Definition 1}, $\delta\left({\sigma_{\tau}^{*}}(t)I\right):=0$ is the well-posed variational principle.

Summarising, the well-posed variational principle is 
\begin{equation}
\delta\left({\sigma_{\tau}^{*}}(t)I\right):=0
\label{}
\end{equation}
under the boundary condition $(\ref{well-posed boundary condition under healthy embeddings})$. 

\subsubsection{The quasi-canonical embeddings: $\tilde{\sigma}_{1}(t)$ and  $\tilde{\sigma}_{3}(t)$}\label{Sec04:04:02}
Let us consider the case of $\tilde{\sigma}_{3}(t)$. The same considerations are applicable to $\tilde{\sigma}_{1}(t)$. 

The quasi-canonical embedding $\tilde{\sigma}_{3}(t)$ leads to ${\tilde{\sigma}^{*}_{3}}(t)(T^{*}M\times\mathbb{R},\omega)=(T^{*}M|_{\Xi,\Psi}\times T^{*}M|_{Q,P}\times \mathbb{R},\omega_{Q,P}(t)+\omega_{\Psi,\Xi}(t)=dQ^{i}(t)\wedge dP_{i}(t)+d\Xi^{a}(t)\wedge d\Psi_{a}(t))$. For $T^{*}M|_{\Xi,\Psi}\times T^{*}M|_{Q,P}\times\mathbb{R}$, we can introduce a function $L=L(\Xi^{a}(t),R^{a}(t),Q^{i}(t),\dot{Q}^{i}(t),t)$ defined in $TM|_{\Xi,R}\times TM|_{Q,\dot{Q}}\times\mathbb{R}$ such that
\begin{equation}
\frac{\partial L}{\partial R^{a}(t)}=\Psi_{a}(t)={constant},
\label{}
\end{equation}
but, of course, 
\begin{equation}
\frac{\partial \Psi_{a}(t)}{\partial R^{b}(t)}=0.
\label{degenerate part in tilde case}
\end{equation}
This indicates that invertible map $\kappa$ does not exist even if we restrict the entire phase space $T^{*}M\times \mathbb{R}$ to the subspace ${\tilde{\sigma}^{*}_{3}}(t)(T^{*}M\times\mathbb{R})$. However, if we restrict $T^{*}M\times \mathbb{R}$ to $T^{*}M|_{Q,P}\times \mathbb{R}$, then $P_{i}$s and $\dot{Q}^{i}$s correspond to one another in a one-to-one manner. Then we can introduce an invertible map $\kappa|_{\tilde{\sigma}_{3}(t)}$ between $TM|_{Q,\dot{Q}}\times \mathbb{R}$ and $T^{*}M|_{Q,P}\times \mathbb{R}$ as follows:
\begin{equation}
\kappa|_{\tilde{\sigma}_{3}(t)}:TM|_{Q,\dot{Q}}\times \mathbb{R}\rightarrow T^{*}M|_{Q,P}\times \mathbb{R};\dot{Q}^{i}(t)\mapsto P_{i}(t)
\label{}
\end{equation}
in a well-defined manner. In addition, under the same restriction of the phase space $T^{*}M\times \mathbb{R}$, an invertible map $\mathfrak{O}|_{\tilde{\sigma}_{3}(t)}:\mathfrak{G}[T(TM|_{Q,\dot{Q}}\times \mathbb{R})]\rightarrow\mathfrak{G}[T(T^{*}M|_{Q,P}\times \mathbb{R})];X_{t}|_{Q,\dot{Q}}\mapsto {_{*}X}_{t}|_{Q,P}$ is introduced in a well-defined manner.

Now, let us consider the map $\iota$. The pullback of the total Lagrangian by $\tilde{\sigma}_{3}(t)$ corresponding to the total Hamiltonian $H_{T}$ is defined as follows:
\begin{equation}
\begin{split}
L_{T}=&P_{i}(t)\dot{Q}^{i}(t)+\Psi_{a}(t)\dot{\Xi}^{a}(t)+\Theta_{\eta}(t)\dot{\Theta}^{\eta}(t)\\
&-H_{T}(\Xi^{a}(t),\Psi_{a}(t)=\epsilon_{a},\Theta^{\eta}(t)=\epsilon^{\eta},\Theta_{\eta}(t)=\epsilon_{\eta},\zeta^{\alpha},Q^{i}(t),P_{i}(t))\\
=&P_{i}(t)\dot{Q}^{i}(t)-H_{T}(\Xi^{a}(t),\Psi_{a}(t)=\epsilon_{a},\Theta^{\eta}(t)=\epsilon^{\eta},\Theta_{\eta}(t)=\epsilon_{\eta},\zeta^{\alpha},Q^{i}(t),P_{i}(t))+\frac{d}{dt}\left[\Psi_{a}(t)\Xi^{a}(t)\right]
\end{split}
\label{H_{T} existing both 1st and 2nd constraints}
\end{equation}
in the space $TM|_{Q,\dot{Q}}\times T^{*}M|_{Q,P}\times TM|_{\Xi,\dot{\Xi}}\times T^{*}M|_{\Xi,\Psi}\times\mathbb{R}$. 
Remark that the condition~$(\ref{coordinates for v-p space in 2nd-class})$ on $Q^{i}$s and $P_{i}$s holds in the subspace $TM|_{Q,\dot{Q}}\times T^{*}M|_{Q,P}\times\mathbb{R}$ as well. $\zeta^{\alpha}$s are Lagrange multipliers and $\alpha$ runs from 1 to the number of the {\it{primary}} first-class constraints. Hereinafter, we abbreviate the argument "$t$" in $Q^{a}(t)$s, $P_{a}(t)$s, $\Xi^{a}(t)$s and $\Psi_{a}(t)$s.

\hyperlink{lemma01}{Lemma 1}, or its proof, indicates that the first-class constraint themselves form canonical momenta. Since $P_{i}$s and $\dot{Q}^{i}$s have the one-to-one correspondence by virtue of ${\text{det}}(\partial P_{i}/\partial \dot{Q}^{j})\neq0$ in Eq~$(\ref{coordinates for v-p space in 2nd-class})$, varying this in the action integral form, we get
\begin{equation}
\begin{split}
\delta\left({\tilde{\sigma}_{3}^{*}}(t)I\right)=&\int^{t_{2}}_{t_{1}}\left[\dot{Q}^{i}-\frac{\partial H_{T}}{\partial P_{i}}\right]\delta P_{i}dt+\left[-\dot{P}_{i}-\frac{\partial H_{T}}{\partial Q^{i}}\right]\delta Q^{i}dt\\
&-\frac{\partial H_{T}}{\partial \Xi^{a}}\delta\Xi^{a}dt-\frac{\partial H_{T}}{\partial \zeta^{\alpha}}\delta\zeta^{\alpha}dt+\left[P_{i}\delta Q^{i}+\Psi_{a}\delta\Xi^{a}\right]^{t_{2}}_{t_{1}}.
\end{split}
\label{action integral existing both 1st and 2nd constraints}
\end{equation}
Note, here, that $\partial H_{T}/\partial \zeta^{\alpha}$s correspond to the pullback of the primary first-class constraints by $\tilde{\sigma}_{3}(t)$. Therefore, we get
\begin{equation}
\begin{split}
\delta\left({\tilde{\sigma}_{3}^{*}}(t)I\right)=&\int^{t_{2}}_{t_{1}}\left[-\frac{d}{dt}\left(\frac{\partial L_{T}}{\partial\dot{Q}^{i}}\right)+\frac{\partial L_{T}}{\partial Q^{i}}\right]\delta Q^{i}dt+\frac{\partial L_{T}}{\partial \Xi^{a'}}\delta\Xi^{a'}dt+\left[\frac{\partial L_{T}}{\partial\dot{Q}^{i}}\delta Q^{i}+\Psi_{a'}\delta\Xi^{a'}\right]^{t_{2}}_{t_{1}}\\
&+\int^{t_{2}}_{t_{1}}\frac{\partial L_{T}}{\partial \Xi^{\alpha}}\delta\Xi^{\alpha}dt+\frac{\partial L_{T}}{\partial \zeta^{\alpha}}\delta\zeta^{\alpha}dt+\left[\Psi_{\alpha}\delta\Xi^{\alpha}\right]^{t_{2}}_{t_{1}}
\end{split}
\label{first variation of pullbacked I}
\end{equation}
in the subspace $TM|_{Q,\dot{Q}}\times TM|_{\Xi,\dot{\Xi}}\times T^{*}M|_{\Xi,\Psi}\times\mathbb{R}$, where we used the one-to-one correspondence between $\dot{Q}^{i}$s and $P_{i}$s which is described by $\kappa|_{\tilde{\sigma}_{3}(t)}$. $a'$s are belonging to the set of indices that eliminates $\alpha^{}$s from $a=1,2,\cdots,r$. Notice that in this subspace the symplectic structure breaks down: $\omega_{Q,P}(t)$ satisfies the definition $(\ref{symplectic 2-form})$, but so is not $\omega_{\Xi,\Psi}(t)$ as mention in Sect.~\ref{Sec04:03:04}. Here, we define the {\it{"effective first-order variation"}} of the action integral as follows:
\begin{equation}
\delta_{{\textrm{effective}}}\left({\tilde{\sigma}_{3}^{*}}(t)I\right):=\int^{t_{2}}_{t_{1}}\left[-\frac{d}{dt}\left(\frac{\partial L_{T}}{\partial\dot{Q}^{i}}\right)+\frac{\partial L_{T}}{\partial Q^{i}}\right]\delta Q^{i}dt+\frac{\partial L_{T}}{\partial \Xi^{a'}}\delta\Xi^{a'}dt+\left[\frac{\partial L_{T}}{\partial\dot{Q}^{i}}\delta Q^{i}+\Psi_{a'}\delta\Xi^{a'}\right]^{t_{2}}_{t_{1}}.
\label{effective 1sr-order variation of action integral}
\end{equation}
The coordinate variables $Q^{i}$ and $\Xi^{a'}$ are now independent. The reason why we split out the terms concerning primary first-class constraints will be revealed soon. Therefore, to vanish the effective first-order variation, we have to fix all the position variables $Q^{i}$s at both $t=t_{1}$ and $t=t_{2}$ as boundary conditions:
\begin{equation}
\delta Q^{i}(t_{2})=\delta Q^{i}(t_{1}):=0
\label{B.C. quasi-healthy for pd.o.f}
\end{equation}
and, since $\Psi_{a'}$s are constant, we have to impose
\begin{equation}
\delta\Xi^{a'}(t_{2})=\delta\Xi^{a'}(t_{1}).
\label{B.C. quasi-healthy for gd.o.f}
\end{equation}
Taking into account the equations of motion $\dot{\Xi}^{a'}=\{\Xi^{a'},H_{T}\}$, appropriate boundary conditions for these variables $\Xi^{a'}$s have to be either
\begin{equation}
\delta\Xi^{a'}(t_{1}):=0,
\label{well-posed boundary condition under unhealthy embeddings 2}
\end{equation}
then $\delta\Xi^{a'}(t_{2})=0$ s are automatically satisfied, or
\begin{equation}
\delta\Xi^{a'}(t_{2}):=0
\label{well-posed boundary condition under unhealthy embeddings 3}
\end{equation}
then $\delta\Xi^{a'}(t_{1})=0$ s are automatically satisfied, since each solution of $\dot{\Xi}^{a'}=\{\Xi^{a'},H_{T}\}$ has one integral constant, respectively. That is, each solution with a given integral constant on a boundary determines the value on the other boundary. Therefore, Eq.~$(\ref{B.C. quasi-healthy for gd.o.f})$ is satisfied and become zero if either Eq.~$(\ref{well-posed boundary condition under unhealthy embeddings 2})$ or Eq.~$(\ref{well-posed boundary condition under unhealthy embeddings 3})$ is imposed. In this work, we adopt the first choice; this choice is none other than setting initial conditions. Here, notice that for $L_{T}$ on $TM|_{Q,\dot{Q}}\times TM|_{\Xi,\dot{\Xi}}\times\mathbb{R}$ the Frobenius integrability~$(\ref{integrability in 1st system in Lagrange formulation})$ under Eq.~$(\ref{the number of constraints})$ has to be imposed. That is, 
\begin{equation}
\theta_{I}(X_{T})\approx\delta_{I}^{\ i}\zeta^{\alpha}\tau^{a}_{\alpha}K^{(1)}_{ij}\frac{\partial\eta^{j}}{\partial\dot{\Xi}^{a}}:\approx0
\label{restricted CI in LF}
\end{equation}
where $I$ runs the range of indices eliminating the ones for the second-class constraint coordinates, $\tau^{a}_{\alpha}$s are zero-eigenvalue vectors of the kinetic matrix, and $\eta^{i}\approx\left({K^{(1)}}^{-1}\right)^{ij}S_{j}$. Remark that the $(i,j)$-block of the kinetic matrix is non-degenerate. In contrast, for ${\tilde{\sigma}_{3}^{*}}(t)H_{T}$ on $T^{*}M|_{Q,P}\times T^{*}M|_{\Xi,\Psi}\times \mathbb{R}$, the Frobenius integrability~$(\ref{integrability in 1st system in Hamilton formulation})$ is automatically satisfied under Eq.~$(\ref{the number of constraints})$ since $\Xi^{a}$s and $\Psi_{a}$s form a part of the entire canonical coordinate system and all $\Psi_{a}$s are canonical momenta with respect to $\Xi^{a}$s, respectively, on the ground of the proofs of \hyperlink{lemma01}{Lemma 1} and \hyperlink{theorem01}{Theorem 1}. Therefore, 
\begin{equation}
{_{*}\theta}_{I}({_{*}X}_{T})\approx-\delta_{Ia}\zeta^{\alpha}\frac{\partial {\Psi}_{\alpha}}{\partial \Xi^{a}}-\delta_{Ii}\zeta^{\alpha}\frac{\partial \Psi_{\alpha}}{\partial Q^{i}}\approx0.
\label{restricted CI in HF}
\end{equation}

In order to introduce the map $\iota$ in a well-defined manner, the following two conditions have to be taken into account. The first is that if the consistency condition $(\ref{gauge fixing})$ is imposed then $\tilde{\sigma}_{3}(t)$ turns into $\sigma_{3}(t)$. The second is that the dimension of a parameter space that is spanned by all independent integral constants is up to $2n-2s-r$. Then the map $\iota$ can be introduced as follows:
\begin{equation}
\iota|_{\tilde{\sigma}_{3}}:M|_{Q,\Xi}[t_{1}]\times M|_{Q}[t_{2}]\rightarrow C|_{\tilde{\sigma}_{3}}; (Q^{i}(t_{1}),\Xi^{a'}(t_{1}),Q^{i}(t_{2}))\mapsto c^{A},
\label{}
\end{equation}
$\delta\Xi^{a'}(t_{2})=0\ $s are automatically satisfied, where $M|_{Q}$ and $M|_{Q,\Xi}$ are the configuration subspace of $T^{*}M|_{Q,P}$ and $T^{*}M|_{Q,P}\times T^{*}M|_{\Xi,\Psi}$, respectively, and $C|_{\tilde{\sigma}_{3}}$ is the parameter space spanned by the independent integral constants restricted by all the constraints. $A$ runs from 1 to the sum of twice number of $Q^{i}$s and $\Xi^{a'}$s. Then map $\iota$ is invertible. Therefore, based on \hyperlink{definition01}{Definition 1}, $\delta_{{\textrm{effective}}}\left({\tilde{\sigma}_{3}^{*}}(t)I\right):=0$ is the well-posed variational principle. 

Under the well-posed variational principle $\delta_{{\textrm{effective}}}\left({\tilde{\sigma}_{3}^{*}}(t)I\right):=0$ with the boundary conditions Eqs.~$(\ref{B.C. quasi-healthy for pd.o.f})$ and $(\ref{well-posed boundary condition under unhealthy embeddings 2})$, the original first-order variation of the action integral $(\ref{first variation of pullbacked I})$ becomes as follows:
\begin{equation}
\delta\left({\tilde{\sigma}_{3}^{*}}(t)I\right)=\int^{t_{2}}_{t_{1}}\frac{\partial L_{T}}{\partial \Xi^{\alpha}}\delta\Xi^{\alpha}dt+\frac{\partial L_{T}}{\partial \zeta^{\alpha}}\delta\zeta^{\alpha}dt+\left[\Psi_{\alpha}\delta\Xi^{\alpha}\right]^{t_{2}}_{t_{1}}.
\label{reduced 1st-order variation of the action integral}
\end{equation}
Applying the variational principle: $\delta\left({\tilde{\sigma}_{3}^{*}}(t)I\right)=0$, we derive
\begin{equation}
\frac{\partial L_{T}}{\partial \Xi^{\alpha}}=-\frac{\partial H_{T}}{\partial \Xi^{\alpha}}=0
\label{}
\end{equation}
if the following equations are identically satisfied:
\begin{equation}
\frac{\partial L_{T}}{\partial \zeta^{\alpha}}=-\frac{\partial H_{T}}{\partial \zeta^{\alpha}}=\Psi_{\alpha}(t)=0.
\label{}
\end{equation}
In fact, the boundary conditions $\delta\Xi^{\alpha}(t_{2})=\delta\Xi^{\alpha}(t_{1})$s are not satisfied since $\dot{\Xi}^{\alpha}=\{\Xi^{\alpha},H_{T}\}\approx\zeta^{\alpha}+f(Q^{i},P_{i})$s leads to 
\begin{equation}
\delta\Xi^{\alpha}(t_{2})-\delta\Xi^{\alpha}(t_{1})=\delta\int^{t_{2}}_{t_{1}}\left[\zeta^{\alpha}+f(Q^{i},P_{i})\right]dt=\delta\int^{t_{2}}_{t_{1}}\zeta^{\alpha}dt,
\label{}
\end{equation}
where $f$ is the definite function determined by $H_{T}$ and we used the fact that the integral on the interval $t_{1}\leq t\leq t_{2}$ of $f$ has a definite value since $Q^{i}$s and $P_{i}$s are physical degrees of freedom and the equations of motion for these variables are already derived by virtue of the well-posed variational principle $\delta_{{\textrm{effective}}}\left({\tilde{\sigma}_{3}^{*}}(t)I\right):=0$. This result indicates that the configurations corresponding to the {\it{primary}} first-class constraints cannot be fixed on the boundaries. Therefore, we have to impose $\Psi_{\alpha}(t)=constant=0$ in advance and this means that the existence of first-class constraints restricts the possible embedding $\tilde{\sigma}_{3}(t)$. Therefore, Eq.~$(\ref{reduced 1st-order variation of the action integral})$ becomes as follows:
\begin{equation}
\delta\left({\tilde{\sigma}_{3}^{*}}(t)I\right)=\int^{t_{2}}_{t_{1}}\frac{\partial L_{T}}{\partial \Xi^{\alpha}}\delta\Xi^{\alpha}dt.
\label{modified reduced 1st-order variation of the action integral}
\end{equation}
The variational principle, $\delta\left({\tilde{\sigma}_{3}^{*}}(t)I\right):=0$, derives the equations of motion, $\partial L_{T}/\partial \Xi^{\alpha}=0$, {\it{without any boundary condition}}. Under this assumption, if the effective first-order variation vanishes: $\delta_{{\textrm{effective}}}\left({\tilde{\sigma}_{3}^{*}}(t)I\right):=0$, the variational principle is applied to the entire phase space, and vice versa. As another aspect, it would be convenient to introduce the {\it{"effective-total Hamiltonian"}} as follows:
\begin{equation}
H_{{\rm{effective}}}:=H_{T}|_{\Psi_{\alpha}:=0}.
\label{}
\end{equation} 
For this $H_{{\rm{effective}}}$, repeating the same consideration going back to Eq.~$(\ref{H_{T} existing both 1st and 2nd constraints})$, Eq.~$(\ref{effective 1sr-order variation of action integral})$ is directly derived. Of course, the appropriate boundary conditions are given by Eqs.~$(\ref{B.C. quasi-healthy for pd.o.f})$ and $(\ref{well-posed boundary condition under unhealthy embeddings 2})$.

There is a remark. Eq.~$(\ref{H_{T} existing both 1st and 2nd constraints})$ can be rewritten as follows:
\begin{equation}
L'_{T}(\Xi^{a},\Psi_{a},Q^{i},P_{i}):=P_{i}\dot{Q}^{i}-\tilde{\sigma}_{3}^{*}(t)H_{T}
\label{rewritten H_{T} existing both 1st and 2nd constraints}
\end{equation}
where
\begin{equation}
L'_{T}=L_{T}-\frac{d}{dt}\left[\Psi^{a}\Xi_{a}\right],
\label{L'T}
\end{equation}
and we abbreviatied $\Theta^{\eta}$s and $\Theta_{\eta}$s. $L'_{T}$ in Eq.~$(\ref{L'T})$ has the arbitrariness of continuous infinite since $\Psi^{a}$s can be regarded as continuous parameters. In other words, $L'_{T}$ is parametrized by $\Psi^{a}$s. The arbitrariness is not the one deriving from a canonical transformation on $T^{*}M|_{Q,P}\times\mathbb{R}$. That is, for a total Hamiltonian $H_{T}$, the corresponding Lagrangian is not uniquely determined, unlike the case of canonical embeddings. It is another aspect of the absence of the inverse map $\mathfrak{O}^{-1}$ in the entire space $T^{*}M\times\mathbb{R}$. 

Summarising, the well-posed variational principle is 
\begin{equation}
\delta_{{\textrm{effective}}}\left({\tilde{\sigma}_{3}^{*}}(t)I\right):=0
\label{well-defined first variation}
\end{equation}
under the boundary condition Eqs.~$(\ref{B.C. quasi-healthy for pd.o.f})$ and $(\ref{well-posed boundary condition under unhealthy embeddings 2})$. The important result is that the configurations corresponding to the {\it{primary}} first-class constraint coordinates never be fixed on the boundaries until some gauge fixing conditions are imposed.

\subsubsection{Invalid canonical and quasi-canonical embeddings: $\sigma^{(1)}_{3}(t)$, $\sigma^{(2)}_{3}(t)$, and $\tilde{\sigma}_{3}^{(1)}(t)$}\label{Sec04:04:03}
The embeddings $\sigma^{(1)}_{3}(t)$ and $\sigma^{(2)}_{3}(t)$ are somewhat special; these are canonical but do not introduce any well-posed variational principle. That is, map $\iota$ is not introduced in any well-defined manner, unlike maps $\kappa$ and $\mathfrak{O}$. Let us consider the case of $\sigma^{(1)}_{3}(t)$. The same considerations are applicable to $\sigma^{(2)}_{3}(t)$. 

The same considerations as in Sects.~\ref{Sec04:04:01} and \ref{Sec04:04:02} lead to the pullback of total Lagrangian by $\sigma^{(1)}_{3}(t)$ as follows:
\begin{equation}
L_{T}=P_{i}\dot{Q}^{i}+\Theta_{\alpha}\dot{\Theta}^{\alpha}-H_{T}(\Xi^{a}=\epsilon^{a},\Psi_{a}=\epsilon_{a},\Theta^{\alpha},\Theta_{\alpha},Q^{i},P_{i}).
\label{}
\end{equation}
Therefore, the first-order variation is computed as follows:
\begin{equation}
\begin{split}
\delta\left({\sigma^{*}}^{(1)}_{3}(t)I\right)=&\int^{t_{2}}_{t_{1}}\left[-\frac{d}{dt}\left(\frac{\partial L_{T}}{\partial \dot{Q}^{a}}\right)+\frac{\partial L_{T}}{\partial Q^{a}}\right]\delta Q^{a}dt+\left[\frac{\partial L_{T}}{\partial \dot{Q}^{a}}\delta Q^{a}\right]^{t_{2}}_{t_{1}}\\
&+\int^{t_{2}}_{t_{1}}\left[-\dot{\Theta}_{\alpha}-\frac{\partial H_{T}}{\partial \Theta^{\alpha}}\right]\delta\Theta^{\alpha}dt+\left[\dot{\Theta}^{\alpha}-\frac{\partial H_{T}}{\partial \Theta_{\alpha}}\right]\delta\Theta_{\alpha}dt+\left[\Theta_{\alpha}\delta\Theta^{\alpha}\right]^{t_{2}}_{t_{1}}
\end{split}
\label{}
\end{equation}
in the symplectic submanifold $(TM|_{Q,\dot{Q}}\times T^{*}M|_{\Theta}\times \mathbb{R},\omega(t)=\omega_{Q,P}(t)+\omega_{\Theta}(t))$ with $\omega_{Q,P}(t)\neq0$ and $\omega_{\Theta}(t)\neq0$. In this case, in general, the map $\iota$ does not exist since we cannot fix the integral constants in the solutions of $-\dot{\Theta}_{\alpha}-\partial H_{T}/\partial \Theta^{\alpha}=0$ through the boundaries. Therefore, the embedding $\sigma^{(1)}_{3}(t)$ does not have appropriate boundary conditions; to apply the variational principle, fixing $\Theta^{\alpha}$s and $\Theta_{\alpha}$s, we have to use the embedding $\sigma_{3}(t)$. 

For the embedding $\tilde{\sigma}_{3}^{(1)}(t)$, for the same reason to the above embedding $\sigma^{(1)}_{3}(t)$, map $\iota$ does not exist in any well-defined manner.

Finally, notice that these embeddings do not restore the constraint space even if all the constant parameters vanish. This is another reason why these embeddings are ruled out.

\section{Examples}\label{Sec05}
\subsection{A system only with first-class constraints}\label{Sec05:01}
Let us consider the following system\cite{Cawley1979}:
\begin{equation}
L_{1}=\dot{q}^{1}\dot{q}^{3}+\frac{1}{2}q^{2}\left(q^{3}\right)^{2}.
\label{CawleyModel}
\end{equation}
This model has no physical degrees of freedom but is historically crucial; it was proposed as a counter-example for the Dirac conjecture\cite{Dirac1950}.

The kinetic matrix $K^{(1)}$ is 
\begin{equation}
K^{(1)}=\begin{bmatrix}
0 & 0 & 1 \\
0 & 0 & 0 \\
1 & 0 & 0
\end{bmatrix},
\label{}
\end{equation}
where we used the canonical momenta: $p_{1}=\dot{q}^{3},p_{2}=0,p_{3}=\dot{q}^{1}$. There is a primary constraint due to ${\text{rank}K^{(1)}}=2$:
\begin{equation}
\phi^{(1)}:=p_{2}:\approx0.
\label{}
\end{equation}
The total Hamiltonian is derived as follows:
\begin{equation}
\begin{split}
&H_{T}=H+\zeta\phi^{(1)},\\
&H=p_{1}p_{3}-\frac{1}{2}q^{2}(q^{3})^{2}.   
\end{split}
\label{}
\end{equation}
The Dirac procedure generates a secondary and a tertiary constraint as follows:
\begin{equation}
\begin{split}
&\dot{\phi}^{(1)}=\{\phi^{(1)},H_{T}\}\approx \frac{1}{2}(q^{3})^{2}\\
&\therefore \phi^{(2)}:=q^{3}:\approx0,\\
&\phi^{(3)}:=\dot{\phi}^{(2)}=\{\phi^{(2)},H_{T}\}\approx p_{1}\\
&\therefore \phi^{(3)}:=p_{1}:\approx0.
\end{split}
\end{equation}
$\dot{\phi}^{(3)}\approx0$ is automatically satisfied. All $\phi^{(1)}$, $\phi^{(2)}$, and $\phi^{(3)}$ are classified into first-class constraint: all P.b.s among them vanish. \hyperlink{lemma01}{Lemma 1} and its proof indicate that this system has a canonical transformation from the original coordinates to the ones such that a part of their canonical momenta themselves are the first-class constraints. In fact, the symplectic 2-form of the system is computed as follows:
\begin{equation}
\omega=dq^{i}\wedge dp_{i}=d\Xi^{i}\wedge d\Psi_{i},
\label{}
\end{equation}
where $\Xi^{i}$s and $\Psi_{i}$s are defined as follows:
\begin{equation}
\begin{split}
&\Xi^{1}:=q^{1},\Xi^{2}:=q^{2},\Xi^{3}:=-p_{3}\\
&\Psi_{1}:=\phi^{(3)},\Psi_{2}:=\phi^{(1)}, \Psi_{3}:=\phi^{(2)}.
\end{split}
\label{}
\end{equation}
Notice that the second equality in the equation of $\omega$ is the strong equality, not weak equality. (See \hyperlink{remark01}{Remark 1}.) The total Hamiltonian is transformed as follows:
\begin{equation}
H_{T}=-\Psi_{1}\Xi^{3}-\frac{1}{2}\Xi^{2}(\Psi_{3})^{2}+\zeta \Psi_{2}.
\label{}
\end{equation}
This system is always Frobenius integrable as mentioned in the section \ref{Sec04:04:02}. In fact, we can compute as follows:
\begin{equation}
\begin{split}
{_{*}\theta}_{i}({_{*}X_{T}})=&\left(d\Psi_{i}+\frac{\partial H}{\partial\Xi^{i}}dt\right)\left(X_{H}+\zeta X_{\Psi_{2}}+\frac{\partial}{\partial t}\right)\\
=&d\Psi_{i}(X_{H})+d\Psi_{i}(\zeta X_{\Psi_{2}})+\frac{\partial H}{\partial \Xi^{i}}\\
\approx&\{\Psi_{i},H\}+\zeta\{\Psi_{i},\Psi_{2}\}+\frac{\partial H}{\partial \Xi^{i}}\\
=&0,\\
&\therefore {_{*}\theta}_{i}({_{*}X_{T}})\approx0.
\end{split}
\label{}
\end{equation}
Therefore, the system has six integral constants, of which three integral constants are occupied by the consistency conditions of the constraints. Therefore, the remaining three independent integral constants, which are originated from the three equations: $\dot{\Xi}^{1}=-\Xi^{3}$, $\dot{\Xi}^{2}=\zeta$, and $\dot{\Xi}^{3}=-\Xi^{2}\Psi_{3}$, have to be fixed by imposing boundary conditions in the variational principle. 

There are a possible quasi-canonical embedding ${\tilde{\sigma}}_{1}(t)$ and a possible canonical embedding $\sigma_{1}(t)$, which are defined in the section \ref{Sec04:03:02}, respectively.

\subsubsection{The quasi-canonical embedding: ${\tilde{\sigma}}_{1}(t)$}\label{Sec05:01:01}
We derive the pullback of total Lagrangian $L_{T}$ by $\tilde{\sigma}_{1}(t)$ and compute the first-order variation of the action integral for $L_{T}$. $L_{T}$ is defined as follows:
\begin{equation}
\begin{split}
&L_{T}:=\tilde{\sigma}_{1}^{*}(t)\left[\Psi_{i}\dot{\Xi}^{i}-H_{T}\right]\\
&\therefore L_{T}=\Psi_{1}\Xi^{3}+\frac{1}{2}\Xi^{2}(\Psi_{3})^{2}
+\frac{d}{dt}\left(\Psi_{a'}\Xi^{a'}\right),
\end{split}
\label{}
\end{equation}
which is defined in the space $TM|_{\Xi,\dot{\Xi}}\times\mathbb{R}$. Where we abbreviated the pullback operator $\tilde{\sigma}_{1}^{*}(t)$, $a'=1,3$, and we used $\tilde{\sigma}_{1}^{*}(t)\Psi_{2}=0$. Then the effective first-order variation is given as follows:
\begin{equation}
\delta_{{\text{effective}}}(\tilde{\sigma}_{1}^{*}(t)I)=\Psi_{1}\int^{t_{2}}_{t_{1}}dt\delta\Xi^{3}+\left[\Psi_{1}\delta\Xi^{1}+\Psi_{3}\delta\Xi^{3}\right]^{t_{2}}_{t_{1}}.
\label{}
\end{equation}
The appropriate boundary conditions are set as follows:
\begin{equation}
\delta\Xi^{a'}(t_{1})=0,
\label{}
\end{equation}
then $\delta\Xi^{a'}(t_{2})=0$ s are automatically satisfied since each solution of the equations of motion for $\Xi^{a'}$s is only one integral constant, respectively. That is, each solution gives a definite value at the boundary $t=t_{2}$. Then the variational principle for the effective first-order variation, $\delta_{{\text{effective}}}(\tilde{\sigma}_{1}^{*}(t)I):=0$, leads to $\Psi_{1}=0$. In addition, $\delta(\tilde{\sigma}_{1}^{*}(t)I):=0$ derives $\Psi_{3}=0$. Where we abbreviated the pullback operator $\tilde{\sigma}_{1}^{*}(t)$. Remark that the Frobenius integrability condition $(\ref{integrability in 1st system in Hamilton formulation})$ in the section \ref{Sec03:02:02} restricted to $\tilde{\sigma}_{1}^{*}(t)(TM\times \mathbb{R})$ is also satisfied, and there occurs no phase space reduction. It indicates that the six integral constants hold, of which the three constants are occupied by the consistency conditions, or equivalently the embedding $\tilde{\sigma}_{1}^{*}(t)$. This fact can be also led by that Eq.~(\ref{integrability in 1st system in Lagrange formulation}) under Eq.~(\ref{the number of constraints}) given in the section \ref{Sec03:02:01} is always satisfied in this system by virtue of a zero-eigenvector $\tau^{i}=(0,\tau^{2},0)$ and $\eta^{i}\approx(0,\eta^{2},0)$ where $\tau^{2}$ and $\eta^{2}$ are arbitrary function in the space $TM|_{\Xi,\dot{\Xi}}\times\mathbb{R}$. Then the invertible map $\iota$ is $\iota|_{\tilde{\sigma}_{3}}:M|_{\Xi}[t_{1}]\rightarrow C|_{\tilde{\sigma}_{3}}; (\Xi^{1}(t_{1}),\Xi^{3}(t_{1}))\mapsto (c^{1},c^{2})$. In fact, we have the two equations: $\dot{\Xi}^{1}=-\Xi^{3}$ and $\dot{\Xi}^{3}=0$. Therefore, $c_{1}=\Xi^{1}(t_{1})-\Xi^{3}(t_{1})t_{1}$, $c_{2}=\Xi^{3}(t_{1})$. Remark that $\kappa|_{\tilde{\sigma}_{1}(t)}$ and $\mathfrak{O}|_{\tilde{\sigma}_{1}(t)}$ do not exist in this case since this system does not have any dynamics. The remaining one integral constant is assigned for $\dot{\Xi}^{2}=\zeta$; this constant does not determine until a gauge fixing condition is imposed. 

\subsubsection{The canonical embedding: $\sigma_{1}(t)$}\label{Sec05:01:02}
We impose the condition $(\ref{gauge fixing})$ to all $\Xi^{a}$s: $\dot{\Xi}^{1}=-\Xi^{3}:\approx0$, $\dot{\Xi}^{2}=\zeta:\approx0$, and $\dot{\Xi}^{3}=-\Xi^{2}\Psi_{3}:\approx0$. The first and the third equations are automatically satisfied by virtue of $\Xi^{a}:\approx0$ $(a=1,2,3)$. The second equation is satisfied if and only if $\zeta:\approx0$. Remark that all these ingredients are derived in the entire space $T^{*}M|_{\Psi,\Xi}\times \mathbb{R}$: the target space of the embedding $\sigma_{1}(t)$. Then the pullback of total Lagrangian $L_{T}$ by $\sigma_{1}(t)$ is introduced as follows:
\begin{equation}
L_{T}=\sigma_{1}^{*}(t)\left[\Psi_{i}\dot{\Xi}^{i}-H_{T}\right]={constant},
\label{}
\end{equation}
which is defined in the null subspace $\{0\}\times\mathbb{R}$. That is, this system does not describe any dynamics. Therefore, of course, $\kappa|_{\tilde{\sigma}_{1}(t)}$ and $\mathfrak{O}|_{\tilde{\sigma}_{1}(t)}$ do not exist. Map $\iota$ is in the same situation: all the integral constants, which are implied by Eq.~(\ref{integrability in 1st system in Lagrange formulation}) under Eq.~(\ref{the number of constraints}) given in Sect.~\ref{Sec03:02:01} by virtue of the same reason as the section \ref{Sec05:01:01}, are occupied by the consistency conditions for $\Psi_{a}$s and $\Xi^{a}$s, or equivalently fixing the embedding $\sigma_{1}(t)$. 

\subsection{A system only with second-class constraints}\label{Sec05:02}
Let us consider the following system:
\begin{equation}
L_{2}=q^{1}\dot{q}^{2}-q^{2}\dot{q}^{1}-\left(q^{1}\right)^{2}-\left(q^{2}\right)^{2}.
\label{}
\end{equation}
This model is an imitation of the Dirac system for spin $1/2$-particles in field theory\cite{Dirac1928}. 

The kinetic matrix $K^{(1)}$ is 
\begin{equation}
K^{(1)}=\begin{bmatrix}
0 & 0\\
0 & 0
\end{bmatrix},
\label{}
\end{equation}
where we used the canonical momenta: $p_{1}=-q^{2}, p_{2}=q^{1}$. There are two primary constraints due to ${\text{rank}K^{(1)}}=0$:
\begin{equation}
\phi^{(1)}_{1}:=p_{1}+q^{2}:\approx0,\phi^{(1)}_{2}:=p_{2}-q^{1}:\approx0.
\label{}
\end{equation}
The P.b. is $\{\phi^{(1)}_{1},\phi^{(1)}_{2}\}=2$ and otherwise vanish. The total Hamiltonian is derived as follows:
\begin{equation}
\begin{split}
&H_{T}=H+\zeta^{\alpha}\phi^{(1)}_{\alpha},\\
&H=(q^{1})^{2}+(q^{2})^{2}.
\end{split}
\label{}
\end{equation}
The Dirac procedure determines all Lagrange multipliers:
\begin{equation}
\zeta^{1}\approx-q^{2},\zeta^{2}\approx q^{1}.
\label{}
\end{equation}
Then the consistency conditions for $\phi^{(1)},\phi^{(2)}$ are satisfied: $\dot{\phi}^{(1)}\approx0,\dot{\phi}^{(2)}\approx0$. That is, all constraints are classified into second-class constraint and the physical degrees of freedom of the system is $(2\times2-2)/2=1$. \hyperlink{lemma02}{Lemma 2} and its proof indicate that the system has a canonical transformation from the original coordinates to the ones such that a part of the entire canonical coordinates are represented by the linear combination of the second-class constraints. In fact, the symplectic 2-form is computed as follows:
\begin{equation}
\omega=dq^{i}\wedge dp_{i}=d\Theta^{1}\wedge d\Theta_{1}+dQ^{1}\wedge dP_{1},
\label{}
\end{equation}
where $\Xi^{1}$, $\Xi_{1}$, $Q^{1}$ and $P_{1}$ are defined as follows:
\begin{equation}
\begin{split}
&\Theta^{1}:=\frac{1}{\sqrt{2}}\phi^{(1)}_{1},\Theta_{1}:=\frac{1}{\sqrt{2}}\phi^{(1)}_{2},\\
&Q^{1}:=\frac{1}{\sqrt{2}}(q^{1}+p_{2}), P_{1}:=\frac{1}{\sqrt{2}}(p_{1}-q^{2}).
\end{split}
\label{}
\end{equation}
Notice that the second equality in the equation of $\omega$ is the strong equality, not weak equality. (See \hyperlink{remark01}{Remark 1}.) The total Hamiltonian is transformed as follows:
\begin{equation}
H_{T}=\frac{1}{2}(P_{1})^{2}+\frac{1}{2}(Q^{1})^{2}-\frac{1}{2}(\Theta^{1})^{2}-\frac{1}{2}(\Theta_{1})^{2}.
\label{}
\end{equation}
The pullback of total Lagrangian by $\sigma_{2}(t)$ is derived as follows:
\begin{equation}
\begin{split}
L_{T}=&\sigma^{*}_{2}(t)\left(\Theta_{1}\dot{\Theta}^{1}+P_{1}\dot{Q}^{1}-H_{T}\right)\\
=&P_{1}\dot{Q}^{1}-\frac{1}{2}(Q^{1})^{2}-\frac{1}{2}(P_{1})^{2}+\frac{1}{2}(\Theta^{1})^{2}+\frac{1}{2}(\Theta_{1})^{2},\\
&\therefore L_{T}=P_{1}\dot{Q}^{1}-\frac{1}{2}(Q^{1})^{2}-\frac{1}{2}(P_{1})^{2}+{constant},
\end{split}
\label{}
\end{equation}
which is defined in the subspace $TM|_{Q,\dot{Q}}\times T^{*}M|_{Q,P}\times\mathbb{R}$. Where, we abbreviated the pullback operator $\sigma^{*}_{2}(t)$, which is defined in Sect.~\ref{Sec04:03:03}, in the second and the last line. Remark that the pullback of $H_{T}$ by $\sigma_{2}(t)$ above is defined in the symplectic submanifold $(T^{*}M|_{Q,P}\times\mathbb{R},\omega_{Q,P}=dQ^{1}\wedge dP_{1})$. This system is Frobenius integrable as mentioned in Sect.~\ref{Sec04:04:01}. In fact, we will show the unique solution for this system.

The first-order variation of the action integral of $L_{T}$ is computed as follows:
\begin{equation}
\delta\left(\sigma^{*}_{2}(t)I\right)=\int^{t_{2}}_{t_{1}}\left[-\dot{P}_{1}-Q^{1}\right]\delta Q^{1}dt+\left[\dot{Q}^{1}-P_{1}\right]\delta P_{1}dt+\left[P_{1}\delta Q^{1}\right]^{t_{2}}_{t_{1}}.
\label{}
\end{equation}
The appropriate boundary conditions are set as follows:
\begin{equation}
\delta Q^{1}(t_{2})=\delta Q^{1}(t_{1})=0.
\label{}
\end{equation}
Then the variational principle leads to the following equations:
\begin{equation}
-\dot{P}_{1}-Q^{1}=0,\dot{Q}^{1}-P_{1}=0.
\label{}
\end{equation}
The second equation gives the explicit form for the canonical momentum $P_{1}$. Therefore, $L_{T}$ is rewritten as follows:
\begin{equation}
L_{T}=\frac{1}{2}(\dot{Q}^{1})^{2}-\frac{1}{2}(Q^{1})^{2}+{constant},
\label{}
\end{equation}
which is now defined in the subspace $TM|_{Q,\dot{Q}}\times\mathbb{R}$. This indicates that the system is always Frobenius integrable as mentioned in Sect.~\ref{Sec03:01} and that two integral constants exist. In fact, $L_{T}$ is none other than describing the one-dimensional harmonic oscillator. The equation of motion is as follows:
\begin{equation}
-\ddot{Q}^{1}-Q^{1}=0.
\label{}
\end{equation}
This equation has the unique solution $Q^{1}(t)=A{\textrm{ exp}}(+it)+B{\textrm{ exp}}(-it)$ and the boundary condition uniquely determines the integral constant $A,B$.

The fundamental maps are given as follows: $\kappa|_{\sigma_{2}(t)}:TM|_{Q,\dot{Q}}\rightarrow TM|_{Q,P};\dot{Q}^{1}\mapsto P_{1}$ and $\mathfrak{O}|_{\sigma_{2}(t)}:X_{t}=Q^{1}(\partial/\partial\dot{Q}^{1})+\dot{Q}^{1}(\partial/\partial Q^{1})+(\partial/\partial t)\mapsto{_{*}X}_{t}=(\partial H_{T}/\partial P_{1})(\partial/\partial Q^{1})-(\partial H_{T}/\partial Q^{1})(\partial/\partial P_{1})+(\partial/\partial t)$, where $H_{T}=(Q^{1})^{2}/2+(P_{1})^{2}/2+constant$; these are introduced in a well-defined manner and invertible. The invertible map $\iota$ is $\iota|_{\sigma_{3}(t)}:M|_{Q}[t_{1}]\times M|_{Q}[t_{2}]\rightarrow C|_{\sigma_{3}(t)};(Q^{1}(t_{1}),Q^{1}(t_{2}))\mapsto(A,B)$ with $A=(Q^{1}(t_{1}){\text{exp}}(it_{2})-Q^{1}(t_{2}){\text{exp}}(it_{1}))/2i{\text{sin}(t_{2}-t_{1})}$ and $B=(Q^{1}(t_{2}){\text{exp}}(-it_{1})-Q^{1}(t_{1}){\text{exp}}(-it_{2}))/2i{\text{sin}(t_{2}-t_{1})}$.

\subsection{A system with first- and second-class constraints}\label{Sec05:03}
Let us consider the following system\cite{BrownModel2022}:
\begin{equation}
L_{3}=\frac{1}{2}\left(q^{1}+\dot{q}^{2}+\dot{q}^{3}\right)^{2}+\frac{1}{2}\left(\dot{q}^{4}-\dot{q}^{2}\right)^{2}+\frac{1}{2}\left(q^{1}+2q^{2}\right)\left(q^{1}+2q^{4}\right).
\label{}
\end{equation}
This model has not only both first and second-class constraints but also physical degrees of freedom. The author in \cite{BrownModel2022} reveals that this system is equivalent to a one-dimensional harmonic oscillator on the ground of the extended Hamiltonian. In this section, however, we use the total Hamiltonian formulation to reveal the dynamics of this system since the extended Hamiltonian formulation has a series of controversies\cite{Cawley1979,Frenkel1980,SuganoKimura1983}. We will derive the same dynamics in \cite{BrownModel2022}. The kinetic matrix $K^{(1)}$ and primary constraints are computed as follows:
\begin{equation}
K^{(1)}=
\begin{bmatrix}
0 & 0 & 0 & 0 \\
0 & 2 & 1 & -1 \\
0 & 1 & 1 & 0 \\
0 & -1 & 0 & 1
\end{bmatrix},
\label{}
\end{equation}
\begin{equation}
\phi^{(1)}_{1}:=p_{1}:\approx0,\phi^{(1)}_{2}:=p_{2}-p_{3}+p_{4}:\approx0,
\label{}
\end{equation}
where $p_{1}=0$, $p_{2}=q^{1}+2\dot{q}^{2}+\dot{q}^{3}-\dot{q}^{4}$, $p_{3}=q^{1}+\dot{q}^{2}+\dot{q}^{3}$, $p_{4}=\dot{q}^{4}-\dot{q}^{2}$. The rank of $K^{(1)}$ is 2. The P.b.s of these constraints vanish. The total Hamiltonian and higher-order constraints are derived as follows:
\begin{equation}
\begin{split}
&H_{T}=H+\zeta^{\alpha}\Phi^{(1)}_{\alpha},\\
&H:=\frac{1}{2}(p_{3})^{2}+\frac{1}{2}(p_{4})^{2}-q^{1}p_{3}-\frac{1}{2}(q^{1}+2q^{2})(q^{1}+2q^{4}),\\
&\Phi^{(1)}_{1}:=\phi^{(1)}_{1}-\frac{1}{2}\phi^{(1)}_{2}=p_{1}-\frac{1}{2}(p_{2}-p_{3}+p_{4}):\approx0,\\
&\Phi^{(1)}_{2}:=\frac{1}{3}(\phi^{(1)}_{1}+\phi^{(1)}_{2})=\frac{1}{3}(p_{1}+p_{2}-p_{3}+p_{4}):\approx0
\end{split}
\label{}
\end{equation}
where $\zeta^{\alpha}$s are Lagrange multipliers, and
\begin{equation}
\begin{split}
&\dot{\Phi}^{(1)}_{1}=\{\Phi^{(1)}_{1},H_{T}\}\approx p_{3}\\
&\therefore \Phi^{(2)}_{1}:=p_{3}:\approx0,\\
&\dot{\Phi}^{(1)}_{2}=\{\Phi^{(1)}_{2},H_{T}\}\approx \frac{1}{3}p_{3}+q^{1}+q^{2}+q^{4}\\
&\therefore \Phi^{(2)}_{2}:=\frac{1}{3}p_{3}+q^{1}+q^{2}+q^{4}:\approx0.
\end{split}
\label{}
\end{equation}
The P.b.s among these constraints are $\{\Phi^{(2)}_{2},\Phi^{(1)}_{2}\}=1$ and otherwise vanish. That is, $\Phi^{(1)}_{1},\Phi^{(2)}_{1}$, and, $\Phi^{(1)}_{2},\Phi^{(2)}_{2}$ are classified into first-class constraint, and, second-class constraint, respectively. The consistency condition for $\Phi^{(2)}_{1}$ is automatically satisfied and the Lagrange multiplier $\zeta^{1}$ remains arbitrary. For $\Phi^{(2)}_{2}$, the consistency condition determines a Lagrange multiplier $\zeta^{2}$ as $-p_{4}$ in the weak equality. Therefore, the degrees of freedom of the system is $(2\times4-2-2\times2)/2=1$. \hyperlink{theorem01}{Theorem 1} and its proof indicate that the system has a canonical transformation from the original coordinates to the ones such that a part of the entire canonical coordinates is composed as linear combinations of the first- and second-class constraints. In fact, the symplectic 2-form of the system is computed as follows:
\begin{equation}
\omega=dq^{i}\wedge dp_{i}=d\Xi^{1}\wedge d\Psi_{1}+d\Xi^{2}\wedge d\Psi_{2}+d\Theta^{1}\wedge d\Theta_{1}+dQ^{1}\wedge dP_{1}
\label{}
\end{equation}
where each variable is defined as follows:
\begin{equation}
\begin{split}
&\Xi^{1}:=2q^{1}+\frac{2}{3}p_{3}-q^{2}-q^{4}, \Xi^{2}:=q^{3}+\frac{1}{3}p_{1}+q^{2},\\
&\Psi_{1}:=\Phi^{(1)}_{1}=\frac{1}{3}p_{1}-\frac{1}{6}(p_{2}-p_{3}+p_{4}), \Psi_{2}:=\Phi^{(2)}_{1}=p_{3},\\
&\Theta^{1}:=\Phi^{(2)}_{2}=\frac{1}{3}p_{3}+q^{1}+q^{2}+q^{4}, \Theta_{1}:=\Phi^{(1)}_{2}=\frac{1}{3}(p_{1}+p_{2}-p_{3}+p_{4}),\\
&Q^{1}:=q^{2}-q^{4}, P_{1}:=\frac{1}{2}(p_{2}-p_{3}-p_{4}).
\end{split}
\label{}
\end{equation}
The P.b.s among these variables are $\{\Xi^{1},\Psi_{1}\}=1$, $\{\Xi^{2},\Psi_{2}\}=1$, $\{\Theta^{1},\Theta_{1}\}=1$, $\{Q^{1},P_{1}\}=1$, and otherwise vanish. Notice that the second equality in the equation of $\omega$ is the strong equality, not weak equality. (See \hyperlink{remark01}{Remark 1}.) Then the total Hamiltonian is transformed as follows:
\begin{equation}
H_{T}=\frac{1}{2}(P_{1})^{2}+\frac{1}{2}(Q^{1})^{2}+\Psi_{1}P_{1}+\zeta^{1}\Psi_{1}+f(\Xi^{1},\Psi_{2},\Theta^{1})+g(\Psi_{1},\Psi_{2},\Theta^{1},\Theta_{1}),
\label{}
\end{equation}
where we set
\begin{equation}
\begin{split}
&f(\Xi^{1},\Psi_{2},\Theta^{1}):=-\frac{1}{18}(\Xi^{1})^{2}-\frac{1}{9}\Xi^{1}(-5\Theta^{1}+4\Psi_{2}),\\
&g(\Psi_{1},\Psi_{2},\Theta^{1},\Theta_{1}):=-\frac{1}{18}(5\Theta^{1}-\Psi_{2})^{2}+\frac{1}{2}(3\Theta_{1}-\Psi_{1})(\Theta_{1}-\Psi_{1})-\frac{1}{3}\Psi_{2}(\Theta^{1}-\Psi_{2}).
\end{split}
\label{}
\end{equation}
The system satisfies the Frobenius integrability condition $(\ref{integrability in 1st system in Hamilton formulation})$ under $(\ref{the number of constraints})$ in the section \ref{Sec03:02:02}. It implies that eight integral constants exist and of which the four constants are occupied by the consistency conditions for $\Psi_{1}$, $\Psi_{2}$, $\Theta^{1}$, and $\Theta_{1}$. 

There are a possible quasi-canonical embedding: $\tilde{\sigma}_{3}(t)$ and a possible canonical embedding: $\sigma_{3}(t)$, which are introduced in the section \ref{Sec04:03:04}. These embeddings, $\tilde{\sigma}_{3}(t)$ and $\sigma_{3}(t)$, occupy the four and six integral constants that are equivalent to the consistency conditions for $\Theta^{1}$, $\Theta_{1}$, $\Psi_{1}$, $\Psi_{2}$ and $\Theta^{1}$, $\Theta_{1}$, $\Psi_{1}$, $\Psi_{2}$, $\Xi^{1}$, $\Xi^{2}$, respectively. 

\subsubsection{The quasi-canonical embedding: $\tilde{\sigma}_{3}(t)$}\label{Sec05:03:01}
The pullback of total Lagrangian by $\tilde{\sigma}_{3}(t)$ is given as follows:
\begin{equation}
\begin{split}
L_{T}:=&\tilde{\sigma}^{*}_{3}(t)\left[P_{1}\dot{Q}^{1}+\Psi_{a}\dot{\Xi}^{a}+\Theta_{1}\dot{\Theta}^{1}-H_{T}\right]\\
=&\frac{d}{dt}\left[\Psi_{2}\Xi^{2}\right]+P_{1}\dot{Q}^{1}-H_{T}+{constant},
\end{split}
\label{}
\end{equation}
which is defined in the subspace $TM|_{Q,\dot{Q}}\times T^{*}M|_{Q,P}\times TM|_{\Xi,\dot{\Xi}}\times T^{*}M|_{\Xi,\Psi}\times \mathbb{R}$. Where we used $\tilde{\sigma}^{*}_{3}(t)\Psi_{1}=0$. Remark that $H_{T}$ above is defined in the symplectic submanifold $(T^{*}M|_{Q,P}\times T^{*}M|_{\Xi,\Psi}\times\mathbb{R},\omega_{Q,P}+\omega_{\Xi,\Psi}=dQ^{1}\wedge dP_{1}+d\Xi^{a}\wedge d\Psi_{a})$. Where we abbreviated the pullback operator $\tilde{\sigma}^{*}_{3}(t)$. The effective first-order variation of the action integral of $L_{T}$ is computed as follows:
\begin{equation}
\delta_{\text{effective}}\left(\tilde{\sigma}^{*}_{3}(t)I\right)=\int^{t_{2}}_{t_{1}}\left[-\dot{P}_{1}-Q^{1}\right]\delta Q^{1}dt+\left[-P_{1}+\dot{Q}^{1}\right]\delta P_{1}dt+\left[\Psi_{2}\delta\Xi^{2}+P_{1}\delta Q^{1}\right]^{t_{2}}_{t_{1}}
\label{}
\end{equation}
The appropriate boundary conditions are set as follows:
\begin{equation}
\delta Q^{1}(t_{2})=\delta Q^{1}(t_{1})=0
\label{B.C.3-2-1}
\end{equation}
and
\begin{equation}
\delta\Xi^{2}(t_{1})=0.
\label{B.C.3-2-2}
\end{equation}
Then the variational principle leads to the following equations:
\begin{equation}
\begin{split}
&-\dot{P}_{1}-Q^{1}=0,\\
&-P_{1}+\dot{Q}^{1}=0.
\end{split}
\label{EoM3-2}
\end{equation}

First, we can verify the following facts that $\kappa|_{\tilde{\sigma}_{3}(t)}:TM|_{Q,\dot{Q}}\rightarrow TM|_{Q,P};\dot{Q}^{1}\mapsto P_{1}$ and $\mathfrak{O}|_{\tilde{\sigma}_{3}(t)}:\mathfrak{O}[T(TM|_{Q,\dot{Q}}\times\mathbb{R})]\rightarrow\mathfrak{O}[T(T^{*}M|_{Q,P}\times\mathbb{R})];X_{t}=Q^{1}(\partial/\partial\dot{Q}^{1})+\dot{Q}^{1}(\partial/\partial Q^{1})+(\partial/\partial t)\mapsto{_{*}X}_{t}=(\partial H_{T}/\partial P_{1})(\partial/\partial Q^{1})-(\partial H_{T}/\partial Q^{1})(\partial/\partial P_{1})+(\partial/\partial t)$ are introduced in a well-defined manner and invertible, where $H_{T}=(Q^{1})^{2}/2+(P_{1})^{2}/2+f(\Xi^{1},\Psi_{2},\Theta^{1})+{constant}$ in $T^{*}M|_{Q,P}\times TM|_{\Xi,\Psi}\times\mathbb{R}$. This $H_{T}$ satisfies the Frobenius integrability condition $(\ref{integrability in 1st system in Hamilton formulation})$ in the section \ref{Sec03:02:02} under $(\ref{the number of constraints})$ in the section \ref{Sec03:02:01}. This indicates that the system has six integral constants. Second, we can also verify that the Frobenius integrable condition $(\ref{integrability in 1st system in Lagrange formulation})$ under $(\ref{the number of constraints})$ in the section \ref{Sec03:02:01} of $L_{T}$ is satisfied. That is, the kinetic matrix restricted to $TM|_{Q,\dot{Q}}\times TM|_{\Xi,\dot{\Xi}}\times \mathbb{R}$:
\begin{equation}
K^{(1)}=\begin{bmatrix}
1 & 0 & 0\\
0 & 0 & 0\\
0 & 0 & 0
\end{bmatrix}
\label{}
\end{equation}
leads to two zero-eigenvalue vectors: $\tau_{1}^{i}=(0,\tau^{2},0)$ and $\tau_{2}^{i}=(0,0,\tau^{3})$, where $\tau^{2}$ and $\tau^{3}$ are arbitrary functions in $TM|_{Q,\dot{Q}}\times TM|_{\Xi,\dot{\Xi}}\times \mathbb{R}$. $\eta^{i}$s are computed as follows: $\eta^{i}\approx(Q^{1},\eta^{2},\eta^{3})$, where $\eta^{2}$ and $\eta^{3}$ are arbitrary functions in $TM|_{Q,\dot{Q}}\times TM|_{\Xi,\dot{\Xi}}\times \mathbb{R}$. Therefore, the statement holds. This leads to the existence of six integral constants and of which the two constants are occupied by the consistency conditions for $\Psi_{i}$s or equivalently the embedding $\tilde{\sigma}_{3}(t)$ without $\Theta^{1}$ and $\Theta_{1}$ since we are now in the subspace $TM|_{Q,\dot{Q}}\times TM|_{\Xi,\dot{\Xi}}\times \mathbb{R}$. The three constants of the remaining others are determined by the boundary conditions. In fact, we can convince the result by solving the equations derived from the well-posed variational principle. The equations $(\ref{EoM3-2})$ describe a one-dimensional harmonic oscillator: $-\ddot{Q}^{1}-Q^{1}=0$ and the boundary condition $(\ref{B.C.3-2-1})$ determines all the integral constants in the solution: $Q^{1}(t)=A{\text{ exp}}(+it)+B{\text{ exp}}(-it)$. For $\Xi^{2}$, we have the equation $\dot{\Xi}^{2}=-4\Xi^{1}/9+2\Theta^{1}/9+5\Psi_{2}/9\approx-4\Xi^{1}/9$ and the solution of this equation occupies one integral constant. Therefore, the invertible map $\iota$ is $\iota|_{\tilde{\sigma}_{3}(t)}:M|_{Q,\Xi}[t_{1}]\times M|_{Q}[t_{2}]\rightarrow C|_{\tilde{\sigma}_{3}(t)};(Q^{1}(t_{1}),\Xi^{2}(t_{1}),Q^{1}(t_{2}))\mapsto(A,B,C)$ with $A=(Q^{1}(t_{1}){\text{exp}}(it_{2})-Q^{1}(t_{2}){\text{exp}}(it_{1}))/2i{\text{sin}(t_{2}-t_{1})}$, $B=(Q^{1}(t_{2}){\text{exp}}(-it_{1})-Q^{1}(t_{1}){\text{exp}}(-it_{2}))/$\\$2i{\text{sin}(t_{2}-t_{1})}$, and $C=\Xi^{2}(t_{1})$. The remaining one integral constant is assigned for $\dot{\Xi}^{1}=-\zeta^{1}+P_{1}-2\Theta_{1}+\Psi_{1}\approx-\zeta^{1}+P_{1}$; this constant does not determine until a gauge fixing condition is imposed. Remark that these equations for $\Xi^{1}$ and $\Xi^{2}$ are considered in the original space: the target space of the embedding $\tilde{\sigma}_{3}(t)$.

\subsubsection{The canonical embedding: $\sigma_{3}(t)$}\label{Sec05:03:02}
First of all, we find an appropriate Lagrange multiplier $\zeta^{1}$; $\Xi^{1}$ is static if and only if $\zeta^{1}\approx-P_{1}$. The pullback of total Lagrangian by $\sigma_{3}(t)$ is given as follows:
\begin{equation}
\begin{split}
L_{T}:=&\sigma^{*}_{3}(t)\left[P_{1}\dot{Q}^{1}+\Psi_{a}\dot{\Xi}^{a}+\Theta_{1}\dot{\Theta}^{1}-H_{T}\right]\\
=&P_{1}\dot{Q}^{1}-H_{T}+{constant}
\end{split}
\label{}
\end{equation}
in the subspace $TM|_{Q,\dot{Q}}\times T^{*}M|_{Q,P}\times\mathbb{R}$. Remark that $H_{T}$ above is defined in the symplectic submanifold $(T^{*}M|_{Q,P}\times\mathbb{R},\omega_{Q,P}=dQ^{1}\wedge dP_{1})$. Where we abbreviated the pullback operator $\sigma^{*}_{3}(t)$. The first-order variation is computed as follows:
\begin{equation}
\delta(\sigma^{*}_{3}(t)I)=\int^{t_{2}}_{t_{1}}\left[-\dot{P}_{1}-Q^{1}\right]\delta Q^{1}dt+\left[\dot{Q}^{1}-P_{1}\right]\delta P_{1}dt+\left[P_{1}\delta Q^{1}\right]^{t_{2}}_{t_{1}}
\label{}
\end{equation}
where we used the pullback of $\zeta^{1}\approx-P_{1}$ by $\sigma_{3}(t)$: $\sigma^{*}_{3}(t)\zeta^{1}=-P_{1}+{constant}$. Then the appropriate boundary conditions is set as follows:
\begin{equation}
\delta Q^{1}(t_{2})=\delta Q^{1}(t_{1})=0.
\label{B.C.3-4}
\end{equation}
Then the variational principle leads to the following equations:
\begin{equation}
\begin{split}
&-\dot{P}_{1}-Q^{1}=0,\\
&\dot{Q}^{1}-P_{1}=0.
\end{split}
\label{}
\end{equation}
Under this construction, we can verify the facts that $\kappa|_{\sigma_{3}(t)}$ and $\mathfrak{O}|_{\sigma_{3}(t)}$ are the same to the case of $\tilde{\sigma}_{3}(t)$ but $H_{T}$ turns into $H_{T}=(Q^{1})^{2}/2+(P_{1})^{2}/2+{constant}$ in $T^{*}M|_{Q,P}\times\mathbb{R}$. Then this system is Frobenius integrable as mentioned in the section \ref{Sec03:01}. This indicates that two integral constants exist and which are determined by the boundary conditions. In fact, combining the equations derived from the well-posed variational principle, we get the equation for a one-dimensional harmonic oscillator. That is, the invertible map $\iota$ is $\iota|_{\sigma_{3}(t)}:M|_{Q}[t_{1}]\times M|_{Q}[t_{2}]\rightarrow C|_{\sigma_{3}(t)};(Q^{1}(t_{1}),Q^{1}(t_{2}))\mapsto(A,B)$ with $A=(Q^{1}(t_{1}){\text{exp}}(it_{2})-Q^{1}(t_{2}){\text{exp}}(it_{1}))/2i{\text{sin}(t_{2}-t_{1})}$ and $B=(Q^{1}(t_{2}){\text{exp}}(-it_{1})-Q^{1}(t_{1}){\text{exp}}(-it_{2}))/2i{\text{sin}(t_{2}-t_{1})}$.

\subsection{A system with second-order time derivatives}\label{Sec05:04}
Let us consider the following system:
\begin{equation}
L_{4}=-\frac{1}{2}q\ddot{q}-\frac{1}{2}q^{2}.
\label{}
\end{equation}
This model is a modification of the model described by $L=-q\ddot{q}/2$ in Ref.~\cite{DyerHinterbichler2009}. The authors introduced it for the purpose of revealing the relations between boundary conditions and counter-terms. Also see Ref.~\cite{KKKM}. For this system, applying the consideration given in Sect.~\ref{Sec02:02:02}, there exists the counter-term $W$ given as follows:
\begin{equation}
W=\frac{1}{2}q\dot{q}+C(q),
\label{}
\end{equation}
where $C(q)$ is an arbitrary function of $q$. Then $L_{4}$ becomes as follows:
\begin{equation}
L'_{4}=L_{4}+\frac{dW}{dt}=\frac{1}{2}(\dot{q})^{2}-\frac{1}{2}q^{2}+\frac{\partial C}{\partial q}\dot{q}.
\label{}
\end{equation}
This is none other than the Lagrangian for a one-dimensional harmonic oscillator. In fact, the first-order variation of $L'_{4}$:
\begin{equation}
\delta I'=\int^{t_{2}}_{t_{1}}(-\ddot{q}-q)\delta qdt+\frac{d}{dt}\left[\left(\dot{q}+\frac{\partial C}{\partial q}\right)\delta q\right]^{t_{2}}_{t_{1}}
\label{}
\end{equation}
and the well-posed variational principle under the Dirichlet boundary condition $\delta q(t_{2})=\delta q(t_{1})=0$ leads to an equation: $-\ddot{q}-q=0$. The solution is, of course, $q(t)=A{\text{exp}}(+it)+B{\text{exp}}(-it)$, where $A,B$ are integral constants that are determined by the boundary conditions.

In this section, for this trivial system, we try to apply the two different methodologies which are briefly mentioned in Sect.~\ref{Sec02:03}, to introduce the well-defined variational principle without any counter-term. That is, (i) the methodology introduced by Sato, Sugano, Ohta, and Kimura \cite{SatoSuganoOhtaKimura1989,SatoSuganoOhtaKimura1989-2}, let us call it the {\it{"SSOK method"}}, and (ii) the methodology introduced by Pons \cite{Pons1989}, let us call it the {\it{"Pons method"}}; both methodologies are based on Refs.~\cite{Ostrogradsky1850,Woodard2015,Dirac1950,Dirac1958}. The former method is already applied in Sect.~\ref{Sec02:03} and the latter method is explained briefly in Sect.~\ref{Sec03:03}.

\subsubsection{The analysis by the SSOK method}\label{Sec05:04:01}
The configuration space, denote $M$, in this method is coordinated by $Q_{(1)}:=q$ and $Q_{(2)}:=\dot{q}$. Then the corresponding canonical momenta are given as follows: $P^{(1)}:=\partial L_{4}/\partial Q_{(2)}-(d/dt)(\partial L_{4}/\dot{Q}_{(2)})=Q_{(2)}/2$ and $P^{(2)}:=\partial L_{4}/\dot{Q}_{(2)}=-Q_{(1)}/2$. Therefore, the rank of the kinetic matrix $K^{(2)}=\partial P^{(2)}/\partial \dot{Q}_{(2)}$ is zero, and there is a primary constraint: $\phi^{(1)}:=P^{(2)}+Q_{(1)}/2:\approx0$. Then the total Hamiltonian is $H_{T}=P^{(1)}Q_{(2)}+P^{(2)}\dot{Q}_{(2)}-L_{4}+\zeta\phi^{(1)}=P^{(1)}Q_{(2)}+(Q_{(1)})^{2}/2+\zeta\phi^{(1)}$, where $\zeta$ is a Lagrange multiplier. This Legendre transformation is, in particular, called Ostrogradski transformation\cite{Ostrogradsky1850,Woodard2015}. The consistency condition for $\phi^{(1)}$ generates a secondary constraint: $\phi^{(2)}:=-P^{(1)}+Q_{(2)}/2:\approx0$, but it expected from the definition of $P^{(1)}$\cite{SatoSuganoOhtaKimura1989,SatoSuganoOhtaKimura1989-2}. The consistency condition for $\phi^{(2)}$ determines the Lagrange multiplier $\zeta$ as $Q_{(1)}$ in the weak equality. Therefore, the procedure stops here. The P.b.s between $\phi^{(1)}$ and $\phi^{(2)}$ is $\{\phi^{(1)},\phi^{(2)}\}=-1$ and otherwise vanish. This system has two second-class constraints, and this indicates that we have to use the canonical embedding $\sigma_{2}(t)$ given in Sect.~\ref{Sec04:03:03} to introduce the well-posed variational principle. 

The symplectic 2-form is computed as follows: $\omega=dQ_{(i)}\wedge dP^{(i)}=d\Theta^{1}\wedge d\Theta_{1}+dQ\wedge dP$, where $\Theta^{1}=\phi^{(2)}$, $\Theta_{1}=\phi^{(1)}$, $Q=P^{(1)}+Q_{(2)}/2$, and $P=P^{(2)}-Q_{(1)}/2$. The P.b.s are $\{\Theta^{1},\Theta_{1}\}=1$, $\{Q,P\}=1$, and otherwise vanish. The original variables $Q_{(1)}$, $Q_{(2)}$, $P^{(1)}$, and $P_{(2)}$ are expressed by using the transformed variables as follows: $Q_{(1)}=\Theta^{1}+Q$, $Q_{(2)}=\Theta_{1}-P$, $P^{(1)}=(Q-\Theta^{1})/2$, and $P^{(2)}=(P+\Theta_{1})/2$. Then the total Hamiltonian is transformed as follows: $H_{T}=P^{2}/2+Q^{2}/2-2\Theta_{1}P-(\Theta^{1})^{2}/2+3(\Theta_{1})^{2}/2$ and is defined in the symplectic manifold $(T^{*}M|_{Q,P}\times T^{*}M|_{\Theta}\times\mathbb{R},\omega)$. The pullback of $H_{T}$ by $\sigma_{2}(t)$ is $\sigma^{*}_{2}(t)H_{T}=P^{2}/2+Q^{2}/2-2\Theta_{1}P+{constant}$ in the symplectic submanifold $(T^{*}M|_{Q,P}\times\mathbb{R},\sigma^{*}_{2}(t)\omega=dQ\wedge dP)$, and this is a Frobenius integrable system. That is, there are two integral constants. The pullback of total Lagrangian is $L_{T}:=\sigma^{*}_{2}(t)[P\dot{Q}+\Theta_{1}\dot{\Theta}^{1}-H_{T}]=P\dot{Q}-P^{2}/2-Q^{2}/2-2\Theta_{1}P+{constant}$ in the subspace $TM|_{Q,\dot{Q}}\times T^{*}M|_{Q,P}\times\mathbb{R}$. Therefore, the first-order variation of the action integral is given as follows:
\begin{equation}
\delta(\sigma^{*}_{2}(t)I)=\int^{t_{2}}_{t_{1}}\left[-\dot{P}-Q\right]\delta Qdt+\left[\dot{Q}-P-2\Theta_{1}\right]\delta Pdt+\left[P\delta Q\right]^{t_{2}}_{t_{1}}.
\label{}
\end{equation}
Under the boundary conditions:
\begin{equation}
\delta Q(t_{2})=\delta Q(t_{1})=0,
\label{}
\end{equation}
the well-posed variational principle $\delta(\sigma^{*}_{2}(t)I):=0$ leads to the equations of motion: $-\dot{P}-Q=0$ and $\dot{Q}-P-2\Theta_{1}=0$. That is, using $\sigma^{*}_{2}(t)\Theta_{1}={constant}$, $-\ddot{Q}-Q=0$; this is none other than the equation of a one-dimensional harmonic oscillator. Remark, here, that $L_{T}$ is now defined in the subspace $TM|_{Q,\dot{Q}}\times T^{*}M|_{Q,P}\times\mathbb{R}\simeq TM|_{Q,\dot{Q}}\times\mathbb{R}$ and Frobenius integrable. In fact, the solution is $Q(t)=A{\text{exp}}(+it)+B{\text{exp}}(-it)$, and this includes the two integral constants: $A$ and $B$. The boundary conditions fix these constants. 

Finally, the maps $\kappa$, $\mathfrak{O}$, and $\iota$ are introduced as follows:
$\kappa|_{\sigma_{2}(t)}:TM|_{Q,\dot{Q}}\rightarrow TM|_{Q,P};\dot{Q}\mapsto P$ and $\mathfrak{O}|_{\sigma_{2}(t)}:\mathfrak{O}[T(TM|_{Q,\dot{Q}}\times\mathbb{R})]\rightarrow\mathfrak{O}[T(T^{*}M|_{Q,P}\times\mathbb{R})];X_{t}=Q(\partial/\partial\dot{Q})+\dot{Q}(\partial/\partial Q)+(\partial/\partial t)\mapsto{_{*}X}_{t}=(\partial H_{T}/\partial P)(\partial/\partial Q)-(\partial H_{T}/\partial Q)(\partial/\partial P)+(\partial/\partial t)$ are introduced in a well-defined manner and invertible, where $H_{T}=P^{2}/2+Q^{2}/2-2\Theta_{1}P+{constant}$ in $T^{*}M|_{Q,P}\times TM|_{\Xi,\Psi}\times\mathbb{R}$. $\iota$ is $\iota|_{\sigma_{2}(t)}:M|_{Q}[t_{1}]\times M|_{Q}[t_{2}]\rightarrow C|_{\sigma_{3}(t)};(Q(t_{1}),Q(t_{2}))\mapsto(A,B)$ with $A=(Q(t_{1}){\text{exp}}(it_{2})-Q(t_{2}){\text{exp}}(it_{1}))/2i{\text{sin}(t_{2}-t_{1})}$ and $B=(Q(t_{2}){\text{exp}}(-it_{1})-Q(t_{1}){\text{exp}}(-it_{2}))/$\\$2i{\text{sin}(t_{2}-t_{1})}$.

\subsubsection{The analysis by the Pons method}\label{Sec05:04:02}
The original Lagrangian $L_{4}$ is represented by using a Lagrange multiplier $\lambda$ as follows:
\begin{equation}
L''_{4}=-\frac{1}{2}q\dot{x}-\frac{1}{2}q^{2}+\lambda(x-\dot{q}).
\label{}
\end{equation}
Regarding $\lambda$ also as a position coordinate of the configuration space, the canonical momenta are derived as follows: $p=-\lambda$, $y=-q/2$, and $\pi=0$. Therefore, the rank of the kinetic matrix $K^{(1)}_{ij}=\partial p_{i}/\partial \dot{q}^{j}$ (where $p_{1}:=p,p_{2}:=y,p_{3}=\pi,q^{1}:=q,q^{2}:=x,q^{3}:=\lambda$) is zero, and there are three primary constraints: $\phi^{(1)}_{1}:=p+\lambda:\approx0$, $\phi^{(1)}_{2}:=y+q/2:\approx0$, and $\phi^{(1)}_{3}:=\pi:\approx0$. The total Hamiltonian is computed as follows: $H_{T}:=q^{2}/2-\lambda x+\zeta^{a}\phi^{(1)}_{a}$, where $\zeta^{\alpha}$s are Lagrange multipliers and $a=1,2,3$. The consistency conditions for $\phi^{(1)}_{a}$ become as follows: $\dot{\phi}^{(1)}_{1}\approx-q-\zeta^{2}/2+\zeta^{3}:\approx0$, $\dot{\phi}^{(1)}_{2}\approx\lambda+\zeta^{1}/2:\approx0$ and $\dot{\phi}^{(1)}_{3}\approx x-\zeta^{1}:\approx0$; there is a secondary constraint: $\phi^{(2)}:=\lambda+x/2$ where we used $\zeta^{1}\approx x$. The consistency condition for $\phi^{(2)}$ restricts the relation between $\zeta^{2}$ and $\zeta^{3}$, and this determines all the multipliers as follows: $\zeta^{2}\approx-q$ and $\zeta^{3}\approx q/2$. The P.b.s among the constraints are computed as follows: $\{\phi^{(1)}_{1},\phi^{(1)}_{2}\}=-1/2$, $\{\phi^{(1)}_{1},\phi^{(1)}_{3}\}=1$, $\{\phi^{(2)},\phi^{(1)}_{2}\}=1/2$, $\{\phi^{(2)},\phi^{(1)}_{3}\}=1$, and otherwise vanish. This system has four second-class constraints, and this indicates that we have to use the canonical embedding $\sigma_{2}(t)$ given in Sect.~\ref{Sec04:03:03} to introduce the well-posed variational principle. 

The symplectic 2-form is computed as follows: $\omega=dq\wedge dp+dx\wedge dy+d\lambda\wedge d\pi=d\Theta^{1}\wedge d\Theta_{1}+d\Theta^{2}\wedge d\Theta_{2}+dQ\wedge dP$, where $\Theta^{1}:=\phi^{(2)}-\phi^{(1)}_{1}$, $\Theta_{1}:=\phi^{(1)}_{2}$, $\Theta^{2}:=(\phi^{(2)}+\phi^{(1)}_{1})/2$, $\Theta_{2}:=\phi^{(1)}_{3}$, $Q:=q/2-y+\pi/2$, and $P:=p+x/2$. Then the total Hamiltonian is transformed as follows: $H_{T}=Q^{2}/2+P^{2}/2-(\Theta^{1})^{2}/2$ in the symplectic manifold $(T^{*}M|_{Q,P}\times T^{*}M|_{\Theta}\times\mathbb{R},\omega)$. The pullback of $H_{T}$ by $\sigma_{2}(t)$ is $\sigma^{*}_{2}(t)H_{T}=Q^{2}/2+P^{2}/2+{constant}$ in the symplectic submanifold $(T^{*}M|_{Q,P}\times\mathbb{R},\sigma^{*}_{2}(t)\omega=dQ\wedge dP)$, and this is a Frobenius integrable system. That is, there are two integral constants. The pullback of total Lagrangian is $L_{T}:=\sigma^{*}_{2}(t)[P\dot{Q}+\Theta_{a}\dot{\Theta}^{a}-H_{T}]=P\dot{Q}-Q^{2}/2-P^{2}/2+{constant}$ in the subspace $TM|_{Q,\dot{Q}}\times T^{*}M|_{Q,P}\times\mathbb{R}$. Therefore, the first-order variation of the action integral is given as follows:
\begin{equation}
\delta(\sigma^{*}_{2}(t)I)=\int^{t_{2}}_{t_{1}}\left[-\dot{P}-Q\right]\delta Qdt+\left[\dot{Q}-P\right]\delta P+\left[P\delta Q\right]^{t_{2}}_{t_{1}}.
\label{}
\end{equation}
Under the boundary conditions
\begin{equation}
\delta Q(t_{2})=\delta Q(t_{1})=0,
\label{}
\end{equation}
the well-posed variational principle $\delta(\sigma^{*}_{2}(t)I):=0$ leads to the equations of motion, resolving as an equation $-\ddot{Q}-Q=0$; this is none other than the equation of a one-dimensional harmonic oscillator. Remark, here, that $L_{T}$ is now defined in the subspace $TM|_{Q,\dot{Q}}\times T^{*}M|_{Q,P}\times\mathbb{R}\simeq TM|_{Q,\dot{Q}}\times\mathbb{R}$ and Frobenius integrable. In fact, the solution is $Q(t)=A{\text{exp}}(+it)+B{\text{exp}}(-it)$, and this includes the two integral constants: $A$ and $B$. The boundary conditions fix these constants. The maps $\kappa$, $\mathfrak{O}$, and $\iota$ are the same to the previous case by just replacing $H_{T}=P^{2}/2+Q^{2}/2-2\Theta_{1}P+{constant}$ by $H_{T}=P^{2}/2+Q^{2}/2+{constant}$.

\section{Summary}\label{Sec06}
In this paper, we constructed a methodology to make the variational principle well-posed in degenerate point particle systems. 

When we applied the variational principle, it was generically possible to consider the first-order variation with respect not only to configurations but also to higher-order time derivative variables. However, when taking into account the compatibility of Lagrange mechanics with Newtonian dynamics, the possible variables for the variation were restricted only to the configurations of a given system. This indicated that position-fixing boundary conditions were necessary for the variational principle to lead to Euler-Lagrange equations even if containing higher-order time derivative terms. In addition, Hamilton-Dirac analysis revealed the stability of higher-order time derivative systems being compatible with Newtonian dynamics: there is no Ostrogradski's instability. 

On the ground of this framework, we investigated the Frobenius integrability conditions for each Lagrange and Hamilton formulation. In particular, we introduced the three fundamental maps: $\iota$, $\kappa$, and $\mathfrak{O}$. Map $\iota$ connected the integral constants in the solutions to the boundary conditions for the variational principle. Maps $\kappa$ and $\mathfrak{O}$ described the correspondence between Lagrange and Hamilton formulation. Armed with these ingredients, we represented the difficulties of making the variational principle well-posed and formulated a set of problems. To resolve these problems, we needed to construct a subspace of the original phase space in which the dynamics lives, the symplectic structure holds, and all the maps $\iota$, $\kappa$, and $\mathfrak{O}$ restricted in this subspace have to be well-defined and invertible. We achieved the purpose by introducing a set of embeddings, canonical and quasi-canonical embeddings, that extract subspaces diffeomorphic to the constraint subspace. A novel theorem with its explicit proof, which states the existence of constraint coordinates, played a fundamental role in this consideration. Applying these embeddings, we resolve the problems. Finally, we applied the methodology to examples. 

Let us summarize the methodology in the following steps. One can use the following methodology;\\
{\it{(1). For a given system, just performing Hamilton-Dirac analysis, reveal the constraint structure. \\
(2). Construct constraint coordinates referring to the proof of Theorem 1 ( or Lemma 1 and/or 2). Then, computing the symplectic 2-form, find a new canonical coordinate system which is indicated by Theorem 1 ( or Lemma 1 and/or 2). Then select a suitable embedding (see Table \ref{embeddings}).
\begin{table}[t]
\centering
\caption{Embeddings for each type of system. {\it{"1st-class system"}} means a system with 1st-class constraint(s). Others are defined in the same manner.}
\begin{tabular}{|l||l|l|l|} \hline
Gauge fixing & 1st-class system & 2nd-class system & 1st- and 2nd-class system \\ \hline \hline
Yes & $\sigma_{1}(t)$ & - & $\sigma_{3}(t)$ \\ \hline
No (or no gauge d.o.f) & $\tilde{\sigma}_{1}(t)$ & $\sigma_{2}(t)$ & $\tilde{\sigma}_{3}(t)$ \\ \hline 
\end{tabular}
\label{embeddings}
\end{table}\\
(3). Consider the pullback of the Legendre transformation of the total Hamiltonian by the selected embedding: the pullback of the total Lagrangian by the selected embedding. \\
-(i) \ If one uses $\sigma_{\tau}(t)$ $(\tau=1,2,3)$, take the first-order variation of its action integral, and just fix the emerged configurations in the boundary term at both end-points. Then the variational principle becomes well-posed. \\
-(ii) If one uses $\tilde{\sigma}_{\tilde{\tau}}(t)$ $(\tilde{\tau}=1,3)$, take the first-order variation of its action integral, under the assumption that the pullback of primary first-class constraint coordinates by $\tilde{\sigma}_{\tilde{\tau}}(t)$ is set to be zero in advance. Then fix the configurations for the physical degrees of freedom at both end-points and the configurations which correspond to higher-order (more than secondary) first-class constraint coordinates at either end-point. Then the variational principle becomes well-posed.}}\\
Remark that, in the case (3)-(ii), we {\it{cannot}} fix the configurations corresponding to {\it{primary}} first-class constraint coordinates on the boundaries; otherwise the boundary conditions becomes over-imposing. To remove this difficulty out, we have to fix the gauge degrees of freedom.

In a previous work\cite{KKKM}, which is established based only on the compatibility of the first-order variation of the action integral to the equations of motion, the well-posed variational principle requires us to fix all configurations on the boundaries that correspond only to the physical degrees of freedom, regardless of the presence of first-class constraints. However, the presented work indicates that configurations corresponding to higher-order (more than secondary) first-class constraints must also be fixed on either end-point. This represents a difference from the previous work and arises from the fact that the previous work did not consider how to determine the integral constants, which are implied by the Frobenius integrability, through boundary conditions, as is assumed in the presented work. 

For future works, mathematical properties of the three fundamental maps $\iota$, $\kappa$, and $\mathfrak{O}$ should be investigated. In particular, revealing the detailed features of the map $\iota$ is important to get a deeper understanding of boundary conditions. The same applies for the canonical embeddings: $\sigma_{\tau}(t)$ and the quasi-canonical embeddings: $\tilde{\sigma}_{\tilde{\tau}}(t)$. In particular, since gauge transformations generically gives rise to some surface terms~\cite{SuganoSaitoKimura1986,SuganoKagraokaKimura1992,SuganoKimura1990,SuganoKagraoka1991,SuganoKagraoka1991-2,SuganoKimura1990-2}, this would affect the determination of the boundary conditions; the quasi-canonical embbedings would be restricted. From the aspects of practical applications for modern physics, the methodology should be extended to field theories. In particular, applications for gravitation are important. For instance, gravitational phenomena for which we cannot neglect boundaries such as black hole physics need to consider appropriate boundary conditions for introducing some counter-term including the so-called Gibbons-Hawking-York term~\cite{York1972,GibbonsHawking1977,York1986,HawkingHorowitz1996,CharpNelson1983,MukhopadhyayPadmanabhan2006,Padmanabhan2014,ParattuChakrabortyMajhiPadmanabhan2016,ParattuChakrabortyPadmanabhan2016,KrishnanRaju2017,Chakraborty2017,DeruelleMerinoOlea2018,OshitaYi-Peng2017,Diego2021,Diego2021-2,GuarnizoCastaneda2010,KhodabakhshiShojaiShirzad2018,PadillaSivanesan2012}, as mentioned also in the previous work\cite{KKKM}. Further, introducing correct counter-terms would play a crucial role in the absence of acausality in higher-order derivative systems as mentioned briefly in Sect.~\ref{Sec02:04}, anti-de Sitter/conformal field thoery (AdS/CFT) correspondence~\cite{EmparanJohnson1999}, and Chern-Simons theory\cite{GallardoMontesinos2011}. We would expect that the methodology gives a new perspective on modern physics. 

\section*{Acknowledgments}
I would like to thank Keisuke Izumi, Keigo Shimada, Mu-In Park, Masahide Yamaguchi, Paolo Gondolo, and Shin'ichi Hirano for their insightful and fruitful discussions. KT is supported by the Tokyo Tech Fund Hidetoshi Kusama Scholarship.
\nocite{*}

\begin{thebibliography}{999}
\bibitem{Teitelboim1992}
M. Henneaux and C. Teitelboim, ISBN 9780691037691, Princeton University Press (1992).
\bibitem{DyerHinterbichler2009}
E. Dyer, K. Hinterbichler, Phys. Rev. D {\textbf{79}}, 024028 (2009) [arXiv:0809.4033 [gr-qc]].
\bibitem{Cawley1979}
R. Cawley, Phys. Rev. Lett. {\textbf{42}}, 413(1979).
\bibitem{Dirac1928}
P.M.A.Dirac, Proceedings of the Royal Society A: Mathematical, Physical and Engineering Sciences. {\textbf{117}}(778) 610 (1928).
\bibitem{Dirac1950}
P. M. A. Dirac, Can.\ J.\ Math. {\bf{2}}, 129 (1950).
\bibitem{Dirac1958}
P. M. A. Dirac, Proc.\ R.\ Soc.\ London Ser. A {\bf{246}}, 326 (1958).
\bibitem{Bergmann1949}
P. G. Bergmann, Phys.\ Rev., {\bf{75}}, 680 (1949).
\bibitem{BergmannBrunings1949}
P. G. Bergmann and J. H. M. Brunings, Rev.\ Mod.\ Phys., {\bf{21}}, 480 (1949).
\bibitem{Bergmann1950}
P. G. Bergmann and R. Penfield and R. Schiller and H. Zatzkis, Phys.\ Rev., {\bf{80}}, 81 (1950).
\bibitem{AndersonBergmann1951}
J. L. Anderson and P. G. Bergmann, Phys.\ Rev., {\bf{83}}, 1018 (1951).
\bibitem{York1972}
J. W. York, Phys.\ Rev.\ Lett. {\bf{28}}, 1082 (1972).
\bibitem{GibbonsHawking1977}
G. W. Gibbons and S. W. Hawking, Phys.\ Rev.\ D {\bf{15}}, 2752 (1977).
\bibitem{York1986}
J. W. York, Found.\ Phys. {\bf{16}}, 249-257 (1986).
\bibitem{HawkingHorowitz1996}
S. W. Hawking and Gary T. Horowitz, Class.\ Quant.\ Grav. {\bf{13}}, 1487-1498 (1996) [arXiv:gr-qc/9501014].
\bibitem{Einstein1916}
A. Einstein, Sitzungsber.\ Preuss.\ Akad.\ Wiss.\ Berlin\ (Math.Phys), 1111-1116 (1916).
\bibitem{Lovelock1971}
D. Lovelock, J.\ Math.\ Phys. {\bf{12}}, 498 (1971).
\bibitem{HehlMcCreaMielkeNeeman1995}
F. W. Hehl, J. D. McCrea, E. W. Mielke, Y. Ne’eman, Phys. Rept., {\bf{258}}, 1-171 (1995).
\bibitem{Rosen1940}
N. Rosen, Phys. Rev. {\bf{57}}, 147 (1940).
\bibitem{Rosen1940-2}
N. Rosen, Phys. Rev. {\bf{57}}, 150 (1940).
\bibitem{HassanRosen2012}
S. F. Hassan, Rachel A. Rosen, JHEP {\bf{02}} 126 (2012) [arXiv:1109.3515 [hep-th]].
\bibitem{Buchdahl1970}
H. A. Buchdahl, Monthly Notices of the Royal Astronomical Society, {\bf{150}}, 1-8 (1970).
\bibitem{SotiriouFaraoni2010}
T. P. Sotiriou and V. Faraoni, Rev.\ Mod.\ Phys. {\bf{82}}, 451 (2010) [arXiv:0805.1726 [gr-qc]].
\bibitem{Horndeski1974}
G. W. Horndeski, Int.\ J.\ Theor.\ Phys. {\bf{10}}, 363 (1974).
\bibitem{DeffayetGaoSteerZahariade2011}
C. Deffayet and X. Gao and D. A. Steer and G. Zahariade, Phys.\ Rev.\ D {\bf{84}} 064039 (2011) [arXiv:1103.3260 [hep-th]].
\bibitem{KobayashiYamaguchiYokoyama2011}
T. Kobayashi and M. Yamaguchi and J. Yokoyama, Prog.\ Theor.\ Phys. {\bf{126}}, 511-529 (2011) [arXiv:1105.5723 [hep-th]].
\bibitem{GleyzesLangloisPiazzaVernizzi2015}
J. Gleyzes and D. Langlois and F. Piazza and F. Vernizzi, Phys.\ Rev.\ Lett. {\bf{21}}, 211101, 114 (2015).
\bibitem{GleyzesLangloisPiazzaVernizzi2015-2}
J. Gleyzes and D.Langlois and F.Piazza and F.Vernizzi, JCAP {\bf{02}}, 018 (2015) [arXiv:1408.1952 [astro-ph.CO]].
\bibitem{LangloisNoui2016}
D. Langlois and K. Noui, JCAP {\bf{02}}, 034 (2016) [arXiv:1510.06930 [gr-qc]].
\bibitem{AchourCrisostomiKoyamaLangloisNouiTasuato2016}
J. B. Achour and M. Crisostomi and K. Koyama and D. Langlois and K. Noui and G. Tasinato, JHEP {\bf{12}}, 100 (2016) [arXiv:1608.08135 [hep-th]].
\bibitem{DeFeliceLangloisMukohyamaNouiWang2018}
A. De Felice, D. Langlois, S. Mukohyama, K. Noui, A. Wang, Phys. Rev. D {\bf{98}}, 084024 (2018).
\bibitem{Moffat2006}
J. W. Moffat, JCAP {\bf{03}}, 004 (2006) [arXiv:gr-qc/0506021].
\bibitem{Ostrogradsky1850}
M. V. Ostrogradski, Mem.\ Acad.\ St.\ Petersbourg {\bf{IV 4}}, 385 (1850).
\bibitem{Woodard2015}
R. P. Woodard, Scholarpedia, {\bf{10}}, 32243 (2015) [arXiv:1506.02210 [hep-th]].
\bibitem{SatoSuganoOhtaKimura1989}
Y. Saito and R. Sugano and T. Ohta and T. Kimura, J.\ Math.\ Phys. {\bf{30}}, 1122 (1989).
\bibitem{SatoSuganoOhtaKimura1989-2}
Y. Saito and R. Sugano and T. Ohta and T. Kimura, J.\ Math.\ Phys. {\bf{34}}, 3775 (1993).
\bibitem{Pons1989}
J. M. Pons, Lett.\ in\ Math.\ Phys., {\textbf{17}}, 181-189 (1989).
\bibitem{KKKM}
K. Izumi, K. Shimada, K. Tomonari and M. Yamaguchi, [arXiv:2303.03805 [hep-th]] (2023).
\bibitem{SuganoKamo1982}
R. Sugano and H. Kamo, Prog.\ Theor.\ Phys. {\bf{67}}, 1966 (1982).
\bibitem{Fitzpatrick2021}
R. Fitzpatrick, ISBN 9781032046624, CRC Press (2021).
\bibitem{MotohashiSuyama2015}
H. Motohashi and T. Suyama, Phys.\ Rev.\ D {\bf{91}}, 085009 (2015) [arXiv:1411.3721 [physics.class-ph]].
\bibitem{MotohashiNouriSuyamaYamaguchiLanglois2016}
H. Motohashi and K. Noui and T. Suyama and M. Yamaguchi and D. Langlois, JCAP {\bf{07}}, 033 (2016) [arXiv:1603.09355 [hep-th]].
\bibitem{MotohashiSuyamaYamaguchi2018}
H. Motohashi, T. Suyama and M. Yamaguchi, J.\ Phys.\ Soc.\ Jpn. {\bf{87}}, 063401 (2018) [arXiv:1711.08125 [hep-th]].
\bibitem{MotohashiSuyamaYamaguchi2018-2}
H. Motohashi, T. Suyama and M. Yamaguchi, JHEP {\bf{06}}, 133 (2018) [arXiv:1804.07990 [hep-th]].
\bibitem{Dirac1938}
P. M. A. Dirac, Proc. Roy. Soc. A{\bf{167}}, 148(1938).
\bibitem{Kuti1993}
K. Jansen, J. Kuti and C. Liu, Nucl. Phys. B Proc. Suppl. {\bf{30}}, 681-684(1993).
\bibitem{Shanmugadhasan1973}
S. Shanmugadhasan, J. Math. Phys. {\textbf{14}}, 677 (1973).
\bibitem{DominiciGomis1980}
D. Dominici and J. Gomis, J. Math. Phys. {\textbf{21}}, 2124 (1980).
\bibitem{Dominici1982}
D. Dominici, J. Math. Phys. {\textbf{23}}, 256 (1982).
\bibitem{MaskawaNakajima1976}
T. Maskawa and H. Nakajima, Prog.\ Theor.\ Phys. {\bf{56}}, 1295 (1976).
\bibitem{Eisenhart2003}
L. P. Eisenhart, ISBN 0486495256, Dover Publications (2003).
\bibitem{SuganoSaitoKimura1986}
R. Sugano and Y. Saito and T. Kimura, Prog.\ Theor.\ Phys. {\bf{76}}, 283 (1986).
\bibitem{SuganoKagraokaKimura1992}
R. Sugano and Y. Kagraoka and T. Kimura, J.\ Math.\ Phys. {\bf{A7}}, 62 (1992).
\bibitem{BrownModel2022}
J. D. Brown, Universe {\textbf{8(3)}}, 171 (2022) [arXiv:2201.06558 [gr-qc]].
\bibitem{Frenkel1980}
A. Frenkel, Phys. Rev. D {\textbf{21}}, 2986(1980).
\bibitem{SuganoKimura1983}
R. Sugano, T. Kimura, Prog. Theor. Phys. {\textbf{69}}, 1241 (1983).
\bibitem{SuganoKimura1990}
R. Sugano and T. Kimura, Phys. Rev. D {\textbf{41}}, 1247 (1990).
\bibitem{SuganoKagraoka1991}
R. Sugano and Y. Kagraoka, Z. Phys. C Particles and Fields {\textbf{52}}, 437 (1991).
\bibitem{SuganoKagraoka1991-2}
R. Sugano and Y. Kagraoka, Z. Phys. C Particles and Fields {\textbf{52}}, 443 (1991).
\bibitem{SuganoKimura1990-2}
R. Sugano and T. Kimura, J. Math. Phys. {\textbf{31}}, 2337 (1990)
\bibitem{CharpNelson1983}
J. M. Charap and J. E. Nelson, J.\ Phys.\ A {bf{16}}, 1661 (1983).
\bibitem{MukhopadhyayPadmanabhan2006}
A. Mukhopadhyay and T. Padmanabhan, Phys.\ Rev.\ D {\bf{74}}, 124023 (2006) [arXiv:hep-th/0608120].
\bibitem{Padmanabhan2014}
T. Padmanabhan, Nod.\ Phys.\ Lett.\ A {\bf{29(08)}}, 1450037 (2014).
\bibitem{ParattuChakrabortyMajhiPadmanabhan2016}
K. Parattu and S. Chakraborty and B. R. Majhi and T. Padmanabhan, Gen.\ Rel.\ Grav. {\bf{48(7)}}, 94 (2016) [arXiv:1501.01053 [gr-qc]].
\bibitem{ParattuChakrabortyPadmanabhan2016}
K. Parattu, S. Chakraborty, T. Padmanabhan, Eur. Phys. J. C {\bf{76}}, 129 (2016) [arXiv:1602.07546 [gr-qc]].
\bibitem{KrishnanRaju2017}
C. Krishnan and A. Raju, Mod.\ Phys.\ Lett.\ A {\bf{32(14)}}, 1750077 (2017).
\bibitem{Chakraborty2017}
S. Chakraborty, Fundam.\ Theor.\ Phys. {\bf{187}}, 43-59 (2017) [arXiv:1607.05986 [gr-qc]].
\bibitem{DeruelleMerinoOlea2018}
N. Deruelle, N. Merino, and R. Olea, Phys. Rev. D {\bf{97(10)}}, 104009 (2018) [arXiv:1709.06478 [gr-qc]].
\bibitem{OshitaYi-Peng2017}
N. Oshita and Yi-Peng Wu, Phys.\ Rev.\ D {\bf{96(4)}}, 044042 (2017).
\bibitem{Diego2021}
Diego S\'aez-Chill\'on G\'omez, Phys. Lett. B {\bf{814}}, 136103 (2021) [arXiv:2011.11568 [gr-qc]].
\bibitem{Diego2021-2}
Diego S\'aez-Chill\'on G\'omez, Phys. Rev. D, {\bf{104(2)}}, 024029 (2021) [arXiv:2103.16319 [gr-qc]].
\bibitem{GuarnizoCastaneda2010}
A. Guarnizo, L. Castaneda, and J. M.
Tejeiro, Gen. Rel. Grav. {\bf{42}}, 2713–2728 (2010) [arXiv:1002.0617 [gr-qc]].
\bibitem{KhodabakhshiShojaiShirzad2018}
H. Khodabakhshi, F. Shojai, and A. Shirzad, Eur. Phys. J. C {\bf{78(12)}}, 1003 (2018) [arXiv:1803.04306 [gr-qc]].
\bibitem{PadillaSivanesan2012}
A. Padilla and V. Sivanesan, JHEP {\bf{08}}, 122 (2012) [arXiv:1206.1258 [gr-qc]].
\bibitem{EmparanJohnson1999}
R. Emparan, C. V. Johnson, and R. C. Myers, Phys. Rev. D {\bf{60}},104001 (1999) [arXiv:hep-th/9903238].
\bibitem{GallardoMontesinos2011}
A. Gallardo and M. Montesinos, J. Phys. A {\bf{44}}, 135402 (2011)[ arXiv:1008.4883 [hep-th]].
\end{thebibliography}

\end{document}